\documentclass[10pt]{IEEEtran}
\usepackage[LGR,T1]{fontenc}
\usepackage[latin9]{inputenc}
\usepackage{color}
\usepackage{array}
\usepackage{units}
\usepackage{amsmath}
\usepackage{amsthm}
\usepackage{amssymb}
\usepackage{graphicx}

\makeatletter

\providecommand{\tabularnewline}{\\}

\theoremstyle{plain}
\newtheorem{thm}{\protect\theoremname}
\theoremstyle{definition}
\newtheorem{defn}{\protect\definitionname}
\theoremstyle{remark}
\newtheorem{rem}{\protect\remarkname}
\theoremstyle{plain}
\newtheorem{cor}{\protect\corollaryname}
\theoremstyle{plain}
\newtheorem{lem}{\protect\lemmaname}
\theoremstyle{plain}
\newtheorem{prop}{\protect\propositionname}

\usepackage{color}

\usepackage{centernot}

\usepackage{cancel}

\usepackage{cite}

\usepackage[mathscr]{eucal}
\usepackage{fixltx2e}
\usepackage{array}
\usepackage{verbatim}
\usepackage{bm}
\usepackage{algorithmic}
\usepackage{algorithm}
\usepackage{verbatim}
\usepackage{textcomp}
\usepackage{mathrsfs}
\usepackage{epstopdf}

\newcommand{\openone}{\leavevmode\hbox{\small1\normalsize\kern-.33em1}}

\catcode`~=11 \def\UrlSpecials{\do\~{\kern -.15em\lower .7ex\hbox{~}\kern .04em}} \catcode`~=13

\allowdisplaybreaks[4]

\newcommand{\calB}{\mathcal{B}}

\newcommand{\calM}{\mathcal{M}}

\newcommand{\calP}{\mathcal{P}}

\newcommand{\calX}{\mathcal{X}}

\DeclareMathAlphabet{\mathbsf}{OT1}{cmss}{bx}{n}
\DeclareMathAlphabet{\mathssf}{OT1}{cmss}{m}{sl}

\DeclareSymbolFont{bsfletters}{OT1}{cmss}{bx}{n}
\DeclareSymbolFont{ssfletters}{OT1}{cmss}{m}{n}
\DeclareMathSymbol{\bsfGamma}{0}{bsfletters}{'000}
\DeclareMathSymbol{\ssfGamma}{0}{ssfletters}{'000}
\DeclareMathSymbol{\bsfDelta}{0}{bsfletters}{'001}
\DeclareMathSymbol{\ssfDelta}{0}{ssfletters}{'001}
\DeclareMathSymbol{\bsfTheta}{0}{bsfletters}{'002}
\DeclareMathSymbol{\ssfTheta}{0}{ssfletters}{'002}
\DeclareMathSymbol{\bsfLambda}{0}{bsfletters}{'003}
\DeclareMathSymbol{\ssfLambda}{0}{ssfletters}{'003}
\DeclareMathSymbol{\bsfXi}{0}{bsfletters}{'004}
\DeclareMathSymbol{\ssfXi}{0}{ssfletters}{'004}
\DeclareMathSymbol{\bsfPi}{0}{bsfletters}{'005}
\DeclareMathSymbol{\ssfPi}{0}{ssfletters}{'005}
\DeclareMathSymbol{\bsfSigma}{0}{bsfletters}{'006}
\DeclareMathSymbol{\ssfSigma}{0}{ssfletters}{'006}
\DeclareMathSymbol{\bsfUpsilon}{0}{bsfletters}{'007}
\DeclareMathSymbol{\ssfUpsilon}{0}{ssfletters}{'007}
\DeclareMathSymbol{\bsfPhi}{0}{bsfletters}{'010}
\DeclareMathSymbol{\ssfPhi}{0}{ssfletters}{'010}
\DeclareMathSymbol{\bsfPsi}{0}{bsfletters}{'011}
\DeclareMathSymbol{\ssfPsi}{0}{ssfletters}{'011}
\DeclareMathSymbol{\bsfOmega}{0}{bsfletters}{'012}
\DeclareMathSymbol{\ssfOmega}{0}{ssfletters}{'012}

\DeclareMathOperator{\supp}{supp}

\def\esssup{{\rm ess\,sup}}

\allowdisplaybreaks[1]
\providecommand{\corollaryname}{Corollary}

\providecommand{\lemmaname}{Lemma}
\providecommand{\remarkname}{Remark}
\providecommand{\theoremname}{Theorem}

\providecommand{\corollaryname}{Corollary}
\providecommand{\definitionname}{Definition}
\providecommand{\lemmaname}{Lemma}
\providecommand{\remarkname}{Remark}
\providecommand{\theoremname}{Theorem}

\providecommand{\corollaryname}{Corollary}
\providecommand{\definitionname}{Definition}
\providecommand{\lemmaname}{Lemma}
\providecommand{\remarkname}{Remark}
\providecommand{\theoremname}{Theorem}

\providecommand{\corollaryname}{Corollary}
\providecommand{\definitionname}{Definition}
\providecommand{\lemmaname}{Lemma}
\providecommand{\remarkname}{Remark}
\providecommand{\theoremname}{Theorem}

\makeatother

\providecommand{\corollaryname}{Corollary}
\providecommand{\definitionname}{Definition}
\providecommand{\lemmaname}{Lemma}
\providecommand{\propositionname}{Proposition}
\providecommand{\remarkname}{Remark}
\providecommand{\theoremname}{Theorem}

\begin{document}
\title{On Exact and $\infty$-R\'enyi Common Informations }
\author{Lei Yu and Vincent Y. F. Tan, \IEEEmembership{Senior Member,~IEEE}
\thanks{ This work was supported by a Singapore Ministry of Education Tier 2 Grant (R-263-000-C83-112). This paper was presented in part at the  2019 IEEE International Symposium on Information Theory (ISIT) \cite{yu2019exact_isit}.} \thanks{ L.~Yu is with the Department of Electrical and Computer Engineering, National University of Singapore (NUS), Singapore 117583 (e-mail: leiyu@nus.edu.sg). V.~Y.~F.~Tan is with the Department of Electrical and Computer Engineering and the Department of Mathematics, NUS, Singapore 119076 (e-mail: vtan@nus.edu.sg).} \thanks{ Communicated by M. Raginsky, Associate Editor for Probability and Statistics. } \thanks{Copyright (c) 2019 IEEE. Personal use of this material is permitted. However, permission to use this material for any other purposes must be obtained from the IEEE by sending a request to pubs-permissions@ieee.org.} }
\maketitle
\begin{abstract}
Recently, two extensions of Wyner's common information\textemdash exact
and R\'enyi common informations\textemdash were introduced respectively
by Kumar, Li, and El Gamal (KLE), and the present authors. The class
of common information problems involves determining the minimum rate
of the common input to two independent processors needed to exactly
or approximately generate a target joint distribution. For the exact
common information problem, exact generation of the target distribution
is required, while for Wyner's and $\alpha$-R\'enyi common informations,
the relative entropy and R\'enyi divergence with order $\alpha$ were
respectively used to quantify the discrepancy between the synthesized
and target distributions. The exact common information is larger than
or equal to Wyner's common information. However, it was hitherto unknown
whether the former is strictly larger than the latter for some joint
distributions. In this paper, we first establish the equivalence between
the exact and $\infty$-R\'enyi common informations, and then provide
single-letter upper and lower bounds for these two quantities. For
doubly symmetric binary sources, we show that the upper and lower
bounds coincide, which implies that for such sources, the exact and
$\infty$-R\'enyi common informations are completely characterized.
Interestingly, we observe that for such sources, these two common
informations are strictly larger than Wyner's. This answers an open
problem posed by KLE. Furthermore, we extend Wyner's, $\infty$-R\'enyi,
and exact common informations to sources with countably infinite
or continuous alphabets, including Gaussian sources. 
\end{abstract}


\section{\label{sec:Introduction}Introduction}

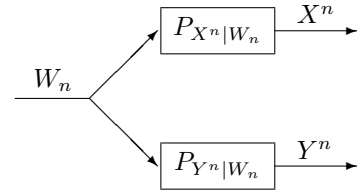
\begin{figure}
\centering\centering \setlength{\unitlength}{0.05cm} 
{ \begin{picture}(100,60) 
\put(5,30){\line(1,0){20}} \put(25,30){\vector(1,1){18}}
\put(25,30){\vector(1,-1){18}} \put(44,42){\framebox(30,12){$P_{X^{n}|W_{n}}$}}
\put(44,6){\framebox(30,12){$P_{Y^{n}|W_{n}}$}} \put(74,48){\vector(1,0){22}}
\put(74,12){\vector(1,0){22}} \put(10,33){%
\mbox{%
$W_{n}$%
}} \put(80,50){%
\mbox{%
$X^{n}$%
}} \put(80,14){%
\mbox{%
$Y^{n}$%
}} \end{picture}}

\label{fig:dss}

\caption{The distributed source simulation problem. For the exact common information
problem, the discrete random variable $W_{n}$ can be arbitrarily
distributed, but for the R\'enyi common information problem, it is restricted
to be uniformly distributed. Hence for the latter case, we use $M_{n}$
to denote the common randomness, in place of the $W_{n}$.}
\end{figure}

How much common randomness is needed to simulate two correlated sources
in a distributed fashion? This problem (depicted in Fig. \ref{fig:dss}),
termed {\em distributed source simulation}, was first studied by
Wyner \cite{Wyner}, who used the normalized relative entropy (Kullback-Leibler
divergence or KL divergence) to measure the discrepancy or ``distance''
between the simulated joint distribution and the joint distribution
of the original correlated sources $\pi_{XY}$. He defined the minimum
rate needed to ensure that the normalized relative entropy vanishes
asymptotically as the \emph{common information} (denoted as $T_{1}\left(\pi_{XY}\right)$)
between the sources $\pi_{XY}$. He also established a single-letter
characterization for the common information, i.e., the common information
between correlated sources $X$ and $Y$  is 
\begin{align}
T_{1}\left(\pi_{XY}\right) & =C_{\mathsf{Wyner}}(\pi_{XY})\\
 & :=\min_{P_{W}P_{X|W}P_{Y|W}:\,P_{XY}=\pi_{XY}}I(XY;W).\label{eqn:CWyner}
\end{align}
For Gray-Wyner's source coding problem subject to the condition that
the total rate of all three messages is $H_{\pi}(XY)$ (the joint
entropy of the correlated sources $\pi_{XY}$), the quantity in \eqref{eqn:CWyner}
was also used to characterize the minimum rate of the common message
\cite{Wyner}.

Recently, the present authors \cite{yu2018wyner,yu2018corrections}
introduced the notion of $\alpha$-R\'enyi common information with $\alpha\in[0,\infty]$,
which is defined as the minimum common rate when the KL divergence
is replaced by more general divergences \textemdash{} the family of
R\'enyi divergences with order $\alpha\in[0,\infty]$. When $\alpha=1$,
R\'enyi common information reduces to Wyner's common information. We
proved that for R\'enyi divergences of order $\alpha\in(0,1]$, the
minimum rate needed to guarantee that the (normalized and unnormalized)
R\'enyi divergences vanish asymptotically is equal to Wyner's common
information. However, for R\'enyi divergences of order $\alpha\in(1,2]$,
we only provided lower and upper bounds. Numerical results show that
our lower and upper bounds coincide for doubly symmetric binary sources
(DSBSes), and for this case, both of them are strictly larger than
Wyner's common information. Furthermore, the common information with
approximation error measured by the total variation (TV) distance
is also equal to Wyner's common information \cite{Cuff,Kumar,yu2018wyner};
and exponential achievability and converse results for this case were
established in \cite{Hayashi06,Cuff,yu2018wyner}.

Kumar, Li, and El Gamal (KLE) \cite{Kumar} extended Wyner's common
information in a different way. They assumed variable-length codes
and exact generation of the correlated sources $(X,Y)\sim\pi_{XY}$,
instead of block codes and approximate simulation of $\pi_{XY}$ as
assumed by Wyner \cite{Wyner} and by us \cite{yu2018wyner,yu2018corrections}.
For such exact generation problem, KLE \cite{Kumar} characterized
the minimum common rate, coined \emph{exact common information}, by
\begin{equation}
T_{\mathrm{Exact}}(\pi_{XY}):=\lim_{n\to\infty}\frac{1}{n}G(\pi_{XY}^{n}).
\end{equation}
where the common entropy 
\begin{equation}
G(\pi_{XY}):=\min_{P_{W}P_{X|W}P_{Y|W}:\,P_{XY}=\pi_{XY}}H(W).\label{eq:commonentropy}
\end{equation}
The exact common information is no smaller than Wyner's common information.
However, it was previously unknown whether they are equal for all
sources $\pi_{XY}$. Even for simple sources, e.g., DSBSes, the exact
common information was still unknown. It is worth noting that the
quantities $G(\pi_{XY})$ and $T_{\mathrm{Exact}}(\pi_{XY})$ were
first considered by Witsenhausen in 1976 \cite[p. 331]{witsenhausen1976values}.
In \cite{witsenhausen1976values}, Witsenhausen studied properties
of Wyner's common information. He provided an example \cite[p. 331]{witsenhausen1976values}
(in the framework of Gray-Wyner's source coding problem \cite{Wyner})
for which Wyner's common information can be attained by a one-shot
coding scheme (i.e., block coding with $n\ge2$ is unnecessary), and
at the same time, zero  error is realized by this one-shot scheme.
For this example, he showed that
\begin{equation}
G(\pi_{XY})=C_{\mathsf{Wyner}}(\pi_{XY}),\label{eq:wrong-1}
\end{equation}
which suggests that one may avoid block coding and also attain zero
error.  In order to better understand the relation between $G(\pi_{XY})$
and $C_{\mathsf{Wyner}}(\pi_{XY})$ for an arbitrary $\pi_{XY}$ (not
specified to that example), Witsenhausen stated the following relation
between $T_{\mathrm{Exact}}(\pi_{XY})$ and $C_{\mathsf{Wyner}}(\pi_{XY})$
for arbitrary $\pi_{XY}$ with finite support:
\begin{equation}
T_{\mathrm{Exact}}(\pi_{XY})\stackrel{ ?  }{=} C_{\mathsf{Wyner}}(\pi_{XY}).\label{eq:wrong}
\end{equation}
However, he did not provide a proof for \eqref{eq:wrong}. In this
paper, we first completely characterize the exact common information
for DSBSes, and then show that for this class of sources, the exact
common information is strictly larger than Wyner's common information.
This implies \eqref{eq:wrong} does not always hold.  Furthermore,
sufficient conditions for \eqref{eq:wrong} to hold (i.e., for equality
of Wyner's common information and the exact common information) were
investigated in \cite{vellambi2016sufficient}.

The exact common information  for \emph{continuous} sources was studied
by Li and El Gamal \cite{li2017distributed}. In \cite{li2017distributed},
Li and El Gamal adopted dyadic decomposition schemes to construct
a discrete common random variable $W$ for continuous random variables
with log-concave probability density functions (pdfs). By using such
schemes, they established the first known upper bound on the exact
common information for continuous sources. Specifically, for a pair
of correlated sources $\left(X,Y\right)\sim\pi_{XY}$ with a log-concave
pdf, they showed that 
\begin{align}
I_{\pi}(X;Y) & \le T_{\mathrm{Exact}}(\pi_{XY})\\
 & \leq G(\pi_{XY})\\
 & \leq I_{\pi}(X;Y)+24\log2\textrm{ nats/symbol},\label{eq:-150}
\end{align}
where $I_{\pi}(X;Y)$ denotes the mutual information between $\left(X,Y\right)\sim\pi_{XY}$.
 This result implies that the exact common information for continuous
sources with log-concave pdfs is finite. Furthermore, it is worth
noting that Li and El Gamal's dyadic decomposition scheme is a one-shot
scheme, i.e., it is valid for the case with blocklength equal to $1$.
For Gaussian sources with correlation coefficient $\rho\in[0,1),$
Li and El Gamal's upper bound in \eqref{eq:-150} reduces to 
\begin{equation}
\frac{1}{2}\log\left[\frac{1}{1-\rho^{2}}\right]+24\log2\textrm{ nats/symbol}.\label{eq:Gaussian2-1}
\end{equation}
It is known that $T_{\mathrm{Exact}}(\pi_{XY})\geq C_{\mathsf{Wyner}}(\pi_{XY})$
\cite{Kumar} and for joint Gaussian sources, $C_{\mathrm{Wyner}}(\pi_{XY})=\frac{1}{2}\log\left[\frac{1+\rho}{1-\rho}\right]$
\cite{xu2013wyner,yu2016generalized}. Hence for joint Gaussian sources,
$T_{\mathrm{Exact}}(\pi_{XY})\geq\frac{1}{2}\log\left[\frac{1+\rho}{1-\rho}\right]$.
Note that there is a large gap between this lower bound and Li and
El Gamal's upper bound in \eqref{eq:Gaussian2-1}. In this paper,
we prove a new upper bound on $T_{\mathrm{Exact}}(\pi_{XY})$ which
is at most $0.72$  bits/symbol larger than the lower bound $\frac{1}{2}\log\left[\frac{1+\rho}{1-\rho}\right]$
and hence much tighter than Li and El Gamal's upper bound, albeit
with the use of a block coding scheme.

\begin{table*}
\caption{\label{tab:Summary-of-Various}Summary of Various Common Informations.}
\centering

\begin{tabular}{|>{\centering}p{0.12\textwidth}|>{\centering}m{0.12\textwidth}|>{\centering}m{0.16\textwidth}|>{\raggedright}m{0.48\textwidth}|}
\hline 
Com. Inf. & Fixed or Variable-Length & \multicolumn{1}{>{\centering}p{0.16\textwidth}|}{Exact or Approx.} & Expressions for Various Alphabets\tabularnewline
\hline 
\hline 
Wyner's CI \cite{Wyner} & Fixed & Approx. ($\frac{1}{n}D$) & 1) Finite $\left(\mathcal{X},\mathcal{Y}\right)$: $C_{\mathsf{Wyner}}(\pi_{XY})$
\cite{Wyner} \newline  2) Countably Infinite/Continuous$^{*}$ $\left(\mathcal{X},\mathcal{Y}\right)$:
$C_{\mathsf{Wyner}}(\pi_{XY})$ {[}P, Cor. \ref{cor:countable}, \ref{cor:continuous}{]}\tabularnewline
\hline 
$\alpha$-R\'enyi CI, $\alpha\in[0,2]$ \cite{yu2018wyner,yu2018corrections} & Fixed & Approx. ($\frac{1}{n}D_{\alpha}$ or $D_{\alpha}$) & Finite $\left(\mathcal{X},\mathcal{Y}\right)$: $\begin{cases}
0 & \alpha=0\\
C_{\mathsf{Wyner}}(\pi_{XY}) & \alpha\in(0,1]\\
\left[\Gamma_{\alpha}^{\mathrm{LB}}(\pi_{XY}),\Gamma_{\alpha}^{\mathrm{UB}}(\pi_{XY})\right] & \alpha\in(1,2]
\end{cases}$ \cite{yu2018wyner,yu2018corrections}\tabularnewline
\hline 
$\infty$-R\'enyi CI {[}P{]} & Fixed & Approx. ($\frac{1}{n}D_{\infty}$ or $D_{\infty}$) & 1) Equivalent for Finite and Countably Infinite/Continuous$^{*}$
$\left(\mathcal{X},\mathcal{Y}\right)$ {[}P, Thm. \ref{thm:equivalence}{]}.
\newline  2) Multiletter Expressions: $\lim_{n\to\infty}\frac{1}{n}G(\pi_{XY}^{n})$
\cite{Kumar} and $\lim_{n\to\infty}\frac{1}{n}\Gamma(\pi_{XY}^{n})$
(for finite $\left(\mathcal{X},\mathcal{Y}\right)$) {[}P, Thm. \ref{thm:equivalence}{]}.\tabularnewline
\cline{1-3} \cline{2-3} \cline{3-3} 
Exact CI \cite{Kumar} & Variable & Exact ($P_{X^{n}Y^{n}}=\pi_{XY}^{n}$) & 3) Singleletter Bounds: $\left[C_{\mathsf{Wyner}}(\pi_{XY}),G(\pi_{XY})\right]$
\cite{Kumar}; $\left[\Gamma^{\mathrm{LB}}(\pi_{XY}),\Gamma^{\mathrm{UB}}(\pi_{XY})\right]$
(for finite $\left(\mathcal{X},\mathcal{Y}\right)$) {[}P, Thm. \ref{thm:singleletter}{]}.
\newline  4) Gaussian Sources: $\left[\frac{1}{2}\log\frac{1+\rho}{1-\rho},\frac{1}{2}\log\frac{1}{1-\rho^{2}}+24\log2\right]$
\cite{Kumar,li2017distributed}; $\left[\frac{1}{2}\log\frac{1+\rho}{1-\rho},\frac{1}{2}\log\frac{1+\rho}{1-\rho}+\frac{\rho}{1+\rho}\right]$
{[}P, Thm. \ref{thm:Gaussian}{]}.\tabularnewline
\hline 
G\'acs-K\"orner's CI \cite{gacs1973common} & Fixed & Approx. ($\mathbb{P}\left(f\left(X^{n}\right)\neq g\left(Y^{n}\right)\right)\leq\varepsilon$
and $f\left(X^{n}\right),g\left(Y^{n}\right)$ almost uniform on $\left[1:e^{nR}\right]$) & Finite $\left(\mathcal{X},\mathcal{Y}\right)$: For $\left(X^{n},Y^{n}\right)\sim\pi_{XY}^{n}$
and any $\varepsilon\in(0,1)$, $\max_{f,g}R=C_{\mathrm{GK}}(\pi_{XY}):=\max_{\hat{f},\hat{g}:\hat{f}(X)=\hat{g}(Y)}H(\hat{f}(X))$
\cite{gacs1973common}\tabularnewline
\hline 
A Variant of G\'acs-K\"orner's CI \cite{gacs1973common} & Variable & Exact ($f\left(X^{n}\right)=g\left(Y^{n}\right)$ a.s.) & Finite $\left(\mathcal{X},\mathcal{Y}\right)$: For $\left(X^{n},Y^{n}\right)\sim\pi_{XY}^{n}$,
$\max_{f,g}\frac{1}{n}H\left(f\left(X^{n}\right)\right)=\frac{1}{n}C_{\mathrm{GK}}(\pi_{XY}^{n})=C_{\mathrm{GK}}(\pi_{XY})$
\cite{gacs1973common}\tabularnewline
\hline 
\end{tabular}

\flushleft Here {[}P{]} refers to the present paper. ``$^{*}$''
means that some regular conditions are required. $\Gamma_{\alpha}^{\mathrm{LB}}(\pi_{XY})$
and $\Gamma_{\alpha}^{\mathrm{UB}}(\pi_{XY})$ were defined in \cite{yu2018corrections}.
In the rightmost box in the rows of ``$\infty$-R\'enyi CI'' and ``Exact
CI'', Points 1) and 2) hold for unnormalized version of $\infty$-R\'enyi
CI (i.e., with $D_{\infty}$ measure). Points 3) and 4) hold for both
normalized and unnormalized versions of $\infty$-R\'enyi CI.
\end{table*}

\subsection{Main Contributions}

Our contributions include the following aspects.
\begin{itemize}
\item We first consider sources with finite alphabets. We establish the
equivalence between the exact common information and $\infty$-R\'enyi
common information. We provide a multi-letter characterization for
the exact and $\infty$-R\'enyi common informations. Using this multi-letter
characterization, we derive single-letter upper and lower bounds. 
\item When specialized to DSBSes, the upper and lower bounds coincide. This
implies that the exact and $\infty$-R\'enyi common informations for
DSBSes are completely solved. Interestingly, we show that they are
both strictly larger than Wyner's common information. This solves
an open problem posed by Kumar, Li, and El Gamal \cite{Kumar}.
\item We extend the exact and $\infty$-R\'enyi common informations, and also
the relative entropy version and the TV distance version of Wyner's
common information to sources with general (countable or continuous)
alphabets, including Gaussian sources. We establish the equivalence
between the exact and $\infty$-R\'enyi common informations for such
general sources. We provide an upper bound on the exact and $\infty$-R\'enyi
common informations for Gaussian sources, which is at least 22.28
bits/symbol smaller than Li and El Gamal's bound \cite{li2017distributed}.
However, it is worth noting that theirs is a one-shot bound that is
obtained by a scheme with blocklength $1$, but ours is an asymptotic
one which requires the blocklength to tend to infinity. Furthermore,
we also completely characterize Wyner's common information for Gaussian
sources which is equal to $C_{\mathrm{Wyner}}(\pi_{XY})$ that was
computed in \cite{xu2013wyner,yu2016generalized} for Gaussian sources.
\item Concerning the innovations in our proofs, they rely on the so-called
\emph{mixture decomposition} or \emph{splitting technique}, which
was previously used in \cite{nummelin1978uniform,athreya1978new,roberts2004general,ho2010interplay,Kumar,vellambi2016sufficient}.
However, in this paper, we combine it with various truncation techniques
to deal with sources with countably infinite alphabets, and also combine
it with truncation, discretization, and Li and El Gamal's dyadic decomposition
techniques \cite{li2017distributed} to deal with sources with continuous
alphabets. Besides the mixture decomposition technique, a superblock
coding approach is also adopted to prove the equivalence between the
exact and $\infty$-R\'enyi common informations. Furthermore, as by-products
of our analyses, various lemmas are derived, e.g., the ``chain rule''
for coupling (Lemma \ref{lem:coupling}), the distributed R\'enyi-covering
lemmas (for sources with finite alphabets and Gaussian sources) (Lemmas
\ref{lem:Renyicovering} and \ref{lem:Renyicovering-1}), and a lemma
on the estimation of conditional mutual information (Lemma \ref{lem:conditionalMI}).
\end{itemize}

\subsection{\label{subsec:Notations}Notations}

We use $P_{X}$ to denote the probability distribution of a random
variable $X$. For brevity, we also use $P_{X}(x)$ to denote the
corresponding probability mass function (pmf) for discrete distributions,
and the corresponding probability density function (pdf) for continuous
distributions. This will also be denoted as $P(x)$ (when the random
variable $X$ is clear from the context). We also use $\pi_{X},\widetilde{P}_{X}$,
$\widehat{P}_{X}$ and $Q_{X}$ to denote various probability distributions
on alphabet $\mathcal{X}$. The set of probability measures on $\mathcal{X}$
is denoted as $\mathcal{P}\left(\mathcal{X}\right)$, and the set
of conditional probability measures on $\mathcal{Y}$ given a variable
in $\mathcal{X}$ is denoted as $\mathcal{P}(\mathcal{Y}|\mathcal{X}):=\left\{ P_{Y|X}:P_{Y|X}(\cdot|x)\in\mathcal{P}(\mathcal{Y}),x\in\mathcal{X}\right\} $.
Furthermore, the support of a distribution $P\in\calP(\calX)$ is
denoted as $\mathrm{supp}(P)=\{x\in\calX:P(x)>0\}$. For two distributions
$P$ and $Q$ defined on the same measurable space, we use $P\ll Q$
to denote that $P$ is absolutely continuous with respect to $Q$.
If $P\ll Q$, we use $\frac{\mathrm{d}P}{\mathrm{d}Q}$ to denote
the Radon\textendash Nikodym derivative of $P$ with respect to $Q$.

The TV distance between two probability mass functions $P$ and $Q$
with a common alphabet $\calX$ is defined as 
\begin{equation}
|P-Q|:=\frac{1}{2}\sum_{x\in\calX}|P(x)-Q(x)|.
\end{equation}

We use $T_{x^{n}}(x):=\frac{1}{n}\sum_{i=1}^{n}1\left\{ x_{i}=x\right\} $
to denote the type (empirical distribution) of a sequence $x^{n}$,
$T_{X}$ and $V_{Y|X}$ to respectively denote a type of sequences
in $\mathcal{X}^{n}$ and a conditional type of sequences in $\mathcal{Y}^{n}$
(given a sequence $x^{n}\in\calX^{n}$). For a type $T_{X}$, the
type class (set of sequences having the same type $T_{X}$) is denoted
by $\mathcal{T}_{T_{X}}$. For a conditional type $V_{Y|X}$ and a
sequence $x^{n}$, the $V_{Y|X}$\emph{-shell of $x^{n}$} (the set
of $y^{n}$ sequences having the same conditional type $V_{Y|X}$
given $x^{n}$) is denoted by $\mathcal{T}_{V_{Y|X}}(x^{n})$. For
brevity, sometimes we use $T(x,y)$ to denote the joint distributions
$T(x)V(y|x)$ or $T(y)V(x|y)$.

For $X\sim P_{X}$, we denote the \emph{entropy} of $X$ as 
\begin{equation}
H_{P}(X)=H(P_{X}):=-\sum_{x\in\mathrm{supp}(P_{X})}P_{X}(x)\log P_{X}(x).
\end{equation}
For $(X,Y)\sim P_{XY}$, we denote the \emph{conditional entropy}
of $X$ given $Y$ as 
\begin{align}
H_{P}(X|Y) & =H(P_{X|Y}|P_{Y})\\
 & :=-\sum_{x,y}P_{XY}(x,y)\log P_{X|Y}(x|y).
\end{align}
For $(X,Y)\sim P_{XY}$, we denote the \emph{mutual information }between
$X$ and $Y$ as 
\begin{align}
I_{P}(X;Y) & =H_{P}(X)-H_{P}(X|Y).
\end{align}
For brevity and when entropies, conditional entropies, and mutual
informations are computed respect to a distribution denoted by ``$P$'',
we omit the subscript and denote them respectively as $H(X)$, $H(X|W)$,
and $I(X;Y)$ instead of the more verbose $H_{P}(X)$, $H_{P}(X|W)$,
and $I_{P}(X;Y)$.

The $\epsilon$-strongly, $\epsilon$-weakly, and $\epsilon$-unified
typical sets \cite{Gamal,Cover,ho2010information,ho2010markov} of
$P_{X}$ are respectively denoted as 
\begin{align}
\mathcal{T}_{\epsilon}^{\left(n\right)}(P_{X}) & :=\big\{ x^{n}\in\mathcal{X}^{n}:\nonumber \\
 & \qquad\left|T_{x^{n}}(x)-P_{X}(x)\right|\leq\epsilon P_{X}(x),\forall x\in\mathcal{X}\big\},\label{eq:typ_set}\\
\mathcal{A}_{\epsilon}^{(n)}\left(P_{X}\right) & :=\big\{ x^{n}\in\mathcal{X}^{n}:\nonumber \\
 & \qquad\left|-\frac{1}{n}\log P_{X}^{n}\left(x^{n}\right)-H(P_{X})\right|\leq\epsilon\big\},\label{eq:}\\
\mathcal{U}_{\epsilon}^{(n)}\left(P_{X}\right) & :=\big\{ x^{n}\in\mathcal{X}^{n}:\nonumber \\
 & \qquad D\left(T_{x^{n}}\|P_{X}\right)+\left|H\left(T_{x^{n}}\right)-H\left(P_{X}\right)\right|\leq\epsilon\big\}.\label{eq:unified}
\end{align}
Note that $\mathcal{T}_{\epsilon}^{\left(n\right)}(P_{X})$ only applies
to sources with finite alphabets, and $\mathcal{U}_{\epsilon}^{(n)}(P_{X})$
applies to sources with countable alphabets. For $\mathcal{A}_{\epsilon}^{(n)}\left(P_{X}\right)$,
if $P_{X}$ is an absolutely continuous distribution, in \eqref{eq:},
$P_{X}^{n}\left(x^{n}\right)$ and $H(P_{X})$ are respectively replaced
with the corresponding pdf and differential entropy. The corresponding
jointly typical sets are defined similarly. The conditionally $\epsilon$-strongly
typical set of $P_{XY}$ is denoted as 
\begin{equation}
\mathcal{T}_{\epsilon}^{\left(n\right)}(P_{XY}|x^{n}):=\left\{ y^{n}\in\mathcal{Y}^{n}:(x^{n},y^{n})\in\mathcal{T}_{\epsilon}^{\left(n\right)}(P_{XY})\right\} ,
\end{equation}
and the conditionally $\epsilon$-weakly and $\epsilon$-unified typical
sets are defined similarly. For brevity, sometimes we write $\mathcal{T}_{\epsilon}^{\left(n\right)}(P_{X}),\mathcal{A}_{\epsilon}^{(n)}\left(P_{X}\right)$
and $\mathcal{U}_{\epsilon}^{(n)}\left(P_{X}\right)$ as $\mathcal{T}_{\epsilon}^{\left(n\right)},\mathcal{A}_{\epsilon}^{(n)}$
and $\mathcal{U}_{\epsilon}^{(n)}$, respectively.

For distributions $P_{X},Q_{X}\in\calP(\calX)$, the {\em relative
entropy} and the {\em R\'enyi divergence of order $1+s\in(0,1)\cup(1,\infty)$}
are respectively defined as\footnote{When the alphabet $\mathcal{X}$ is uncountable, it is understood
that $\frac{P_{X}}{Q_{X}}$ should be replaced by the Radon-Nikodym
derivative $\frac{\mathrm{d}P_{X}}{\mathrm{d}Q_{X}}$ of $P_{X}$
respect to $Q_{X}$. If $P_{X}$ is not absolutely continuous respect
to $Q_{X}$, then as assumed conventionally, the relative entropy
and the R\'enyi divergence of order $1+s\in(1,\infty)$ are defined
as $\infty$. (Note that for this case, the R\'enyi divergence of order
$1+s\in(0,1)$ is well-defined.)} 
\begin{align}
D(P_{X}\|Q_{X}) & :=\sum_{x\in\mathrm{supp}(P_{X})}P_{X}(x)\log\frac{P_{X}(x)}{Q_{X}(x)}\label{eq:-19}\\*
D_{1+s}(P_{X}\|Q_{X}) & :=\frac{1}{s}\log\sum_{x\in\mathrm{supp}(P_{X})}P_{X}(x)^{1+s}Q_{X}(x)^{-s},\label{eq:-40}
\end{align}
and the conditional versions are respectively defined as 
\begin{align}
D(P_{Y|X}\|Q_{Y|X}|P_{X}) & :=D(P_{X}P_{Y|X}\|P_{X}Q_{Y|X})\\*
D_{1+s}(P_{Y|X}\|Q_{Y|X}|P_{X}) & :=D_{1+s}(P_{X}P_{Y|X}\|P_{X}Q_{Y|X}),
\end{align}
where the summations in \eqref{eq:-19} and \eqref{eq:-40} are taken
over the elements in $\mathrm{supp}(P_{X})$. The R\'enyi divergence
of order $1+s\in\{0,\infty\}$ is defined by the continuous extensions
of $D_{1+s}$. The R\'enyi divergence of order $1$ is defined as $D_{1}(P_{X}\|Q_{X}):=\lim_{s\uparrow0}D_{1+s}(P_{X}\|Q_{X})$.
Throughout, $\log$ and $\exp$ are to the natural base $e$ and $s\geq-1$.
It is known that $D_{1}(P_{X}\|Q_{X})=D(P_{X}\|Q_{X})$ so a special
case of the R\'enyi divergence (or the conditional version) is the usual
relative entropy (or the conditional version). The R\'enyi divergence
of order $\infty$ satisfies 
\begin{align}
D_{\infty}(P_{X}\|Q_{X}) & =\log\sup_{x\in\mathrm{supp}(P_{X})}\frac{P_{X}(x)}{Q_{X}(x)}.
\end{align}
If $\mathcal{X}$ is a countable alphabet or $\mathbb{R}$ and we
replace $Q_{X}$ by respectively the counting or the Lebesgue measures,
then the R\'enyi divergence $D_{1+s}(P_{X}\|Q_{X})$ of order $1+s\in[0,\infty]$
reduces to the R\'enyi entropy $-H_{1+s}(P_{X})$ of the same order.

Denote the coupling sets of $(P_{X},P_{Y})$ and $(P_{X|W},P_{Y|W})$
respectively as 
\begin{align}
C(P_{X},P_{Y}) & :=\bigl\{ Q_{XY}\in\mathcal{P}(\mathcal{X}\times\mathcal{Y}):\nonumber \\
 & \qquad Q_{X}=P_{X},Q_{Y}=P_{Y}\bigr\}\\
C(P_{X|W},P_{Y|W}) & :=\bigl\{ Q_{XY|W}\in\mathcal{P}(\mathcal{X}\times\mathcal{Y}|\mathcal{W}):\nonumber \\
 & \qquad Q_{X|W}=P_{X|W},Q_{Y|W}=P_{Y|W}\bigr\}.\label{eq:-33}
\end{align}

For $i,j\in\mathbb{Z}$, and $i\le j$, we define $[i:j]:=\{i,i+1,\ldots,j\}$.
Given a number $a\in[0,1]$, we define $\overline{a}=1-a$. Define
$\left[x\right]^{+}=\max\left\{ x,0\right\} $. Denote $\mathcal{A}^{\mathsf{c}}$
as the complement of the set $\mathcal{A}$. Finally, we write $f(n)\sim g(n)$
if ${\displaystyle \lim_{n\to\infty}\nicefrac{f(n)}{g(n)}=1}$.

\section{Problem Formulations }

\subsection{\label{subsec:R=0000E9nyi-Common-Information}R\'enyi Common Information }

Consider the distributed source simulation setup depicted in Fig.
\ref{fig:dss}. Two terminals both have access to a uniformly distributed
common randomness $M_{n}$. Given a target distribution $\pi_{XY}$,
one of terminals uses $M_{n}$ and his own local randomness to generate
$X^{n}$ and the other one uses $M_{n}$ and his own local randomness
to generate $Y^{n}$ such that the the generated (or synthesized)
distribution $P_{X^{n}Y^{n}}$ is close to the product distribution
$\pi_{XY}^{n}$ under R\'enyi divergence measures. We wish to find the
limit on the least amount of common randomness satisfying such a requirement.
More specifically, given a target distribution $\pi_{XY}$, we wish
to minimize the alphabet size of a random variable $M_{n}$ that is
uniformly distributed over\footnote{For simplicity, we assume that $e^{nR}$ and similar expressions are
integers.} $\calM_{n}:=[1:e^{nR}]$ ($R$ is a positive number known as the
{\em rate}), such that the generated (or synthesized) distribution
\begin{align}
 & P_{X^{n}Y^{n}}(x^{n},y^{n})\nonumber \\
 & :=\frac{1}{|{\cal M}_{n}|}\sum_{m\in{\cal M}_{n}}P_{X^{n}|M_{n}}(x^{n}|m)P_{Y^{n}|M_{n}}(y^{n}|m)
\end{align}
forms a good approximation to the product distribution $\pi_{XY}^{n}$.
\begin{defn}
A \emph{fixed-length $(n,R)$-code} consists of  a pair of random
mappings $P_{X^{n}|W_{n}}:{\cal W}_{n}\to\mathcal{X}^{n},P_{Y^{n}|W_{n}}:{\cal W}_{n}\to\mathcal{Y}^{n}$
for some countable set ${\cal W}_{n}$ such that $\frac{1}{n}\log\left|{\cal W}_{n}\right|\le R$.
\end{defn}
In the R\'enyi common information problem \cite{yu2018wyner}, the unnormalized
R\'enyi divergence $D_{1+s}(P_{X^{n}Y^{n}}\|\pi_{XY}^{n})$ and the
normalized R\'enyi divergence $\frac{1}{n}D_{1+s}(P_{X^{n}Y^{n}}\|\pi_{XY}^{n})$
are adopted to measure the discrepancy between $P_{X^{n}Y^{n}}$ and
$\pi_{XY}^{n}$.
\begin{defn}
\cite{yu2018wyner} The \emph{unnormalized and normalized R\'enyi common
informations $T_{1+s}(\pi_{XY})$ and $\widetilde{T}_{1+s}(\pi_{XY})$
of order $1+s\in[0,\infty]$ }between two sources with joint distribution
$\pi_{XY}$ are defined as 
\begin{align}
 & T_{1+s}(\pi_{XY})\nonumber \\
 & :=\inf\left\{ \begin{array}{l}
R:\exists\left\{ \textrm{fixed-length }(n,R)\textrm{ code}\right\} _{n=1}^{\infty}\textrm{ s.t.}\\
\qquad\lim_{n\to\infty}D_{1+s}(P_{X^{n}Y^{n}}\|\pi_{XY}^{n})=0
\end{array}\right\} 
\end{align}
and 
\begin{align}
 & \widetilde{T}_{1+s}(\pi_{XY})\nonumber \\
 & :=\inf\left\{ \begin{array}{l}
R:\exists\left\{ \textrm{fixed-length }(n,R)\textrm{ code}\right\} _{n=1}^{\infty}\textrm{ s.t.}\\
\qquad\lim_{n\to\infty}\frac{1}{n}D_{1+s}(P_{X^{n}Y^{n}}\|\pi_{XY}^{n})=0
\end{array}\right\} .
\end{align}
\end{defn}
It is clear that 
\begin{equation}
\widetilde{T}_{1+s}(\pi_{XY})\le{T}_{1+s}(\pi_{XY}).\label{eqn:stronger}
\end{equation}

If $s=0$, then the unnormalized and normalized\emph{ }R\'enyi common
informations respectively reduce to the unnormalized and normalized
versions of Wyner's common informations \cite{Wyner}.

\subsection{Exact Common Information }

In the formulation of the R\'enyi common information problem, fixed-length
block codes and approximate generation of the target distribution
$\pi_{XY}^{n}$ are assumed. In contrast, in the exact common information
problem \cite{Kumar}, KLE considered variable-length codes and exact
generation of $\pi_{XY}^{n}$. The target is also to find the limit
on the least amount of common randomness satisfying such a requirement,
but the amount here is quantified in term of per-letter expected codeword
length, rather than the exponent of alphabet size described in the
previous subsection.

Define $\left\{ 0,1\right\} ^{*}:=\bigcup_{n\ge1}\left\{ 0,1\right\} ^{n}$
as the set of finite-length strings of symbols from a binary alphabet
$\left\{ 0,1\right\} $. Denote the alphabet of the common random
variable $W_{n}$ as ${\cal W}_{n}$, which can be any countable set
(without loss of generality, one can assume ${\cal W}_{n}=\mathbb{N}$).
Consider a prefix-free  code $f:{\cal W}_{n}\to\left\{ 0,1\right\} ^{*}$.
Then for each symbol $w\in{\cal W}_{n}$ and the code $f$, let $\ell_{f}(w)$
denote the length of the codeword $f\left(w\right)$. 
\begin{defn}
The expected codeword length $L_{f}(W_{n})$ for compressing the random
variable $W_{n}$ by a uniquely decodable code $f$ is defined as
$L_{f}(W_{n}):=\mathbb{E}\left[\ell_{f}(W_{n})\right]$.
\end{defn}
\begin{defn}
A \emph{variable-length  $(n,R)$-code} consists of $(P_{W_{n}},f,P_{X^{n}|W_{n}},P_{Y^{n}|W_{n}})$,
i.e., consists of a distribution $P_{W_{n}}$ on for some countable
set ${\cal W}_{n}$, a pair of random mappings $P_{X^{n}|W_{n}}:{\cal W}_{n}\to\mathcal{X}^{n},P_{Y^{n}|W_{n}}:{\cal W}_{n}\to\mathcal{Y}^{n}$,
and a prefix-free code $f:{\cal W}_{n}\to\left\{ 0,1\right\} ^{*}$
such that the expected codeword length for $W_{n}$ satisfies $L_{f}(W_{n})/{n}\le R$.
\end{defn}
By using variable-length codes, $W_{n}$ is transmitted to two terminals
with error free. The generated (or synthesized) distribution for such
setting is 
\begin{align}
 & P_{X^{n}Y^{n}}(x^{n},y^{n})\nonumber \\
 & :=\sum_{w\in{\cal W}_{n}}P_{W_{n}}(w)P_{X^{n}|W_{n}}(x^{n}|w)P_{Y^{n}|W_{n}}(y^{n}|w),
\end{align}
which is required to be $\pi_{XY}^{n}$ exactly. 
\begin{defn}
\cite{Kumar} The \emph{exact common information $T_{\mathrm{Exact}}(\pi_{XY})$
}between two sources with joint distribution $\pi_{XY}$ is defined
as the minimum asymptotic rate required to ensure $P_{X^{n}Y^{n}}=\pi_{XY}^{n}$
for all $n\ge1$, i.e.,
\begin{align}
 & T_{\mathrm{Exact}}(\pi_{XY})\nonumber \\
 & :=\inf\left\{ \begin{array}{l}
R:\exists\left\{ \textrm{variable-length }(n,R^{(n)})\textrm{ code}\right\} _{n=1}^{\infty}\textrm{ s.t.}\\
\qquad P_{X^{n}Y^{n}}=\pi_{XY}^{n},\forall n\ge1\\
\qquad R\ge\limsup_{n\to\infty}R^{(n)}
\end{array}\right\} .
\end{align}
\end{defn}
By observing that the expected codeword length $L_{f}(W_{n})$ satisfies
$H(W_{n})\leq L_{f}(W_{n})<H(W_{n})+1$, it is easy to verify that
$\frac{1}{n}\left(L_{f}(W_{n})-H(W_{n})\right)\to0$ as $n\to\infty$.
Based on such an argument, KLE \cite{Kumar} provided the following
multi-letter characterization of the exact common information:

\begin{equation}
T_{\mathrm{Exact}}(\pi_{XY})=\lim_{n\to\infty}\frac{1}{n}\min_{P_{W}P_{X^{n}|W}P_{Y^{n}|W}:P_{X^{n}Y^{n}}=\pi_{XY}^{n}}H(W).\label{eq:-106}
\end{equation}
Hence a variable-length synthesis code can be represented by $(P_{W_{n}},P_{X^{n}|W_{n}},P_{Y^{n}|W_{n}})$,
where the dependence on the variable-length compression code $f$
is omitted.

\section{\label{sec:Main-Results-for}Main Results for Sources with Finite
Alphabets}

\subsection{Equivalence and Multi-letter Characterization }

We first establish the equivalence between the exact and $\infty$-R\'enyi
common informations, and characterize them using a multi-letter expression.
The proof of Theorem \ref{thm:equivalence} is given in Appendix \ref{sec:equivalence}.
\begin{thm}[Equivalence]
\label{thm:equivalence} For a source with distribution $\pi_{XY}$
defined on a finite alphabet, 
\begin{equation}
T_{\mathrm{Exact}}(\pi_{XY})=T_{\infty}(\pi_{XY})=\lim_{n\to\infty}\frac{1}{n}\Gamma(\pi_{XY}^{n}),\label{eq:equivalence}
\end{equation}
where\footnote{Note that per Subsection \ref{subsec:Notations}, the conditional
entropy $H(X^{n}Y^{n}|W)$ is computed with respect to $P_{W}P_{X^{n}|W}P_{Y^{n}|W}$.
Hence in fact, $H(X^{n}Y^{n}|W)=H(X^{n}|W)+H(Y^{n}|W)$. } 
\begin{align}
\Gamma(\pi_{XY}^{n}) & :=\inf_{\substack{\substack{P_{W}P_{X^{n}|W}P_{Y^{n}|W}:}
\\
P_{X^{n}Y^{n}}=\pi_{XY}^{n}
}
}\max_{\substack{\substack{Q_{X^{n}Y^{n}|W}\in}
\\
C(P_{X^{n}|W},P_{Y^{n}|W})
}
}-H(X^{n}Y^{n}|W)\nonumber \\
 & \qquad-\sum_{w}P(w)\sum_{x^{n},y^{n}}Q(x^{n},y^{n}|w)\log\pi^{n}\left(x^{n},y^{n}\right).
\end{align}
\end{thm}
\begin{rem}
By setting $W=\left(W_{1},W_{2}\right)$ and $P_{W}P_{X^{n}|W}P_{Y^{n}|W}=\left(P_{W_{1}}P_{X^{n_{1}}|W_{1}}P_{Y^{n_{1}}|W_{1}}\right)\left(P_{W_{2}}P_{X^{n_{2}}|W_{2}}P_{Y^{n_{2}}|W_{2}}\right)$,
it is easy to verify that $\Gamma(\pi_{XY}^{n})$ is subadditive in
$n$, i.e., $\Gamma(\pi_{XY}^{n})\leq\Gamma(\pi_{XY}^{n_{1}})+\Gamma(\pi_{XY}^{n_{2}})$
for all $n_{1}+n_{2}=n$.
\end{rem}
\begin{rem}
\label{rem:By-a-similar}By using a proof similar to that for the
converse part of Theorem \ref{thm:equivalence}, one can show the
following lower bound on the normalized $\infty$-R\'enyi common information.
\begin{align}
\widetilde{T}_{\infty}(\pi_{XY}) & \geq\lim_{\epsilon\downarrow0}\lim_{n\to\infty}\frac{1}{n}\Gamma_{\epsilon}(\pi_{XY}^{n}),\label{eq:-124}
\end{align}
where 
\begin{align}
 & \Gamma_{\epsilon}(\pi_{XY}^{n})\nonumber \\
 & :=\inf_{\substack{\substack{P_{W}P_{X^{n}|W}P_{Y^{n}|W}:}
\\
\frac{1}{n}D_{\infty}\left(P_{X^{n}Y^{n}}\|\pi_{XY}^{n}\right)\le\epsilon
}
}\max_{\substack{\substack{Q_{X^{n}Y^{n}|W}\in}
\\
C(P_{X^{n}|W},P_{Y^{n}|W})
}
}-H(X^{n}Y^{n}|W)\nonumber \\
 & \qquad-\sum_{w}P(w)\sum_{x^{n},y^{n}}Q(x^{n},y^{n}|w)\log\pi^{n}\left(x^{n},y^{n}\right).\label{eq:GammaEpsilon}
\end{align}
It is easy to verify that given $\epsilon>0$, $\Gamma_{\epsilon}(\pi_{XY}^{n})$
is subadditive in $n$, i.e., $\Gamma_{\epsilon}(\pi_{XY}^{n})\leq\Gamma_{\epsilon}(\pi_{XY}^{n_{1}})+\Gamma_{\epsilon}(\pi_{XY}^{n_{2}})$
for all $n_{1}+n_{2}=n$. Hence the limit in \eqref{eq:-124} exists. 
\end{rem}
\begin{rem}
A similar equivalence as the first equality in \eqref{eq:equivalence}
has been found by Kumar, Li, and El Gamal in \cite[Remark on Page 164]{Kumar}.
They showed that the exact common information is equal to a variant
of the $\infty$-R\'enyi common information in which variable-length
codes are allowed. Our equivalence enhances their equivalence for
the direction of $T_{\mathrm{Exact}}(\pi_{XY})\geq T_{\infty}(\pi_{XY})$.
Such a difference enables us to derive the converse part of the multiletter
characterization given in \eqref{eq:equivalence}.
\end{rem}

\subsection{Single-letter Bounds}

Define the maximal cross-entropy over couplings $C(P_{X|W=w},P_{Y|W=w'})$
as\footnote{Note that the maximization in \eqref{eq:maximalcrossentropy} is an
optimal transport problem \cite{rachev1998mass,villani2003topics}.
Hence Kantorovich duality can be used to bound the maximal cross-entropy
if it is required. For more details about the maximal cross-entropy,
please refer to \cite[Section III.A]{yu2018exact}.}
\begin{align}
 & \mathcal{H}(P_{X|W=w},P_{Y|W=w'}\|\pi_{XY})\nonumber \\
 & :=\sup_{Q_{XY}\in C(P_{X|W=w},P_{Y|W=w'})}\sum_{x,y}Q(x,y)\log\frac{1}{\pi\left(x,y\right)}.\label{eq:maximalcrossentropy}
\end{align}
Here for the finite alphabet case, the supremum is in fact a maximum.

Define 
\begin{align}
\Gamma^{\mathrm{UB}}(\pi_{XY}) & :=\Gamma(\pi_{XY})\nonumber \\
 & =\min_{P_{W}P_{X|W}P_{Y|W}:P_{XY}=\pi_{XY}}\Bigl\{-H(XY|W)\nonumber \\
 & \qquad+\sum_{w}P(w)\mathcal{H}(P_{X|W=w},P_{Y|W=w}\|\pi_{XY})\Bigr\},\label{eq:UB}
\end{align}
and 
\begin{align}
\Gamma^{\mathrm{LB}}(\pi_{XY}) & :=\inf_{P_{W}P_{X|W}P_{Y|W}:P_{XY}=\pi_{XY}}\Bigl\{-H(XY|W)\nonumber \\
 & \qquad+\inf_{Q_{WW'}\in C(P_{W},P_{W})}\sum_{w,w'}Q(w,w')\nonumber \\
 & \qquad\times\mathcal{H}(P_{X|W=w},P_{Y|W=w'}\|\pi_{XY})\Bigr\}.\label{eq:LB}
\end{align}
For \eqref{eq:UB}, it suffices to restrict the size of the alphabet
of $W$ such that $|\mathcal{W}|\le|\mathcal{X}||\mathcal{Y}|$. 
This is because 
\begin{align}
 & -H(XY|W)+\sum_{w}P(w)\mathcal{H}(P_{X|W=w},P_{Y|W=w}\|\pi_{XY})
\end{align}
is a linear function of $P_{W}$. Hence by standard cardinality bounding
techniques (e.g., the support lemma in \cite[Appendix C]{Gamal}),
there exists an optimal distribution $P_{W}P_{X|W}P_{Y|W}$ with $P_{XY}=\pi_{XY}$
and $|\supp(W)|\le|\mathcal{X}||\mathcal{Y}|$ attaining the minimization
in \eqref{eq:UB}.

By utilizing the multi-letter expression in Theorem \ref{thm:equivalence},
we provide single-letter lower and upper bounds for the exact and
$\infty$-R\'enyi common informations. The proof of Theorem \ref{thm:singleletter}
is given in Appendix \ref{sec:singleletter}. 
\begin{thm}[Single-letter Bounds]
\label{thm:singleletter} The exact and $\infty$-R\'enyi common informations
for a source with distribution $\pi_{XY}$ defined on a finite alphabet
satisfy 
\begin{align}
\max\left\{ \Gamma^{\mathrm{LB}}(\pi_{XY}),C_{\mathsf{Wyner}}(\pi_{XY})\right\}  & \leq\widetilde{T}_{\infty}(\pi_{XY})\\
 & \le T_{\infty}(\pi_{XY})\\
 & =T_{\mathrm{Exact}}(\pi_{XY})\\
 & \leq\Gamma^{\mathrm{UB}}(\pi_{XY}).
\end{align}
\end{thm}
Note that the only difference between the upper and lower bounds is
that in the lower bound, the minimization operation is taken over
all couplings of $\left(P_{W},P_{W}\right)$, but in the upper bound,
it is not (or equivalently, the expectation in \eqref{eq:UB} can
be seen as being taken under the equality coupling of $\left(P_{W},P_{W}\right)$,
namely $P_{W}(w)1\{w'=w\}$). The upper bound $\Gamma^{\mathrm{UB}}(\pi_{XY})$
and lower bound $\Gamma^{\mathrm{LB}}(\pi_{XY})$ are consistent with
the bounds for $\alpha$-R\'enyi common information for $\alpha\in[0,\infty]$
\cite{yu2018wyner,yu2018corrections}.

The $\infty$-R\'enyi common information code we adopt in the proof
is a truncated i.i.d.\ code. For such a code, the codewords are independent
and each codeword is drawn according to a distribution $P_{W^{n}}$
which is generated by truncating a product distribution $Q_{W}^{n}$
onto some (strongly) typical set. Truncated i.i.d. codes are rather
useful (i.e., strictly better than i.i.d. codes without truncation)
for $\infty$-R\'enyi-approximate synthesis (but achieve  the same performance
as i.i.d. codes for Wyner's synthesis, i.e., $1$-R\'enyi-approximate
synthesis). This follows from the following argument. Observe that
for both $\infty$-R\'enyi-approximate synthesis and Wyner's synthesis,
$X^{n}\to W_{n}\to Y^{n}$ forms a Markov chain. Hence given $W_{n}=w$,
the support of $P_{X^{n}|W_{n}}\left(\cdot|w\right)P_{Y^{n}|W_{n}}\left(\cdot|w\right)$
is a product set, which in turn implies that the support of $P_{X^{n}Y^{n}}$
is the union of a family of product sets. Such a requirement leads
to the fact that the support of $P_{X^{n}Y^{n}}$ includes not only
a jointly typical set, but also other joint type classes, which is
termed by us as the\emph{ type overflow phenomenon}. Wyner's synthesis
(under the relative entropy measure) only requires the sequences in
a typical set to be well-simulated. However, $\infty$-R\'enyi-approximate
synthesis requires \emph{all} the sequences in the support of $P_{X^{n}Y^{n}}$
to be well-simulated. Hence the type overflow phenomenon does not
affect Wyner's synthesis asymptotically, but plays a critical role
in minimizing the rate of $\infty$-R\'enyi-approximate synthesis (or
exact synthesis). Truncated i.i.d. coding is an efficient approach
to control the possible types of the output sequence of a code (or
more precisely, to mitigate the effects of type overflow). Furthermore,
truncated i.i.d. codes have also been used by the present authors
\cite{yu2019renyi,yu2018wyner,yu2018corrections} to study $\alpha$-R\'enyi
common informations, and by Vellambi and Kliewer \cite{vellambi2016sufficient,vellambi2018new}
to study sufficient conditions for equality of the exact and Wyner's
common informations. 

The maximal cross-entropy in \eqref{eq:maximalcrossentropy} has the
following intuitive interpretation.  Consider a joint distribution
$\pi_{XY}$, a pair of distributions $\left(P_{X},P_{Y}\right)$,
and a sequence of pairs of types $\{(T_{X}^{(n)},T_{Y}^{(n)})\in\mathcal{P}_{n}\left(\mathcal{X}\right)\times\mathcal{P}_{n}\left(\mathcal{Y}\right)\}_{n\in\mathbb{N}}$
such that $(T_{X}^{(n)},T_{Y}^{(n)})\to\left(P_{X},P_{Y}\right)$
as $n\to\infty$. The minimum of the exponents of probabilities
$\pi_{XY}^{n}\left(x^{n},y^{n}\right)$ such that $T_{x^{n}}=T_{X}^{(n)},T_{y^{n}}=T_{Y}^{(n)}$
satisfy that 
\begin{align}
 & \lim_{n\to\infty}\min_{\substack{\left(x^{n},y^{n}\right):\\
T_{x^{n}}=T_{X}^{(n)},\\
T_{y^{n}}=T_{Y}^{(n)}
}
}-\frac{1}{n}\log\pi_{XY}^{n}\left(x^{n},y^{n}\right)\nonumber \\
 & =\lim_{n\to\infty}\min_{\substack{\left(x^{n},y^{n}\right):\\
T_{x^{n}}=T_{X}^{(n)},\\
T_{y^{n}}=T_{Y}^{(n)}
}
}\sum_{x,y}T_{x^{n},y^{n}}(x,y)\log\frac{1}{\pi\left(x,y\right)}\\
 & =\mathcal{H}(P_{X},P_{Y}\|\pi_{XY}).
\end{align}

Based on the type overflow argument and the intuitive explanation
of the maximal cross-entropy given above, our bounds are easy to comprehend
intuitively. The exact synthesis requires that there exists a sequence
of variable-length codes with asymptotic rate $R$ satisfying $\frac{P_{Y^{n}|X^{n}}(y^{n}|x^{n})}{\pi_{Y|X}^{n}(y^{n}|x^{n})}=1$
for all $(x^{n},y^{n})\in\mathcal{X}^{n}\times\mathcal{Y}^{n}$. By
using the mixture decomposition technique, the exact synthesis problem
can be relaxed to the $\infty$-R\'enyi-approximate synthesis problem,
which requires that there exists a sequence of fixed-length codes
with asymptotic rate $R$ satisfying 
\begin{equation}
\frac{P_{X^{n}Y^{n}}(x^{n},y^{n})}{\pi_{XY}^{n}(x^{n},y^{n})}\leq1+o(1)\label{eq:-35-1}
\end{equation}
for all $(x^{n},y^{n})\in\supp\left(P_{X^{n}Y^{n}}\right)$; see Lemma
\ref{lem:renyilarger}. By using truncated i.i.d. codes, to mitigate
the effect of type overflow we can restrict $(W^{n},X^{n})\in\mathcal{T}_{\epsilon}^{\left(n\right)}\left(P_{WX}\right)$
and $(W^{n},Y^{n})\in\mathcal{T}_{\epsilon}^{\left(n\right)}\left(P_{WY}\right)$.
Suppose that $M_{n}$ is the message for $\infty$-R\'enyi-approximate
synthesis.  Then for sufficiently large $n$ and sufficiently small
$\epsilon$,
\begin{align}
 & P_{X^{n}Y^{n}}(x^{n},y^{n})\nonumber \\
 & \approx\sum_{m}P_{M_{n}}(m)P_{X|W}^{n}(x^{n}|w^{n}(m))P_{Y|W}^{n}(y^{n}|w^{n}(m))\\
 & \approx N(x^{n},y^{n})e^{-nR}e^{-nH(X|W)}e^{-nH(Y|W)},\label{eq:-38-1}
\end{align}
where $N(x^{n},y^{n})$ denotes the number of codewords $w^{n}(m)$
that cover $x^{n}$ and $y^{n}$ (i.e., that are jointly typical with
$x^{n}$ and jointly typical with $y^{n}$). On the other hand,
\begin{align}
 & \min_{(x^{n},y^{n})\in\supp\left(P_{X^{n}Y^{n}}\right)}\pi_{XY}^{n}(x^{n},y^{n})\nonumber \\
 & \approx\min_{\left(w^{n},x^{n},y^{n}\right):T_{w^{n}x^{n}}\approx P_{WX},T_{w^{n}y^{n}}\approx P_{WY}}\pi_{XY}^{n}(x^{n},y^{n})\\
 & \approx e^{-n\sum_{w}P_{W}(w)\mathcal{H}(P_{X|W=w},P_{Y|W=w}\|\pi_{XY})}.\label{eq:-39-1}
\end{align}
Substituting \eqref{eq:-38-1} and \eqref{eq:-39-1} into \eqref{eq:-35-1}
and observing that $N(x^{n},y^{n})\ge1$ for $(x^{n},y^{n})\in\supp\left(P_{X^{n}Y^{n}}\right)$,
we obtain 
\begin{equation}
R\gtrsim-H(XY|W)+\sum_{w}P(w)\mathcal{H}(P_{X|W=w},P_{Y|W=w}\|\pi_{XY}).\label{eq:-161}
\end{equation}
Taking the minimum over all distributions $P_{W}P_{X|W}P_{Y|W}$ such
that $P_{XY}=\pi_{XY}$, we obtain the upper bound $\Gamma^{\mathrm{UB}}(\pi_{XY})$.
We make this argument precise in Appendix \ref{sec:singleletter}.

\subsection{Doubly Symmetric Binary Sources}

A doubly symmetric binary source (DSBS) is a source $\left(X,Y\right)$
with distribution 
\begin{equation}
\pi_{XY}:=\left[\begin{array}{cc}
\alpha_{0} & \beta_{0}\\
\beta_{0} & \alpha_{0}
\end{array}\right]\label{eq:-10}
\end{equation}
where $\alpha_{0}=\frac{1-p}{2},\beta_{0}=\frac{p}{2}$ with $p\in(0,\frac{1}{2})$.
This is equivalent to $X\sim\mathrm{Bern}(\frac{1}{2})$ and $Y=X\oplus E$
with $E\sim\mathrm{Bern}(p)$ independent of $X$; or $X=W\oplus A$
and $Y=W\oplus B$ with $W\sim\mathrm{Bern}(\frac{1}{2})$, $A\sim\mathrm{Bern}(a)$,
and $B\sim\mathrm{Bern}(a)$ mutually independent, where $a:=\frac{1-\sqrt{1-2p}}{2}\in(0,\frac{1}{2})$
or equivalently, $\alpha_{0}=\frac{1}{2}\left(a^{2}+(1-a)^{2}\right),\beta_{0}=a(1-a)$.
Here we do not lose any generality by restricting $p$ or $a\in(0,\frac{1}{2})$,
since otherwise, we can set $X\oplus1$ to $X$.

\begin{figure*}
\centering \includegraphics[width=0.7\textwidth]{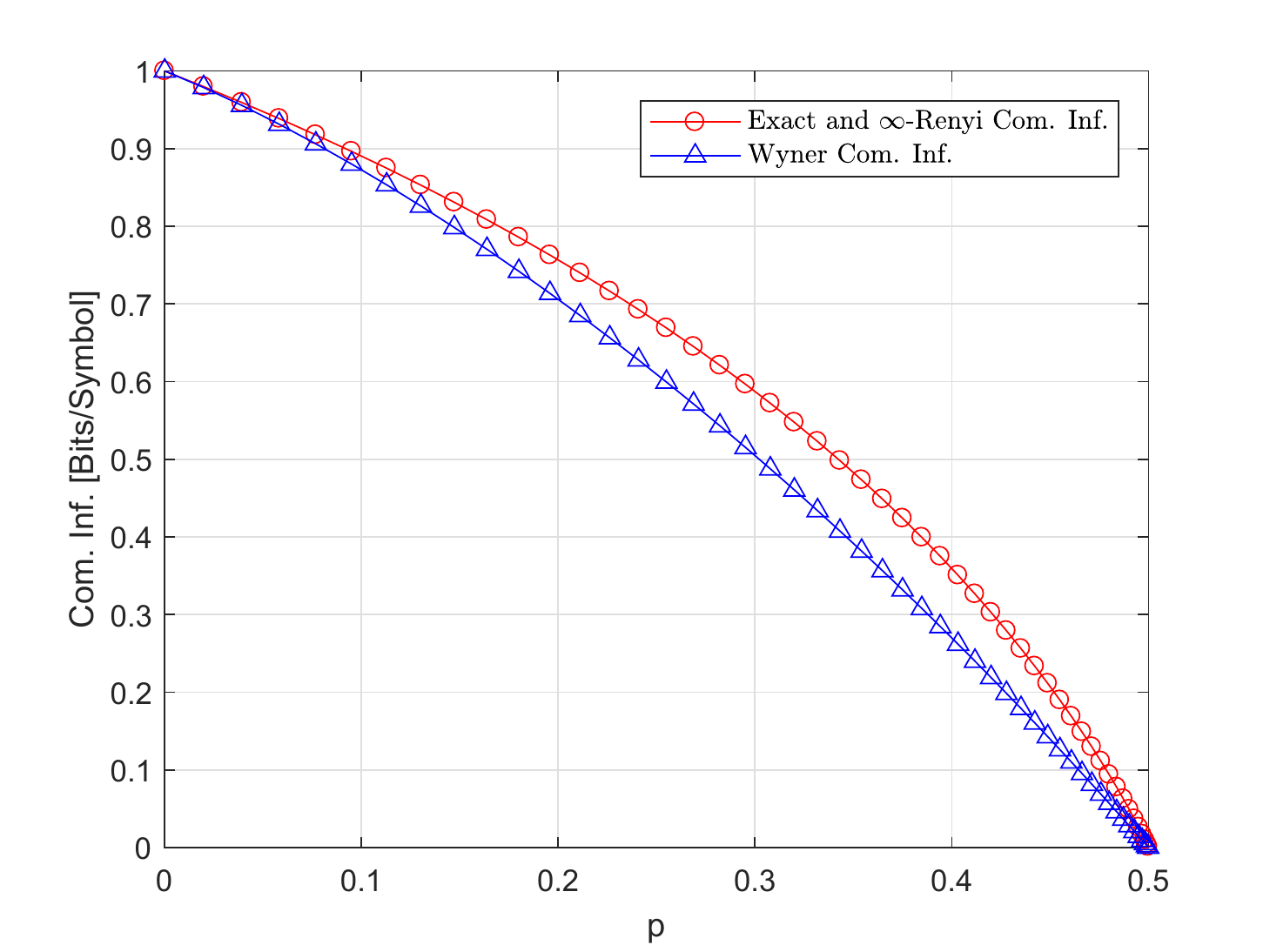}

\caption{\label{fig:Common-informations-for}Illustrations of the exact and
$\infty$-R\'enyi common informations \eqref{eq:-61} and Wyner's common
information \eqref{eq:-24} for DSBSes $(X,Y)$ such that $X\sim\mathrm{Bern}(\frac{1}{2})$
and $Y=X\oplus E$ with $E\sim\mathrm{Bern}(p)$ independent of $X$. }
\end{figure*}

By utilizing the lower and upper bounds in Theorem \ref{thm:singleletter},
we completely characterize the exact and $\infty$-R\'enyi common informations
for DSBSes. The proof of Theorem \ref{thm:DSBS} is given in Appendix
\ref{sec:DSBS}. 
\begin{thm}
\label{thm:DSBS}For a DSBS $\left(X,Y\right)$ with distribution
$\pi_{XY}$ given in \eqref{eq:-10}, 
\begin{align}
 & \widetilde{T}_{\infty}(\pi_{XY})=T_{\infty}(\pi_{XY})=T_{\mathrm{Exact}}(\pi_{XY})\nonumber \\
 & =-2H_{2}(a)-(1-2a)\log\left[\frac{1}{2}\left(a^{2}+(1-a)^{2}\right)\right]\nonumber \\
 & \qquad-2a\log\left[a(1-a)\right],\label{eq:-61}
\end{align}
where 
\begin{align}
H_{2}(a) & :=-a\log a-(1-a)\log(1-a)\label{eq:binaryentropy}
\end{align}
denotes the binary entropy function.
\end{thm}
\begin{cor}
\label{cor:For-a-DSBS}For a DSBS $\left(X,Y\right)$ with distribution
$\pi_{XY}$ given in \eqref{eq:-10}, 
\begin{equation}
\widetilde{T}_{\infty}(\pi_{XY})=T_{\infty}(\pi_{XY})=T_{\mathrm{Exact}}(\pi_{XY})>C_{\mathrm{Wyner}}(\pi_{XY})
\end{equation}
for the parameter $a\in(0,\frac{1}{2})$.
\end{cor}
\begin{rem}
For this case, the exact common information is strictly larger than
Wyner's common information. This answers an open problem posed by
KLE \cite{Kumar}. 
\end{rem}
\begin{IEEEproof}
For DSBSes, Wyner \cite{Wyner} showed that 
\begin{align}
 & T_{1}(\pi_{XY})=C_{\mathrm{Wyner}}(\pi_{XY})\nonumber \\
 & =-2H_{2}(a)-\left(a^{2}+(1-a)^{2}\right)\log\left[\frac{1}{2}\left(a^{2}+(1-a)^{2}\right)\right]\nonumber \\
 & \qquad-2a(1-a)\log\left[a(1-a)\right].\label{eq:-24}
\end{align}
Hence 
\begin{align}
 & T_{\infty}(\pi_{XY})-C_{\mathrm{Wyner}}(\pi_{XY})\nonumber \\
 & =\left(\left(a^{2}+(1-a)^{2}\right)-(1-2a)\right)\log\left[\frac{1}{2}\left(a^{2}+(1-a)^{2}\right)\right]\nonumber \\
 & \qquad+\left(2a(1-a)-2a\right)\log\left[a(1-a)\right]\\
 & =2a^{2}\log\left[\frac{\frac{1}{2}\left(a^{2}+(1-a)^{2}\right)}{a(1-a)}\right]>0.
\end{align}
We obtain the desired result. 
\end{IEEEproof}
The exact, $\infty$-R\'enyi, and Wyner's common informations for DSBSes
are illustrated in Fig.~\ref{fig:Common-informations-for}.

\subsection{\textmd{\normalsize{}Sufficient Conditions for Equality of Exact
and Wyner's Common Informations}}

In Corollary \ref{cor:For-a-DSBS}, we showed that for a DSBS, the
exact common information is strictly larger than Wyner's common information.
Now we study sufficient conditions for equality of exact and Wyner's
common informations. Obviously, if $\Gamma^{\mathrm{UB}}(\pi_{XY})=C_{\mathsf{Wyner}}(\pi_{XY})$,
then the exact and Wyner's common informations are equal. We first
introduce a condition on $\pi_{XY}$.

Condition $(*)$: There exists some optimal distribution $P_{W}P_{X|W}P_{Y|W}$
attaining $C_{\mathrm{Wyner}}(\pi_{XY})$ such that for any $w\in\supp\left(P_{W}\right)$,
$\pi_{XY}$ when restricted to $\mathcal{A}_{w}:=\supp\left(P_{X|W=w}\right)\times\supp\left(P_{Y|W=w}\right)$
is a product distribution, i.e., $\pi_{XY}\left(\cdot|\mathcal{A}_{w}\right)$
is a product distribution for each $w\in\supp\left(P_{W}\right)$.
\begin{thm}
\textup{\label{thm:sufficient}If $\pi_{XY}$ satisfies Condition
$(*)$, then the exact and Wyner's common informations are equal,
i.e., 
\begin{equation}
T_{\mathrm{Exact}}(\pi_{XY})=C_{\mathrm{Wyner}}(\pi_{XY}).
\end{equation}
}
\end{thm}
\begin{rem}
Theorem \ref{thm:sufficient} generalizes the sufficient conditions
given in \cite{Kumar,vellambi2016sufficient,vellambi2018new}.
\end{rem}
Theorem \ref{thm:sufficient} follows from the following lemma.
\begin{lem}
\textup{\label{thm:sufficient-1}$\Gamma^{\mathrm{UB}}(\pi_{XY})=C_{\mathrm{Wyner}}(\pi_{XY})$
if and only if $\pi_{XY}$ satisfies Condition $(*)$. }
\end{lem}
\begin{rem}
Lemma \ref{thm:sufficient-1} implies that if the upper bound $\Gamma^{\mathrm{UB}}(\pi_{XY})$
is tight for the exact common information (i.e., $T_{\mathrm{Exact}}(\pi_{XY})=\Gamma^{\mathrm{UB}}(\pi_{XY})$),
then Condition $(*)$ is necessary and sufficient for $T_{\mathrm{Exact}}(\pi_{XY})=C_{\mathrm{Wyner}}(\pi_{XY})$.
\end{rem}
\begin{IEEEproof}
``If'' Part: Suppose that $\pi_{XY}$ satisfies Condition $(*)$.
Then by \cite[Proposition 2]{yu2018exact}, we obtain that for any
$w\in\supp\left(P_{W}\right)$,
\begin{align}
 & \mathcal{H}(P_{X|W=w},P_{Y|W=w}\|\pi_{XY})\nonumber \\
 & =\sum_{x,y}P(x|w)P(y|w)\log\frac{1}{\pi\left(x,y\right)}.
\end{align}
After taking the expectation respect to $P_{W}$, we obtain
\begin{equation}
\sum_{w}P(w)\mathcal{H}(P_{X|W=w},P_{Y|W=w}\|\pi_{XY})=H(XY).
\end{equation}
Therefore, substituting the distribution $P_{W}P_{X|W}P_{Y|W}$ into
$\Gamma^{\mathrm{UB}}(\pi_{XY})$, we obtain that $\Gamma^{\mathrm{UB}}(\pi_{XY})\leq C_{\mathsf{Wyner}}(\pi_{XY}).$
Since $\Gamma^{\mathrm{UB}}(\pi_{XY})\ge C_{\mathsf{Wyner}}(\pi_{XY})$,
we obtain that $\Gamma^{\mathrm{UB}}(\pi_{XY})=C_{\mathsf{Wyner}}(\pi_{XY}).$

``Only If'' Part:  Suppose that $\Gamma^{\mathrm{UB}}(\pi_{XY})=C_{\mathsf{Wyner}}(\pi_{XY}).$
For a distribution $\pi_{XY}$, denote $P_{W}P_{X|W}P_{Y|W}$ as an
optimal distribution attaining $\Gamma^{\mathrm{UB}}(\pi_{XY})$.
 Then we have that for any $w\in\supp\left(P_{W}\right)$, $\supp\left(P_{X|W=w}\right)\times\supp\left(P_{Y|W=w}\right)\subseteq\supp(\pi_{XY})$,
otherwise, $\sum_{w}P(w)\mathcal{H}(P_{X|W=w},P_{Y|W=w}\|\pi_{XY})=\infty$
which contradicts the optimality of $P_{W}P_{X|W}P_{Y|W}$. On the
other hand, we have that
\begin{align}
 & \Gamma^{\mathrm{UB}}(\pi_{XY})\nonumber \\
 & =-H(XY|W)+\sum_{w}P(w)\mathcal{H}(P_{X|W=w},P_{Y|W=w}\|\pi_{XY})\label{eq:-162}\\
 & \geq-H(XY|W)+H(XY)\label{eq:-163}\\
 & \geq C_{\mathsf{Wyner}}(\pi_{XY})\label{eq:-164}
\end{align}
By assumption, the inequalities in \eqref{eq:-163} and \eqref{eq:-164}
are equalities. Hence $P_{W}P_{X|W}P_{Y|W}$ also attains $C_{\mathsf{Wyner}}(\pi_{XY})$
and the following equality holds:
\begin{equation}
\sum_{w}P(w)\mathcal{H}(P_{X|W=w},P_{Y|W=w}\|\pi_{XY})=H(XY).\label{eq:-105}
\end{equation}
Equation \eqref{eq:-105} implies
\begin{align}
 & \mathcal{H}(P_{X|W=w},P_{Y|W=w}\|\pi_{XY})\nonumber \\
 & =\sum_{x,y}P(x|w)P(y|w)\log\frac{1}{\pi\left(x,y\right)}
\end{align}
for every $w\in\supp\left(P_{W}\right)$. By \cite[Proposition 2]{yu2018exact},
for every $w\in\supp\left(P_{W}\right)$, $\pi_{XY}$ is product on
the set $\supp\left(P_{X|W=w}\right)\times\supp\left(P_{Y|W=w}\right)$.
 Hence $\pi_{XY}$ satisfies Condition $(*)$.
\end{IEEEproof}
The following is a special case of Condition $(*)$.
\begin{defn}
A joint distribution $\pi_{XY}$ is pseudo-product if  for some $A\subseteq\mathcal{X}\times\mathcal{Y}$,
\begin{equation}
\pi_{XY}(x,y)=\begin{cases}
\alpha(x)\beta(y) & (x,y)\in A\\
0 & \textrm{otherwise}
\end{cases}
\end{equation}
where $\alpha:\mathcal{X}\to\mathbb{R}_{>0}$ and $\beta:\mathcal{Y}\to\mathbb{R}_{>0}$
are two positive functions such that $\sum_{(x,y)\in A}\alpha(x)\beta(y)=1$.
\end{defn}
\begin{rem}
In general, a pseudo-product distribution may not be a product distribution.
For example, 
\begin{equation}
\frac{1}{\alpha_{0}\beta_{0}+\alpha_{0}\beta_{1}+\alpha_{1}\beta_{0}}\left[\begin{array}{cc}
\alpha_{0}\beta_{0} & \alpha_{0}\beta_{1}\\
\alpha_{1}\beta_{0} & 0
\end{array}\right]\label{eq:-10-1-1-1-1}
\end{equation}
is a pseudo-product distribution but not a product distribution. However,
if $\supp(\pi_{XY})$ is a product set, then a pseudo-product distribution
$\pi_{XY}$ is a product distribution.
\end{rem}
Obviously, pseudo-product distributions satisfy Condition $(*)$.
Hence for pseudo-product distributions, the exact and Wyner's common
informations are equal.

\section{Extension to Sources with General Alphabets}

In Section \ref{sec:Main-Results-for}, we derived exact and $\infty$-R\'enyi
common informations for sources with finite alphabets. In this section,
we generalize the results to sources with countably infinite alphabets
and a certain class of continuous sources. Furthermore, note that
for Wyner's common information, till date, only the case of sources
with finite alphabets was studied by Wyner \cite{Wyner}, and there
is no characterization\footnote{More precisely, there is no converse result derived for sources with
countably infinite alphabets and continuous sources. As for the achievability
part, several existing results on channel resolvability (e.g., \cite{Hayashi06,Cuff,yu2019renyi})
can be applied to obtain achievability results for the Wyner's common
information problem.} for sources with countably infinite alphabets and continuous sources.
Hence in this section, before generalizing the exact and $\infty$-R\'enyi
common informations, we first generalize \emph{Wyner's common information}
to such sources. In the proofs of the converse parts, the mixture-decomposition
technique is used extensively. We show that for sources with countably
infinite alphabets and a certain class of continuous sources (including
Gaussian sources), Wyner's common information remains $C_{\mathrm{Wyner}}(\pi_{XY})$.
Moreover, for a source with countably infinite alphabets, we show
that Wyner's common information can be obtained by computing the common
information for an alphabet-truncated version of the source and then
taking limits to enlarge the domain of the truncated alphabet.

\subsection{Wyner's Common Information}

Wyner \cite{Wyner} only characterized the common information for
sources with finite alphabets. Here we extend his results to sources
with countably infinite alphabets and continuous sources. The proof
of Theorem \ref{thm:GeneralWyner} is given in Appendix \ref{sec:GeneralWyner}.
\begin{thm}[Wyner's Common Information for General Sources]
\label{thm:GeneralWyner} Let $(X,Y)$ be a source with distribution
$\pi_{XY}$ defined on the product of two arbitrary alphabets (i.e.,
on the product of two arbitrary measurable spaces). Then we have 
\begin{equation}
\widetilde{C}_{\mathrm{Wyner}}(\pi_{XY})\leq\widetilde{T}_{1}(\pi_{XY})\leq T_{1}(\pi_{XY})\leq\widehat{C}_{\mathrm{Wyner}}(\pi_{XY}),\label{eqn:Wyner}
\end{equation}
where\footnote{For this arbitrary alphabet case, the distributions $P_{W}P_{X|W}P_{Y|W}$
in the infimizations in the definitions of $\widetilde{C}_{\mathrm{Wyner}}(\pi_{XY})$
and $C_{\mathrm{Wyner}}(\pi_{XY})$ (for the latter, see \eqref{eqn:CWyner})
are restricted to satisfy that the mutual information $I\left(XY;W\right)$
exists; if there is no such distribution, then $\widetilde{C}_{\mathrm{Wyner}}(\pi_{XY}):=\infty$
and $C_{\mathrm{Wyner}}(\pi_{XY}):=\infty$. Here we say the mutual
information $I\left(U;V\right)$ of two random variables $U$ and
$V$ exists if $P_{UV}\ll P_{U}P_{V}$ and the integral $\int_{\mathcal{U}\times\mathcal{V}}\left|\log\frac{\mathrm{d}P_{UV}}{\mathrm{d}\left(P_{U}P_{V}\right)}\right|\mathrm{d}P_{UV}<\infty$.
The mutual information always exists for distributions with finite
alphabets but does not always exist for other distributions. Hence
here we need to add this constraint. Similarly, the distribution
$P_{W}P_{X|W}P_{Y|W}$ in the infimization in the definition of $\widehat{C}_{\mathrm{Wyner}}(\pi_{XY})$
is restricted to satisfy that $D_{1+s}(P_{X|W}P_{Y|W}\|P_{XY}|P_{W})$
exists for some $s>0$; if there is no such distribution, then $\widehat{C}_{\mathrm{Wyner}}(\pi_{XY}):=\infty$.
Here we say $D_{1+s}\left(P_{U|V}\|P_{U}|P_{V}\right)$ exists if
$P_{UV}\ll P_{U}P_{V}$ and the integral $\frac{1}{s}\log\int_{\mathcal{U}\times\mathcal{V}}\left(\frac{\mathrm{d}P_{UV}}{\mathrm{d}\left(P_{U}P_{V}\right)}\right)^{s}\mathrm{d}P_{UV}<\infty$.} 
\begin{align}
\widetilde{C}_{\mathrm{Wyner}}(\pi_{XY}) & :=\lim_{\epsilon\downarrow0}\inf_{\substack{P_{W}P_{X|W}P_{Y|W}:\\
D\left(P_{XY}\|\pi_{XY}\right)\le\epsilon
}
}I\left(XY;W\right)\label{eq:-53}
\end{align}
\textup{and 
\begin{align}
 & \widehat{C}_{\mathrm{Wyner}}(\pi_{XY})\nonumber \\
 & :=\inf_{\substack{P_{W}P_{X|W}P_{Y|W}:\\
P_{XY}=\pi_{XY}
}
}\lim_{s\downarrow0}D_{1+s}(P_{X|W}P_{Y|W}\|P_{XY}|P_{W}).\label{eq:-53-2}
\end{align}
}
\end{thm}
Since $D(P_{X|W}P_{Y|W}\|P_{XY}|P_{W})=I\left(XY;W\right)$ and for
any fixed $\left(P,Q\right)$, $D_{1+s}(P\|Q)$ is non-decreasing
in $s$, we know that 
\begin{equation}
\widetilde{C}_{\mathrm{Wyner}}(\pi_{XY})\leq C_{\mathrm{Wyner}}(\pi_{XY})\le\widehat{C}_{\mathrm{Wyner}}(\pi_{XY}).\label{eq:-134}
\end{equation}
Note that for any fixed $\left(P,Q\right)$, the R\'enyi divergence
$D_{1+s}(P\|Q)$ is continuous in $s\in[-1,0]\cup\left\{ s\in(0,\infty]:D_{1+s}(P\|Q)<\infty\right\} $.
However, van Erven and Harremo\"es in \cite{Erven} showed that there
exists a pair of distributions $\left(P,Q\right)$ such that the R\'enyi
divergence $D_{1+s}(P\|Q)$ is not continuous at $s=0$. Hence we
do not know if the inequalities in \eqref{eq:-134} are equalities
in general.
\begin{prop}
\label{prop:GeneralWyner-1}The following are sufficient conditions
to ensure $\widehat{C}_{\mathrm{Wyner}}(\pi_{XY})=C_{\mathrm{Wyner}}(\pi_{XY}).$
\begin{enumerate}
\item There exists a joint distribution $P_{W}P_{X|W}P_{Y|W}$ that attains
$C_{\mathrm{Wyner}}(\pi_{XY})$ and satisfies
\begin{equation}
D_{1+s}(P_{X|W}P_{Y|W}\|P_{XY}|P_{W})<\infty\label{eq:-166}
\end{equation}
for some $s>0$. 
\item There exists a sequence of joint distributions $P_{WXY}^{(k)}:=P_{W}^{(k)}P_{X|W}^{(k)}P_{Y|W}^{(k)}$
such that they attain $C_{\mathrm{Wyner}}(\pi_{XY})$ asymptotically,
i.e., $P_{XY}^{(k)}=\pi_{XY}$, $\lim_{k\to\infty}I_{P^{(k)}}(XY;W)=C_{\mathrm{Wyner}}(\pi_{XY})$,
and for every $k$, there exists some $s_{k}>0$ satisfying 
\begin{equation}
D_{1+s_{k}}(P_{X|W}^{(k)}P_{Y|W}^{(k)}\|P_{XY}^{(k)}|P_{W}^{(k)})<\infty.\label{eq:-165}
\end{equation}
\end{enumerate}
\end{prop}
\begin{IEEEproof}
Here we only prove Statement 2). Statement 1) follows similarly.

Suppose that there exists a sequence of joint distributions $P_{W}^{(k)}P_{X|W}^{(k)}P_{Y|W}^{(k)}$
satisfying the conditions given in Statement 2). Then \eqref{eq:-165}
implies that given $P_{W}^{(k)}P_{X|W}^{(k)}P_{Y|W}^{(k)}$, the conditional
R\'enyi divergence $D_{1+s}(P_{X|W}^{(k)}P_{Y|W}^{(k)}\|P_{XY}^{(k)}|P_{W}^{(k)})$
is continuous in $s\in[-1,s_{k}]$. Hence 
\begin{equation}
\lim_{s\downarrow0}D_{1+s}(P_{X|W}^{(k)}P_{Y|W}^{(k)}\|P_{XY}^{(k)}|P_{W}^{(k)})=I_{P^{(k)}}(XY;W).
\end{equation}
Therefore, by the definition of $\widehat{C}_{\mathrm{Wyner}}(\pi_{XY})$,
for all $k$,
\begin{align}
\widehat{C}_{\mathrm{Wyner}}(\pi_{XY}) & \leq\lim_{s\downarrow0}D_{1+s}(P_{X|W}^{(k)}P_{Y|W}^{(k)}\|P_{XY}^{(k)}|P_{W}^{(k)})\\
 & =I_{P^{(k)}}(XY;W).\label{eqn:Wyner-1-2}
\end{align}
By assumption, $I_{P^{(k)}}(XY;W)\to C_{\mathrm{Wyner}}(\pi_{XY})$
as $k\to\infty$. Hence 
\begin{equation}
\widehat{C}_{\mathrm{Wyner}}(\pi_{XY})\leq C_{\mathrm{Wyner}}(\pi_{XY}).\label{eqn:Wyner-1-2-1}
\end{equation}
\end{IEEEproof}
Observe that the requirements \eqref{eq:-166} and \eqref{eq:-165}
are respectively equivalent to 
\begin{equation}
\int_{\mathcal{W}\times\mathcal{X}\times\mathcal{Y}}\left(\frac{\mathrm{d}\left(P_{X|W}P_{Y|W}\right)}{\mathrm{d}P_{XY}}\right)^{s}\mathrm{d}\left(P_{W}P_{X|W}P_{Y|W}\right)<\infty\label{eq:-167}
\end{equation}
 and 
\begin{equation}
\int_{\mathcal{W}\times\mathcal{X}\times\mathcal{Y}}\left(\frac{\mathrm{d}\left(P_{X|W}^{(k)}P_{Y|W}^{(k)}\right)}{\mathrm{d}P_{XY}^{(k)}}\right)^{s}\mathrm{d}\left(P_{W}^{(k)}P_{X|W}^{(k)}P_{Y|W}^{(k)}\right)<\infty\label{eq:-168}
\end{equation}
for $s>0$. Note that for $s=0$, \eqref{eq:-167} and \eqref{eq:-168}
are satisfied. Hence we conjecture that the conditions given in Proposition
\ref{prop:GeneralWyner-1} hold for a large class of sources.

For the finite alphabet case, it is easy to verify that $\widetilde{C}_{\mathrm{Wyner}}(\pi_{XY})=C_{\mathrm{Wyner}}(\pi_{XY})=\widehat{C}_{\mathrm{Wyner}}(\pi_{XY}).$
Hence $\widetilde{T}_{1}(\pi_{XY})=T_{1}(\pi_{XY})=C_{\mathrm{Wyner}}(\pi_{XY})$
for this case. The result $\widetilde{T}_{1}(\pi_{XY})=C_{\mathrm{Wyner}}(\pi_{XY})$
was first proven by Wyner \cite{Wyner}. The case concerning sources
with countably infinite alphabets and the case concerning a certain
class of continuous sources are considered in the following corollaries.
The proofs are given in Appendices \ref{sec:countable} and \ref{sec:continuous}.
\begin{cor}
\label{cor:countable} Let $(X,Y)$ be a source with distribution
$\pi_{XY}$ defined on the product of two countably infinite alphabets.\textup{
Assume $H_{\alpha}(\pi_{XY})$ exists (and hence is finite) for some
$\alpha\in[0,1)$.} Then we have 
\begin{equation}
\widetilde{T}_{1}(\pi_{XY})=T_{1}(\pi_{XY})=C_{\mathrm{Wyner}}(\pi_{XY}).\label{eqn:stronger-1-2-1}
\end{equation}
\end{cor}
\begin{rem}
\label{rem:In-our-proof,}In our proof, we show that 
\begin{align}
\widetilde{C}_{\mathrm{Wyner}}(\pi_{XY}) & =C_{\mathrm{Wyner}}(\pi_{XY})\\
 & =\lim_{k\to\infty}C_{\mathrm{Wyner}}(\pi_{XY}^{(k)})\\
 & =\lim_{k\to\infty}C_{\mathrm{Wyner}}(\pi_{\left[X\right]_{k}\left[Y\right]_{k}})
\end{align}
where 
\begin{equation}
\pi_{XY}^{(k)}(x,y):=\frac{\pi_{XY}(x,y)1\left\{ (x,y)\in[-k,k]^{2}\right\} }{\pi_{XY}([-k,k]^{2})}
\end{equation}
and $\pi_{\left[X\right]_{k}\left[Y\right]_{k}}$ with $\left[z\right]_{k}:=z$,
if $\left|z\right|\le k$, and $k+1$, otherwise, denote distributions
induced by truncation operations. That is to say, we can compute Wyner's
common information for countably-infinite-valued sources by computing
the common information for their truncated versions and then taking
limit in $k$.
\end{rem}
\begin{cor}
\label{cor:continuous} Assume $\pi_{XY}$ is an absolutely continuous
distribution on $\mathbb{R}^{2}$ such that\textup{\emph{ }}$\widehat{C}_{\mathrm{Wyner}}(\pi_{XY})=C_{\mathrm{Wyner}}(\pi_{XY})$
(e.g., at least one of the conditions given in Proposition \ref{prop:GeneralWyner-1}
is satisfied) and its pdf\footnote{For brevity, we use the same notation $\pi_{XY}$ to denote both an
absolutely continuous distribution and the corresponding pdf. } $\pi_{XY}$ is log-concave\footnote{A pdf $\pi_{XY}$ is log-concave if $\log\pi_{XY}$ is concave.}
and differentiable. Assume $I(X;Y)$ exists (and hence is finite).
For $d>0$, define 
\begin{equation}
L_{d}:=\sup_{\left(x,y\right)\in[-d,d]^{2}}\left|\frac{\partial}{\partial x}\log\pi_{XY}\left(x,y\right)\right|+\left|\frac{\partial}{\partial y}\log\pi_{XY}\left(x,y\right)\right|,\label{eq:-78}
\end{equation}
and 
\begin{equation}
\epsilon_{d}:=1-\pi_{XY}\left([-d,d]^{2}\right).\label{eq:-6}
\end{equation}
Assume that $\epsilon_{d}\log\left(dL_{d}\right)\to0$ as $d\to+\infty$.
Then we have 
\begin{equation}
\widetilde{T}_{1}(\pi_{XY})=T_{1}(\pi_{XY})=C_{\mathrm{Wyner}}(\pi_{XY}).\label{eqn:stronger-1-2-1-1}
\end{equation}
\end{cor}
\begin{rem}
\label{rem:If-the-pdf}If the pdf $\pi_{XY}$ is \emph{not} differentiable,
then Corollary \ref{cor:continuous} still holds if 1) the pdf $\pi_{XY}$
is continuous (this is also implied by the log-concavity of the pdf
$\pi_{XY}$) and 2)  the definition of $L_{d}$ in \eqref{eq:-78}
is replaced with 
\begin{equation}
L_{d}:=\sup_{\Delta\ge0}\frac{1}{\Delta}\log\sup_{\substack{\left(x,y\right),\left(\hat{x},\hat{y}\right)\in[-d,d]^{2}:\\
\left|x-\hat{x}\right|,\left|y-\hat{y}\right|\leq\Delta
}
}\frac{\pi_{XY}\left(x,y\right)}{\pi_{XY}\left(\hat{x},\hat{y}\right)}.\label{eq:-27-1}
\end{equation}
This claim follows since in our proof of Corollary \ref{cor:continuous},
the assumption of differentiability of the pdf $\pi_{XY}$ is used
to upper bound the RHS of \eqref{eq:-27-1} by using $L_{d}$  (see
Lemma \ref{lem:RatioBounds}); however, adopting the definition of
$L_{d}$ in \eqref{eq:-27-1} avoids this complicated derivation,
since it directly relates $L_{d}$ to the RHS of \eqref{eq:-27-1}.
\end{rem}
Now we consider bivariate Gaussian sources $(X,Y)$. Without loss
of any generality, we assume that the correlation coefficient $\rho$
between $X$ and $Y$ is nonnegative; otherwise, we can set $-X$
to $X$. For this case, 
\begin{equation}
C_{\mathrm{Wyner}}(\pi_{XY})=\frac{1}{2}\log\left[\frac{1+\rho}{1-\rho}\right]
\end{equation}
and it is attained by the joint Gaussian distribution $P_{W}P_{X|W}P_{Y|W}$
with $P_{W}=\mathcal{N}(0,\rho),P_{X|W}(\cdot|w)=\mathcal{N}(w,1-\rho),P_{Y|W}(\cdot|w)=\mathcal{N}(w,1-\rho)$
\cite{xu2013wyner,yu2016generalized}. Using the formula for R\'enyi
divergences between Gaussian distributions derived in \cite{gil2013renyi},
we obtain that for $0<s\leq\sqrt{\frac{1+\rho}{2\rho}}$, 
\begin{align}
 & D_{1+s}(P_{X|W}P_{Y|W}\|P_{XY}|P_{W})\nonumber \\
 & =\frac{1}{2}\log\left[\frac{1+\rho}{1-\rho}\right]-\frac{1}{2s}\log\left(1-\frac{2s^{2}\rho}{1+\rho}\right).
\end{align}
Hence Gaussian sources satisfy the sufficient condition 1) given in
Proposition \ref{prop:GeneralWyner-1}, which in turn implies $\widehat{C}_{\mathrm{Wyner}}(\pi_{XY})=C_{\mathrm{Wyner}}(\pi_{XY})$.
Furthermore, it is easy to verify that other conditions given in Corollary
\ref{cor:continuous} are also satisfied by Gaussian sources. Hence
we obtain the following result.
\begin{cor}
\label{cor:Gaussian}For a bivariate Gaussian source $(X,Y)$ with
correlation coefficient $\rho\in[0,1),$ we have 
\begin{equation}
\widetilde{T}_{1}(\pi_{XY})=T_{1}(\pi_{XY})=C_{\mathrm{Wyner}}(\pi_{XY})=\frac{1}{2}\log\left[\frac{1+\rho}{1-\rho}\right].\label{eq:Gaussian-1}
\end{equation}
\end{cor}
\begin{IEEEproof}
The last equality in \eqref{eq:Gaussian-1} was proven in \cite{xu2013wyner,yu2016generalized}.
The first two equalities in \eqref{eq:Gaussian-1} are implied by
Corollary \ref{cor:continuous}, since it is easy to verify that the
hypotheses as stated in Corollary \ref{cor:continuous} are satisfied
by Gaussian sources.
\end{IEEEproof}

If we replace the relative entropy measure with the TV-distance, we
can define the TV-distance version of Wyner's common information as
\begin{align}
 & T_{\mathrm{TV}}(\pi_{XY}):=\inf\Big\{ R:\;\lim_{n\to\infty}\left|P_{X^{n}Y^{n}}-\pi_{XY}^{n}\right|=0\Big\}.\label{eq:-170}
\end{align}
By replacing the relative entropy with the TV-distance in our proofs,
one can easily obtain the following result. The proof is similar to
the ones for the relative entropy versions, and hence is omitted here.
\begin{thm}
Redefine 
\begin{equation}
\widetilde{C}_{\mathrm{Wyner}}(\pi_{XY}):=\lim_{\epsilon\downarrow0}\inf_{P_{W}P_{X|W}P_{Y|W}:\left|P_{XY}-\pi_{XY}\right|\le\epsilon}I\left(XY;W\right).\label{eq:-53-3}
\end{equation}
Then Theorem \ref{thm:GeneralWyner}  as well as Corollaries \ref{cor:countable},
\ref{cor:continuous}, and \ref{cor:Gaussian} hold \emph{mutatis
mutandis} for the TV-distance version of Wyner's common information.
\end{thm}
\begin{rem}
For the TV-distance version, the upper bound in Theorem \ref{thm:GeneralWyner}
$\widehat{C}_{\mathrm{Wyner}}(\pi_{XY})$ can be tightened to be $C_{\mathrm{Wyner}}(\pi_{XY})$,
by replacing Lemma \ref{lem:oneshotach} in the proof with \cite[Theorem VII.1]{Cuff}.
This in turn implies that the requirement of existence of $H_{\alpha}(\pi_{XY})$
for some $\alpha\in[0,1)$ in Corollary \ref{cor:countable} can be
relaxed to that of existence of $H(\pi_{XY})$, and the requirement
that $\widehat{C}_{\mathrm{Wyner}}(\pi_{XY})=C_{\mathrm{Wyner}}(\pi_{XY})$
in Corollary \ref{cor:continuous} can be removed.
\end{rem}
Since it is difficult to obtain closed-form expressions for the optimal
joint distributions $P_{W}P_{X|W}P_{Y|W}$ attaining $C_{\mathrm{Wyner}}(\pi_{XY})$,
the sufficient conditions for $\widehat{C}_{\mathrm{Wyner}}(\pi_{XY})=C_{\mathrm{Wyner}}(\pi_{XY})$
given in Proposition \ref{prop:GeneralWyner-1} is difficult to verify.
However, for the TV-distance version, the requirement of $\widehat{C}_{\mathrm{Wyner}}(\pi_{XY})=C_{\mathrm{Wyner}}(\pi_{XY})$
in Corollary \ref{cor:continuous} can be removed. Hence for this
case, the conditions in Corollary \ref{cor:continuous} can be easily
verified. For example, besides Gaussian sources, ``Laplacian'' sources\footnote{Note that here $\pi_{XY}(x,y)\propto\exp\left(-\left|ax+by\right|\right)$
is not the common bivariate Laplacian distribution.} $\pi_{XY}(x,y)\propto\exp\left(-\left|ax+by\right|\right)$ with
$a,b\in\mathbb{R}$ also satisfy the required conditions, and hence
for ``Laplacian'' sources, $T_{\mathrm{TV}}(\pi_{XY})=C_{\mathrm{Wyner}}(\pi_{XY})$.

The exponential strong converse holds for the TV-distance version
of Wyner's common information when the alphabet is finite; see \cite{yu2018wyner,yu2018corrections}.
We conjecture that the exponential strong converse also holds when
the alphabet is infinite (countably infinite or uncountable).

\subsection{Exact and $\infty$-R\'enyi Common Informations}

Now we generalize exact and $\infty$-R\'enyi common informations to
sources with countably infinite alphabets and a certain class of continuous
sources.

\subsubsection{Equivalence}

In Theorem \ref{thm:equivalence}, we established the equivalence
between the exact and $\infty$-R\'enyi common informations for sources
with finite alphabets. Now we extend it to the countably infinite
alphabet case.
\begin{thm}[Equivalence]
\label{thm:equivalence-1} Let $(X,Y)$ be a source with distribution
$\pi_{XY}$ defined on the product of two countably infinite alphabets.
Assume $H(\pi_{XY})$ exists (and hence is finite). Then we have
\begin{align}
T_{\mathrm{Exact}}(\pi_{XY}) & =T_{\infty}(\pi_{XY}).
\end{align}
\end{thm}
For sources with discrete (finite or countably infinite) or continuous
alphabets, we have shown $T_{\mathrm{Exact}}(\pi_{XY})\ge T_{\infty}(\pi_{XY})$
in Lemma \ref{lem:exactlarger} in Appendix \ref{sec:equivalence}.
Thus it suffices to prove the reverse inequality. 
\begin{lem}
\label{lem:countable}Let $(X,Y)$ be a source with distribution $\pi_{XY}$
defined on the product of two countably infinite alphabets. Assume
$H(\pi_{XY})$ exists (and hence is finite). Then for a source with
such a distribution $\pi_{XY}$, if there exists a sequence of fixed-length
codes with rate $R$ that generates $P_{X^{n}Y^{n}}$ such that $D_{\infty}(P_{X^{n}Y^{n}}\|\pi_{XY}^{n})\to0$,
then there must exist a sequence of variable-length codes with rate
$R$ that exactly generates $\pi_{XY}^{n}$. That is, $T_{\mathrm{Exact}}(\pi_{XY})\leq T_{\infty}(\pi_{XY})$.
\end{lem}
The proof of Lemma \ref{lem:countable} is given in Appendix \ref{sec:countable-1}.

Until now, we have shown that $T_{\mathrm{Exact}}(\pi_{XY})\ge T_{\infty}(\pi_{XY})$
holds for sources with discrete or continuous alphabets, and $T_{\mathrm{Exact}}(\pi_{XY})\leq T_{\infty}(\pi_{XY})$
holds for sources with discrete alphabets. However, we do not know
whether $T_{\mathrm{Exact}}(\pi_{XY})\leq T_{\infty}(\pi_{XY})$ always
holds for continuous sources. Next we prove that it indeed holds if
continuous sources satisfy certain regularity conditions, and the
optimal (minimum) $\infty$-R\'enyi divergence $D_{\infty}(P_{X^{n}Y^{n}}\|\pi_{XY}^{n})$
converges to zero sufficiently fast. The proof of Lemma \ref{lem:continuous}
is given in Appendix \ref{sec:continuous-1}.
\begin{lem}
\label{lem:continuous} Assume $\pi_{XY}$ is an absolutely continuous
distribution on $\mathbb{R}^{2}$ with $\mathbb{E}\left[X^{2}\right],\mathbb{E}\left[Y^{2}\right]<\infty$.
Without loss of generality, we assume $\mathbb{E}\left[X^{2}\right]=\mathbb{E}\left[Y^{2}\right]=1$.
Assume the pdf of $\pi_{XY}$ is log-concave, and continuously differentiable.
Assume $I(X;Y)$ exists (and hence is finite). For $\epsilon>0$ and
$n\in\mathbb{N}$, define 
\begin{align}
L_{\epsilon,n} & :=\sup_{\left(x,y\right)\in\mathcal{\mathcal{L}}_{\epsilon,n}^{2}}\biggl\{\left|\frac{\partial}{\partial x}\log\pi_{XY}\left(x,y\right)\right|\nonumber \\
 & \qquad+\left|\frac{\partial}{\partial y}\log\pi_{XY}\left(x,y\right)\right|\biggr\},\label{eq:-97}
\end{align}
where 
\begin{equation}
\mathcal{\mathcal{L}}_{\epsilon,n}:=\left\{ x\in\mathbb{R}:|x|\leq\sqrt{n\left(1+\epsilon\right)}\right\} .\label{eq:-36}
\end{equation}
Assume $\log L_{\epsilon,n}$ is sub-exponential in $n$ for fixed
$\epsilon$ (i.e., $\lim_{n\to\infty}\frac{1}{n}\log\log L_{\epsilon,n}=0$
for all fixed $\epsilon>0$). Then for a source with such a distribution
$\pi_{XY}$, if there exists a sequence of fixed-length codes with
rate $R$ that generates $P_{X^{n}Y^{n}}$ such that $D_{\infty}(P_{X^{n}Y^{n}}\|\pi_{XY}^{n})=o\left(\frac{1}{n+\log L_{\epsilon,n}}\right)$
for any $\epsilon>0$, then there must exist a sequence of variable-length
codes with rate $R$ that exactly generates $\pi_{XY}^{n}$. That
is, $T_{\mathrm{Exact}}(\pi_{XY})\leq T_{\infty}'(\pi_{XY})$, where
\begin{align}
T_{\infty}'(\pi_{XY}) & :=\inf\Biggl\{ R:\;D_{\infty}(P_{X^{n}Y^{n}}\|\pi_{XY}^{n})\nonumber \\
 & \qquad=o\left(\frac{1}{n+\log L_{\epsilon,n}}\right),\forall\epsilon>0\Biggr\}.
\end{align}
\end{lem}
\begin{rem}
Similar to Remark \ref{rem:If-the-pdf}, if the pdf $\pi_{XY}$ is
not differentiable, then Corollary \ref{cor:continuous} still holds
if 1) the pdf $\pi_{XY}$ is continuous and 2)  the definition of
$L_{\epsilon,n}$ in \eqref{eq:-97} is replaced with 
\begin{equation}
L_{\epsilon,n}:=\sup_{\Delta\ge0}\frac{1}{\Delta}\log\sup_{\substack{\left(x,y\right),\left(\hat{x},\hat{y}\right)\in\mathcal{\mathcal{L}}_{\epsilon,n}^{2}:\\
\left|x-\hat{x}\right|,\left|y-\hat{y}\right|\leq\Delta
}
}\frac{\pi_{XY}\left(x,y\right)}{\pi_{XY}\left(\hat{x},\hat{y}\right)}.\label{eq:-27-1-2}
\end{equation}
\end{rem}
\begin{rem}
\label{rem:Gaussian}One important example satisfying the conditions
in the lemma above is bivariate Gaussian sources. Consider a bivariate
Gaussian source $\pi_{XY}=\mathcal{N}\left(0,\Sigma_{XY}\right)$
where $\Sigma_{XY}=\left[\begin{array}{cc}
1 & \rho\\
\rho & 1
\end{array}\right]$ with $\rho\in[0,1)$. For this case, 
\begin{align}
L_{\epsilon,n} & =\sup_{\left(x,y\right)\in\mathcal{\mathcal{L}}_{\epsilon,n}^{2}}\left|\frac{x-\rho y}{1-\rho^{2}}\right|+\left|\frac{y-\rho x}{1-\rho^{2}}\right|\\
 & =\frac{2\sqrt{n\left(1+\epsilon\right)}}{1-\rho}.
\end{align}
Hence $\log L_{\epsilon,n}$ is sub-exponential in $n$ for fixed
$\epsilon$. Observe that $\frac{1}{n+\log L_{\epsilon,n}}\sim\frac{1}{n}$.
Hence, by this lemma, if there exists a sequence of fixed-length codes
with rate $R$ that generates $P_{X^{n}Y^{n}}$ such that $D_{\infty}(P_{X^{n}Y^{n}}\|\pi_{XY}^{n})=o\left(\frac{1}{n}\right)$,
then there must exist a sequence of variable-length codes with rate
$R$ that exactly generates $\pi_{XY}^{n}$. 
\end{rem}

\subsubsection{Discrete Sources with Countably Infinite Alphabets}

We now generalize the exact and $\infty$-R\'enyi common informations
to sources with countably infinite alphabets. In the proof of Theorem
\ref{thm:equivalence}, a truncated i.i.d. code was adopted to prove
the achievability part, in which the codewords are i.i.d. with each
drawn according to a set of truncated distributions (obtained by truncating
a set of product distributions into some (strongly) typical sets).
For the countably infinite alphabet case, we need replace strongly
typical sets with unified typical sets (defined in \eqref{eq:unified}).
Then we establish the following result.
\begin{cor}
\label{cor:countable-1-1} Let $(X,Y)$ be a source with distribution
$\pi_{XY}$ defined on the product of two countably infinite alphabets.
Assume $H(\pi_{XY})$ exists (and hence is finite). We have 
\begin{align}
\max\left\{ \widehat{\Gamma}^{\mathrm{LB}}(\pi_{XY}),C_{\mathsf{Wyner}}(\pi_{XY})\right\}  & \leq\widetilde{T}_{\infty}(\pi_{XY})\\
 & \le T_{\infty}(\pi_{XY})\\
 & =T_{\mathrm{Exact}}(\pi_{XY})\\
 & \le\widehat{\Gamma}^{\mathrm{UB}}(\pi_{XY}),
\end{align}
where 
\begin{align}
 & \widehat{\Gamma}^{\mathrm{UB}}(\pi_{XY}):=\lim_{\epsilon\downarrow0}\inf_{\substack{\substack{P_{W}P_{X|W}P_{Y|W}:}
\\
P_{XY}=\pi_{XY}
}
}\sup_{\substack{\substack{Q_{XYW}:}
\\
D\left(Q_{WX}\|P_{WX}\right)\le\epsilon,\\
D\left(Q_{WY}\|P_{WY}\right)\le\epsilon
}
}\nonumber \\
 & \qquad\left\{ -\sum_{w,x,y}P(w)Q\left(x,y|w\right)\log\pi\left(x,y\right)-H(XY|W)\right\} \label{eq:-23}
\end{align}
and 
\begin{align}
 & \widehat{\Gamma}^{\mathrm{LB}}(\pi_{XY}):=\lim_{\epsilon\downarrow0}\inf_{\substack{\substack{P_{W}P_{X|W}P_{Y|W}:}
\\
D\left(P_{XY}\|\pi_{XY}\right)\le\epsilon
}
}\Bigl\{-H(XY|W)\nonumber \\
 & \qquad+\inf_{Q_{WW'}\in C(P_{W},P_{W})}\sum_{w,w'}Q(w,w')\nonumber \\
 & \qquad\times\mathcal{H}(P_{X|W=w},P_{Y|W=w'}\|\pi_{XY})\Bigr\}.\label{eq:-59-2}
\end{align}
\end{cor}
For the finite alphabet case, the $\epsilon$'s in the optimizations
in \eqref{eq:-23} and \eqref{eq:-59-2} can be removed by using the
compactness technique or the splitting technique. For the countably
infinite alphabet case, in general we cannot apply the compactness
technique. However, it may be possible to apply the splitting technique
to remove $\epsilon$'s, similarly as in the proof of Corollary \ref{cor:countable}.
Nevertheless, we need carefully deal with the terms involving $\log\pi\left(x,y\right)$
in \eqref{eq:-23} and \eqref{eq:-59-2}, since a little difference
between $Q_{XY}$ and $\pi_{XY}$ could lead to a large increase of
$\sum_{x,y}Q(x,y)\log\frac{1}{\pi\left(x,y\right)}$.

\subsubsection{Gaussian Sources}

Next we generalize the exact and $\infty$-R\'enyi common informations
to a certain class of continuous sources. We provide an upper bound
on $T_{\mathrm{Exact}}(\pi_{XY})$ and $T_{\infty}(\pi_{XY})$ for
bivariate Gaussian sources $\pi_{XY}$. Without loss of any generality,
we assume that the correlation coefficient $\rho$ between $(X,Y)$
is nonnegative. The proof of Theorem \ref{thm:Gaussian} is given
in Appendix \ref{sec:Gaussian}. 
\begin{thm}
\label{thm:Gaussian} For a Gaussian source $(X,Y)$ with correlation
coefficient $\rho\in[0,1),$ we have 
\begin{align}
\frac{1}{2}\log\left[\frac{1+\rho}{1-\rho}\right] & \leq\widetilde{T}_{\infty}(\pi_{XY})\\
 & \leq T_{\infty}(\pi_{XY})\\
 & =T_{\mathrm{Exact}}(\pi_{XY})\\
 & \leq\frac{1}{2}\log\left[\frac{1+\rho}{1-\rho}\right]+\frac{\rho}{1+\rho}.\label{eq:Gaussian}
\end{align}
\end{thm}
\begin{rem}
For Gaussian sources $(X,Y)$ with correlation coefficient $\rho\in[0,1),$
Li and El Gamal \cite{li2017distributed} provided the following upper
bound 
\begin{equation}
T_{\mathrm{Exact}}(\pi_{XY})\leq\frac{1}{2}\log\left[\frac{1}{1-\rho^{2}}\right]+24\log2.\label{eq:Gaussian2}
\end{equation}
Such an upper bound is a one-shot bound, and hence it is also valid
for the case with blocklength equal to $1$. However, our upper bound
requires blocklength to be infinity. Furthermore, for the asymptotic
case, Li and El Gamal's bound is rather loose, since the difference
between the upper bounds in \eqref{eq:Gaussian2} and \eqref{eq:Gaussian}
is 
\begin{align}
 & \frac{1}{2}\log\left[\frac{1}{1-\rho^{2}}\right]+24\log2-\left(\frac{1}{2}\log\left[\frac{1+\rho}{1-\rho}\right]+\frac{\rho}{1+\rho}\right)\nonumber \\
 & =24\log2-1+\frac{1}{1+\rho}+\log\left[\frac{1}{1+\rho}\right]\\
 & \geq15.44\textrm{ Nats}/\textrm{Symbol}\\
 & =22.28\textrm{ Bits}/\textrm{Symbol}.
\end{align}
Li and El Gamal's bound was proven by using a dyadic decomposition
scheme which decomposes the joint distribution into a sequence of
\emph{uniform} distributions. For such a scheme, even if the source
$\pi_{XY}$ is comprised of two independent components and at least
one of them is not uniform (i.e., $\pi_{XY}=\pi_{X}\pi_{Y}$ but either
$\pi_{X}$ or $\pi_{Y}$ is not uniform), the induced common randomness
rate between them is still strictly positive. This is because for
this case, Li and El Gamal's dyadic decomposition scheme cannot identify
the optimal decomposition $\pi_{XY}=\pi_{X}\pi_{Y}$. Hence the common
randomness rate induced by Li and El Gamal's scheme does not cross
$0$ for $\rho=0$. In addition, it is worth noting that our exact
common information scheme is a mixture of Li and El Gamal's scheme
and an $\infty$-R\'enyi common information scheme. In our scheme, Li
and El Gamal's scheme is invoked with asymptotically vanishing probability,
and hence the performance of our scheme is dominated by the $\infty$-R\'enyi
common information scheme which requires a much lower rate. 
\end{rem}
For the DSBS case, our upper bound is tight. Hence it is natural to
conjecture that for Gaussian sources, the upper bound in \eqref{eq:Gaussian}
is also tight. Similarly to the discrete source case, one can show
the following lower bound on $T_{\mathrm{Exact}}(\pi_{XY})$ and $T_{\infty}(\pi_{XY})$
holds for continuous sources (including Gaussian sources). 
\begin{align}
 & \widehat{\Gamma}^{\mathrm{LB}}(\pi_{XY}):=\lim_{\epsilon\downarrow0}\inf_{\substack{\substack{P_{W}P_{X|W}P_{Y|W}:}
\\
D\left(P_{XY}\|\pi_{XY}\right)\le\epsilon
}
}\Bigl\{-h(XY|W)\nonumber \\
 & \qquad+\inf_{Q_{WW'}\in C(P_{W},P_{W})}\sum_{w,w'}Q(w,w')\nonumber \\
 & \qquad\times\mathcal{H}(P_{X|W=w},P_{Y|W=w'}\|\pi_{XY})\Bigr\},\label{eq:-59-2-5}
\end{align}
where $P_{W}$ is a discrete distribution, given $w,w'$, $P_{X|W=w},P_{Y|W=w'}$
are continuous distributions, and $\mathcal{H}(P_{X|W=w},P_{Y|W=w'}\|\pi_{XY})$
is the maximal (differential) cross-entropy defined in \eqref{eq:maximalcrossentropy}
(with $Q_{XY},\pi_{XY}$ denoting the pdfs rather than pmfs). However,
we do not know how to prove $\widehat{\Gamma}^{\mathrm{LB}}(\pi_{XY})\geq\frac{1}{2}\log\left[\frac{1+\rho}{1-\rho}\right]+\frac{\rho}{1+\rho}$.
Furthermore, it is possible to generalize the upper bound in Theorem
\ref{thm:Gaussian} to other continuous sources by utilizing general
typicality, e.g., \cite{raginsky2013empirical,jeon2014generalized}.

\begin{figure*}
\centering \includegraphics[width=0.7\textwidth]{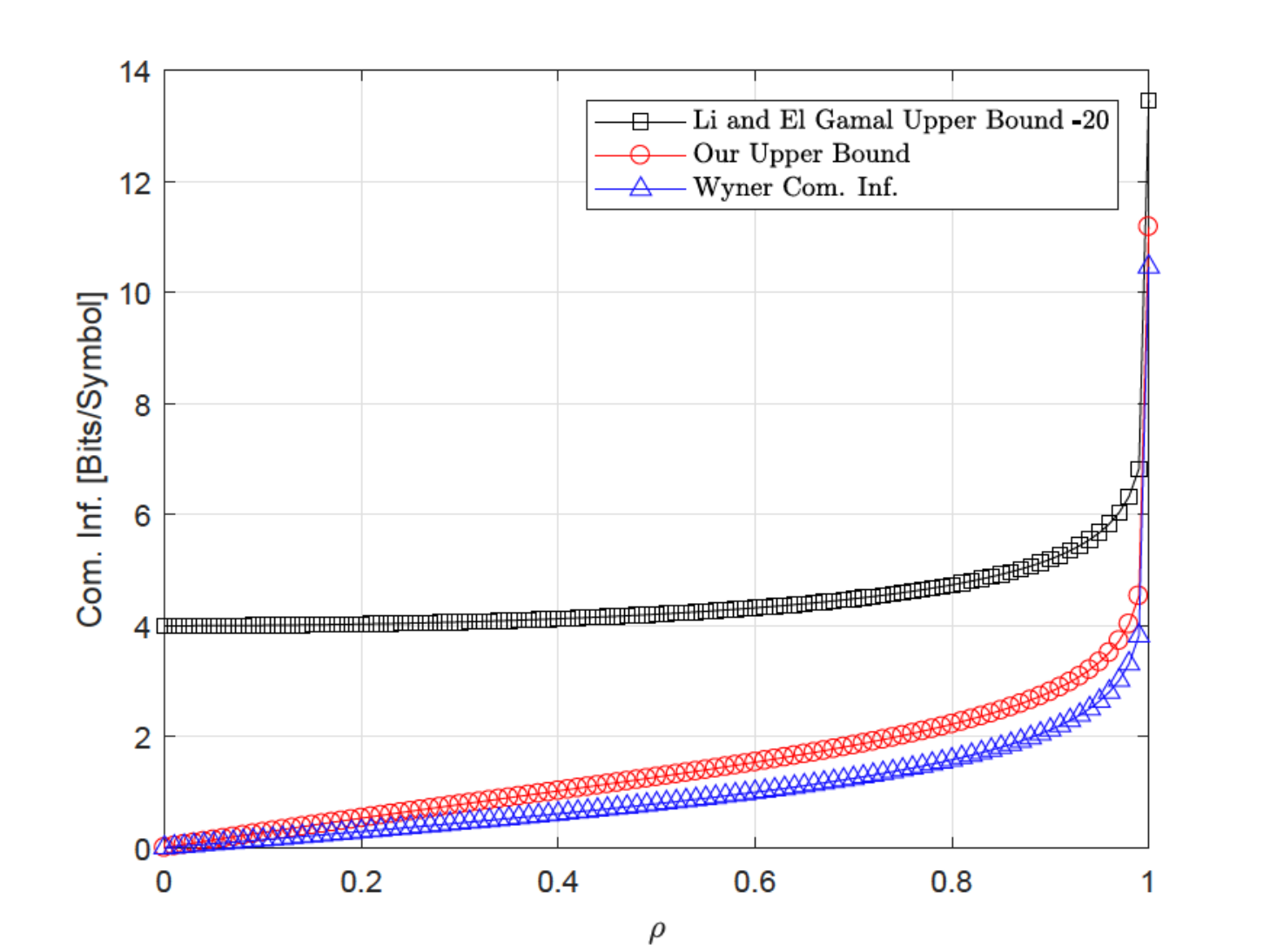}

\caption{\label{fig:Common-informations-for-1}Illustrations of Wyner's common
information \eqref{eq:Gaussian-1}, as well as Li and El Gamal's upper
bound \eqref{eq:Gaussian2} and our upper bound \eqref{eq:Gaussian}
on the exact and $\infty$-R\'enyi common informations for Gaussian
sources with correlation coefficient $\rho\in[0,1)$. For ease of
comparison, here we plot Li and El Gamal's upper bound minus 20 (bits/symbol),
rather than their bound itself, since their bound is much larger than
our bound and Wyner's common information. }
\end{figure*}

For Gaussian sources, Li and El Gamal's upper bound in \eqref{eq:Gaussian2},
our upper bound in \eqref{eq:Gaussian}, and Wyner's common information
in \eqref{eq:Gaussian-1} are illustrated in Fig.~\ref{fig:Common-informations-for-1}.
The exact and $\infty$-R\'enyi common informations are lower bounded
by Wyner's common information. Hence the exact and $\infty$-R\'enyi
common informations are between Wyner's common information and our
bound. The gap between them is $\frac{\rho}{1+\rho}\le0.5$ nats/symbol
or $0.72$ bits/symbol.

\section{Connection to Other Problems}

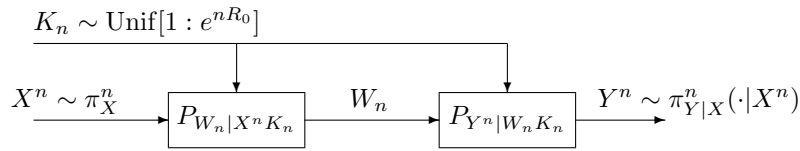
\begin{figure*}
\centering \setlength{\unitlength}{0.06cm} { \begin{picture}(140,35)
\put(-5,10){\vector(1,0){30}} \put(-10,13){%
\mbox{%
$X^{n}\sim\pi_{X}^{n}$%
}} \put(25,4){\framebox(30,12){$P_{W_{n}|X^{n}K_{n}}$}} \put(55,10){\vector(1,0){30}}
\put(65,13){%
\mbox{%
$W_{n}$%
}} \put(85,4){\framebox(30,12){$P_{Y^{n}|W_{n}K_{n}}$}} \put(115,10){\vector(1,0){20}}
\put(120,13){%
\mbox{%
$Y^{n}\sim\pi_{Y|X}^{n}(\cdot|X^{n})$%
}} \put(40,27){\vector(0,-1){11}} \put(-5,30){%
\mbox{%
$K_{n}\sim\mathrm{Unif}[1:e^{nR_{0}}]$%
}} \put(100,27){\vector(0,-1){11}} \put(-5,27){\line(1,0){105}}
\end{picture}}

\caption{\label{fig:dcs}The exact channel synthesis problem. We would like
to design the code $\left(P_{W_{n}|X^{n}K_{n}},P_{Y^{n}|W_{n}K_{n}}\right)$
such that the induced conditional distribution $P_{Y^{n}|X^{n}}$
satisfies $P_{Y^{n}|X^{n}}=\pi_{Y|X}^{n}$.}
\end{figure*}

The exact common information problem is related to (or can be generalized
to) the following problems.
\begin{itemize}
\item Distributed Channel Synthesis
\end{itemize}
In both the exact and TV-approximate senses, the common information
problem is equivalent to the distributed channel simulation problem
(with no shared information). The distributed channel simulation problem
(or the communication complexity problem for generating correlation),
illustrated in Fig. \ref{fig:dcs}, was studied in \cite{bennett2002entanglement,winter2002compression,Cuff,bennett2014quantum,harsha2010communication}.
The distributed exact (resp. TV-approximate) channel simulation problem
refers to determining the minimum communication rate needed to generate
two correlated sources $\left(X^{n},Y^{n}\right)$ respectively at
the encoder and decoder such that the induced joint distribution $P_{X^{n}Y^{n}}$
exactly equals $\pi_{XY}^{n}$ (resp. the TV distance $P_{X^{n}Y^{n}}$
and $\pi_{XY}^{n}$ vanishes asymptotically).

The exact common information problem (or exact correlation generation
problem) is essentially equivalent to the distributed channel simulation
problem with no shared information (or the communication complexity
problem for generating correlation) \cite{bennett2002entanglement,winter2002compression,Cuff,bennett2014quantum,harsha2010communication}
(illustrated in Fig. \ref{fig:dcs} with $R_{0}=0$). This can be
easily obtained by observing that if there exists an exact common
information code $(P_{M_{n}},P_{X^{n}|M_{n}},P_{Y^{n}|M_{n}})$ then
$(P_{M_{n}|X^{n}},P_{Y^{n}|M_{n}})$ forms an exact channel synthesis
code; and vice versa.

In the literature, Bennett \emph{et al.} \cite{bennett2002entanglement}
studied exact syntheses of a target channel when there is \emph{unlimited
shared randomness}, i.e., $R_{0}=\infty$, available at the encoder
and decoder. They showed that the minimum communication rates for
this case is equal to the mutual information $I_{\pi}(X;Y)$ in which
$(X,Y)\sim\pi_{XY}$. Harsha \emph{et al.} \cite{harsha2010communication}
used a rejection sampling scheme to prove a one-shot bound for \emph{exact}
simulation for finitely-supported $(X,Y)$. They showed that the number
of bits of the shared randomness can be limited to $O(\log\log|\mathcal{X}|+\log|\mathcal{Y}|)$
if the expected description length is increased by $O(\log\left(I_{\pi}(X;Y)+1\right)+\log\log|\mathcal{Y}|)$
bits from the lower bound $I_{\pi}(X;Y)$. Li and El Gamal \cite{li2018strong}
used functional representation lemma to prove that if the expected
description length is increased by $\log(I_{\pi}(X;Y)+1)+5$ bits
from $I_{\pi}(X;Y)$, then the number of bits of the shared randomness
can be upper bounded by $\log(|\mathcal{X}|(|\mathcal{Y}|-1)+2)$.
The tradeoff between the communication rate and the shared randomness
rate for exact synthesis of the symmetric binary erasure source (SBES)
was characterized by Kumar, Li, and El Gamal \cite{Kumar}. Recently,
we extend the results and the proof techniques in this paper to study
the tradeoff between the communication rate and the shared randomness
rate for exact synthesis of discrete and continuous memoryless channels.
In particular, we completely characterized the tradeoff for DSBSes.
Furthermore, there are also multiple works, e.g., \cite{bennett2002entanglement,winter2002compression,Cuff},
studying \emph{approximate} syntheses of a target channel, in which
the distance between the generated channel and the target channel
is required to converge to zero asymptotically.

\begin{itemize}
\item Exact $\ensuremath{\alpha}$-R\'enyi Common Informations
\end{itemize}
As shown in \cite{Kumar} (or \eqref{eq:-106}), the exact common
information for $\pi_{XY}$ is equal to
\begin{equation}
\lim_{n\to\infty}\frac{1}{n}\min_{P_{W}P_{X^{n}|W}P_{Y^{n}|W}:P_{X^{n}Y^{n}}=\pi_{XY}^{n}}H(W).\label{eq:-106-1}
\end{equation}
Note that the \emph{$\alpha$-}R\'enyi entropy with $\alpha\in[-\infty,\infty]$
is defined as 
\begin{align}
H_{\alpha}(W) & :=\frac{1}{1-\alpha}\log\sum_{w\in\mathrm{supp}(P_{W})}P_{W}(w)^{\alpha}\label{eq:-40-1}
\end{align}
for $\alpha\notin\{-\infty,1,\infty\}$ and where $H_{-\infty}=\lim_{\alpha\downarrow-\infty}H_{\alpha},H_{\infty}=\lim_{\alpha\uparrow\infty}H_{\alpha}$,
and $H_{1}=\lim_{\alpha\uparrow1}H_{\alpha}$.  The \emph{$\alpha$-}R\'enyi
entropy is a natural generalization of the Shannon entropy. For $\pi_{XY}$,
we define the \emph{common $\alpha$-R\'enyi entropy} with $\alpha\in[-\infty,\infty]$
as 
\begin{equation}
G_{\alpha}(\pi_{XY}):=\min_{P_{W}P_{X|W}P_{Y|W}:P_{XY}=\pi_{XY}}H_{\alpha}(W).\label{eq:-113}
\end{equation}
(The common \emph{$\alpha$-}R\'enyi entropy is a generalization of
the common entropy \cite{witsenhausen1976values,Kumar}; see the definition
of the common entropy in \eqref{eq:commonentropy}). The exact common
information can be generalized to the \emph{exact $\alpha$-R\'enyi
common information }with $\alpha\in[-\infty,\infty]$, which is defined
as 
\begin{equation}
T_{\mathrm{Exact}}^{(\alpha)}(\pi_{XY}):=\lim_{n\to\infty}\frac{1}{n}G_{\alpha}(\pi_{XY}^{n}).\label{eq:-109}
\end{equation}
Here the existence of the limit in \eqref{eq:-109} follows by the
subadditivity of the sequence of $\left\{ G_{\alpha}(\pi_{XY}^{n})\right\} _{n\in\mathbb{N}}$.

Since $H_{\alpha}$ is non-increasing in $\alpha\in[-\infty,\infty]$,
we have that $G_{\alpha}(\pi_{XY})$ and $T_{\mathrm{Exact}}^{(\alpha)}(\pi_{XY})$
are also non-increasing in $\alpha\in[-\infty,\infty]$. Furthermore,
for $\alpha\in\left\{ 0,1,\infty\right\} $, we have the following
characterization of $G_{\alpha}(\pi_{XY})$ and $T_{\mathrm{Exact}}^{(\alpha)}(\pi_{XY})$.
The proof is provided in Appendix \ref{sec:exactrenyi}.
\begin{prop}
\label{prop:exactRenyi}We have
\begin{equation}
G_{\alpha}(\pi_{XY})=\begin{cases}
\log\mathrm{rank}^{+}(\pi_{XY}), & \alpha=0\\
\min_{P_{W}P_{X|W}P_{Y|W}:P_{XY}=\pi_{XY}}H(W), & \alpha=1\\
\min_{Q_{X},Q_{Y}}D_{\infty}(Q_{X}Q_{Y}\|\pi_{XY}), & \alpha=\infty
\end{cases}\label{eq:-159}
\end{equation}
where $\mathrm{rank}^{+}(\mathbf{A})$ denotes the \emph{nonnegative
rank }of a matrix $\mathbf{A}$, i.e., the minimum $k\in\mathbb{N}$
such that there exist nonnegative matrices $\mathbf{U}\in\mathbb{R}_{\ge0}^{k\times|\mathcal{X}|},\mathbf{V}\in\mathbb{R}_{\ge0}^{k\times|\mathcal{Y}|}$
satisfying $\mathbf{U}\mathbf{V}^{\top}=\mathbf{A}$. Furthermore,
for $\alpha=\infty$,
\[
T_{\mathrm{Exact}}^{(\infty)}(\pi_{XY})=\min_{Q_{X},Q_{Y}}D_{\infty}(Q_{X}Q_{Y}\|\pi_{XY}).
\]
\end{prop}
By definition, the exact $0$-R\'enyi common information  corresponds
to the minimum common randomness rate for exact generation of the
target distribution in which the common randomness is only allowed
to be compressed by \emph{fixed-length} codes. By \eqref{eq:-159},
the exact $0$-R\'enyi common information can be expressed as
\begin{align}
T_{\mathrm{Exact}}^{(0)}(\pi_{XY}) & =\lim_{n\to\infty}\frac{1}{n}\log\mathrm{rank}^{+}(\pi_{XY}^{\otimes n}),\label{eq:-169}
\end{align}
where $\pi_{XY}^{\otimes n}$ denotes the Kronecker product of $n$
copies of the matrix $\pi_{XY}$. That is, $T_{\mathrm{Exact}}^{(0)}(\pi_{XY})$
is the exponent of $\mathrm{rank}^{+}(\pi_{XY}^{\otimes n})$ as $n\to\infty$.
By definition, we can easily obtain
\begin{equation}
\log\mathrm{rank}(\pi_{XY})\le T_{\mathrm{Exact}}^{(0)}(\pi_{XY})\leq\log\mathrm{rank}^{+}(\pi_{XY}).
\end{equation}
Yannakakis \cite{yannakakis1991expressing} first related the nonnegative
rank of a matrix to the communication complexity (the minimum number
of communication bits) of distributively computing a matrix (or a
bivariate function). The equivalence between the nonnegative rank
and the exact common information (when only fixed-length codes allowed),
as shown in \eqref{eq:-169}, was previously obtained in \cite{jain2013efficient,braun2016common}.

\begin{itemize}
\item Nonnegative $\alpha$-Rank
\end{itemize}
The class of common information problems can also be cast in the light
of approximate or exact decomposition of a joint distribution. Let
$\mathbf{P}_{W}$ be the diagonal matrix with the probability values
of $P_{W}$ as its diagonal elements and let $\top$ denote the transposition
operation. The exact common information problem is equivalent to decomposing
a joint distribution as a mixture of product conditional distributions
\begin{equation}
\mathbf{P}_{XY}=\mathbf{P}_{X|W}^{\top}\mathbf{P}_{W}\mathbf{P}_{Y|W}\label{eq:-26}
\end{equation}
such that the entropy $H(\mathbf{P}_{W})$ is minimized. Such a decomposition
is closely related to nonnegative matrix factorization (NMF) and the
nonnegative rank \cite{vandaele2016heuristics}. The nonnegative
rank and NMF play a crucial role in many subdisciplines of theoretical
computer science and discrete mathematics, including signal processing,
machine learning, communication complexity, and combinatorial optimization,
e.g., \cite{lovasz1990communication}.

Recall the common \emph{$\alpha$-}R\'enyi entropy defined in \eqref{eq:-113}.
When $\alpha=0$, this quantity is equal to the logarithm of nonnegative
rank of the joint distribution matrix. Inspired by this relationship,
we can generalize the nonnegative rank to the nonnegative $\alpha$-rank
as follows. For a nonnegative matrix (but not zero matrix) $\mathbf{A}$
and $\alpha\in[-\infty,\infty]$, we define the \emph{nonnegative
$\alpha$-rank }of $\mathbf{A}$ as 
\begin{equation}
\mathrm{rank}_{\alpha}^{+}(\mathbf{A}):=\exp\left\{ G_{\alpha}\left(\frac{\mathbf{A}}{\left\Vert \mathbf{A}\right\Vert _{1}}\right)\right\} .\label{eq:-110}
\end{equation}
Here, note that the argument of $G_{\alpha}$ is the normalized version
of the matrix $\mathbf{A}$ because the argument of $G_{\alpha}$
needs to be a joint probability distribution. This makes sense since
any reasonable definition of ``rank'' should satisfy invariance
under scaling operations (with non-zero scale factors). When $\alpha=0$,
the nonnegative $0$-rank defined in \eqref{eq:-110} reduces to the
traditional nonnegative rank, i.e., $\mathrm{rank}_{0}^{+}(\mathbf{A})=\mathrm{rank}^{+}(\mathbf{A})$.
Equivalently, the nonnegative $\alpha$-rank\emph{ }$\mathrm{rank}_{\alpha}^{+}(\mathbf{A})$
can be alternatively expressed as\footnote{One can also define a variant $\overline{\mathrm{rank}}_{\alpha}^{+}(\mathbf{A})$
of the nonnegative $\alpha$-rank by replacing $\left\Vert \mathbf{D}\right\Vert _{\alpha}^{\frac{\alpha}{1-\alpha}}$
with $\left\Vert \mathbf{D}\right\Vert _{\alpha}^{\alpha}$. This
variant can be written as $\overline{\mathrm{rank}}_{\alpha}^{+}(\mathbf{A})=\left(\mathrm{rank}_{\alpha}^{+}(\mathbf{A})\right)^{1-\alpha}$
with $\mathrm{rank}_{\alpha}^{+}(\mathbf{A})$ denoting the nonnegative
$\alpha$-rank defined in \eqref{eq:-110} or \eqref{eq:-110-1-1}.
Hence $\overline{\mathrm{rank}}_{\alpha}^{+}(\mathbf{A})$ and $\mathrm{rank}_{\alpha}^{+}(\mathbf{A})$
are uniquely determined by each other  except for $\alpha\in\left\{ -\infty,1,\infty\right\} $.
}
\begin{align}
\mathrm{rank}_{\alpha}^{+}(\mathbf{A}) & =\min_{\mathbf{U},\mathbf{D},\mathbf{V}}\left\Vert \mathbf{D}\right\Vert _{\alpha}^{\frac{\alpha}{1-\alpha}}\label{eq:-110-1-1}
\end{align}
where the minimization in \eqref{eq:-110-1-1} is taken over all nonnegative
matrices $\mathbf{U}\in\mathbb{R}_{\ge0}^{k\times|\mathcal{X}|},\mathbf{D}\in\mathbb{R}_{\ge0}^{k\times k},\mathbf{V}\in\mathbb{R}_{\ge0}^{k\times|\mathcal{Y}|}$
for some $k\in\mathbb{N}$ such that $\mathbf{D}$ is diagonal and
$\mathbf{U}^{\top}\mathbf{D}\mathbf{V}=\frac{\mathbf{A}}{\left\Vert \mathbf{A}\right\Vert _{1}}$.

Here we only provide the definition of nonnegative $\alpha$-rank.
Investigation on applications of nonnegative $\alpha$-rank is outside
the scope of this paper, which remains to be done in the future.

\section{Concluding Remarks}

\label{sec:concl}In this paper, we established the equivalence between
the exact and $\infty$-R\'enyi common informations; provided single-letter
upper and lower bounds on these two quantities; completely characterized
them for DSBSes; and extended the exact and $\infty$-R\'enyi common
informations, and also Wyner's common information to sources with
general (countable or continuous) alphabets, including Gaussian sources.

For DSBSes, we observed that the exact and $\infty$-R\'enyi common
informations are both strictly larger than Wyner's common information.
This resolves an open problem posed by Kumar, Li, and El Gamal \cite{Kumar}.
For Gaussian sources with correlation coefficient $\rho\in[0,1)$,
we provided an upper bound on the exact and $\infty$-R\'enyi common
informations, which is at most $0.72$ (exactly, $\frac{\rho}{1+\rho}\log_{2}e$)
bits/symbol larger than Wyner's common information, and at least $22.28$
bits/symbol smaller than Li and El Gamal's one-shot bound \cite{li2017distributed}.
We conjectured our upper bound is tight.

Due to the equivalence between the exact common information and exact
channel simulation, we apply our results on the former problem to
the latter problem. In \cite{bennett2002entanglement,winter2002compression,Cuff,bennett2014quantum},
it was shown that when there exists unlimited shared randomness, the
minimum communication rates are the same for TV-approximate and exact
channel simulation problems, and this rate is equal to the mutual
information. However, this is not the case when there is no shared
randomness. Our results imply that with no shared randomness, the
minimum communication rate for TV-approximate channel simulation is
Wyner's common information; however the minimum rate for exact channel
simulation is the exact common information which is larger than Wyner's
common information.

We also connected the common information problem to the distributed
channel synthesis problem. Our results imply that with no shared randomness,
the minimum rate for exact channel simulation is the exact common
information which is larger than Wyner's common information. When
there is randomness shared by the encoder and decoder, the best tradeoff
between the shared randomness rate and the communication rate were
studied in our paper \cite{yu2018exact}. In the future, we are planning
to work on various closely-related problems, e.g., the exact versions
of various coordination problems \cite{Cuff10}.

\appendices{}

\section{\label{sec:equivalence}Proof of Theorem \ref{thm:equivalence}}

\subsection{\label{subsec:Proof-of}Proof of $T_{\mathrm{Exact}}(\pi_{XY})=T_{\infty}(\pi_{XY})$}

One direction of the equivalence $T_{\mathrm{Exact}}(\pi_{XY})=T_{\infty}(\pi_{XY})$
follows from the following lemmas. 
\begin{lem}
\label{lem:renyilarger}\cite{Kumar} If there exists a sequence of
fixed-length synthesis codes with rate $R$ that generates $P_{X^{n}Y^{n}}$
such that $D_{\infty}(P_{X^{n}Y^{n}}\|\pi_{XY}^{n})\to0$, then there
must exist a sequence of variable-length synthesis codes with asymptotic
rate $R$ that exactly generates $\pi_{XY}^{n}$. That is, $T_{\mathrm{Exact}}(\pi_{XY})\leq T_{\infty}(\pi_{XY}).$ 
\end{lem}
This lemma was proven by Kumar, Li, and El Gamal in \cite[Remark on Page 164]{Kumar}
using the following mixture decomposition technique\footnote{The decomposition of a distribution into a mixture of several distributions,
as in \eqref{eq:-100} and \eqref{eq:-100-1}, is termed the \emph{mixture
decomposition (}or\emph{ split}) of a distribution. This mixture decomposition
is rather useful to construct a desired distribution from a given
one. Such an idea originated from Nummelin' work \cite{nummelin1978uniform}
and Athreya and Ney's work \cite{athreya1978new}. In both of \cite{nummelin1978uniform}
and \cite{athreya1978new}, the authors used this splitting technique
to study limiting theorems of recurrent Markov processes. Furthermore,
such a technique was also used to study the mixing rate of Markov
Chain Monte Carlo (MCMC) methods \cite{roberts2004general}, by constructing
a coupling of an original Markov chain and an target Markov chain.
Besides as a tool, the mixture decomposition is also an important
topic in probability and statistics theories that has independent
interest; see \cite{titterington1985statistical} (or more general
decomposition theories \cite{estrada2012distributional}). The mixture
decomposition is also related to other information-theoretic problems.
For example, such a technique was used in the proof of \cite[Theorem 16]{ho2010interplay}.
Furthermore as mentioned in Remark \ref{rem:Given-a-set}, finding
an optimal mixture decomposition (with the minimum coefficient for
the residual part) is equivalent to the  $\infty$-R\'enyi resolvability
problem \cite{yu2019renyi}.} (also termed ``splitting technique''). According to the definition
of $D_{\infty}$, $D_{\infty}(P_{X^{n}Y^{n}}\|\pi_{XY}^{n})\leq\epsilon$
with $\epsilon>0$ implies that $P_{X^{n}Y^{n}}\left(x^{n},y^{n}\right)\leq e^{\epsilon}\pi_{XY}^{n}\left(x^{n},y^{n}\right)$
for all $x^{n},y^{n}$. Define 
\begin{equation}
\widehat{P}_{X^{n}Y^{n}}\left(x^{n},y^{n}\right):=\frac{e^{\epsilon}\pi_{XY}^{n}\left(x^{n},y^{n}\right)-P_{X^{n}Y^{n}}\left(x^{n},y^{n}\right)}{e^{\epsilon}-1},
\end{equation}
then obviously, $\widehat{P}_{X^{n}Y^{n}}\left(x^{n},y^{n}\right)$
is a distribution. Hence $\pi_{XY}^{n}$ can be written as a mixture
distribution 
\begin{align}
 & \pi_{XY}^{n}\left(x^{n},y^{n}\right)\nonumber \\
 & =e^{-\epsilon}P_{X^{n}Y^{n}}\left(x^{n},y^{n}\right)+\left(1-e^{-\epsilon}\right)\widehat{P}_{X^{n}Y^{n}}\left(x^{n},y^{n}\right).\label{eq:-100}
\end{align}
The encoder first generates a Bernoulli random variable $U$ with
$P_{U}(1)=e^{-\epsilon}$, compresses it by using $1$ bit, and transmits
it to the two generators. If $U=1$, then the encoder generates a
uniform random variable $M\sim\mathrm{Unif}[1:e^{nR}]$, and the encoder
and two generators use the fixed-length synthesis codes with rate
$R$ to generate $P_{X^{n}Y^{n}}$. If $U=0$, then the encoder generates
$\left(X^{n},Y^{n}\right)\sim\widehat{P}_{X^{n}Y^{n}}$, and uses
a variable-length compression code with rate $\leq\log|\mathcal{X}||\mathcal{Y}|$
to generate $\widehat{P}_{X^{n}Y^{n}}$. The distribution generated
by such a mixed code is $e^{-\epsilon}P_{X^{n}Y^{n}}\left(x^{n},y^{n}\right)+\left(1-e^{-\epsilon}\right)\widehat{P}_{X^{n}Y^{n}}\left(x^{n},y^{n}\right)$,
i.e., $\pi_{XY}^{n}\left(x^{n},y^{n}\right)$. The total code rate
is no larger than $\frac{1}{n}+e^{-\epsilon}R+\left(1-e^{-\epsilon}\right)\log|\mathcal{X}||\mathcal{Y}|$,
which converges to $R$ upon taking the limit in $n\to\infty$ and
the limit in $\epsilon\to0$. 

The mixture decomposition (or split) of a distribution  in \eqref{eq:-100}
can be generalized to general distributions. 
\begin{lem}[Mixture Decomposition of General Distributions]
\label{lem:mixturedecomposition} Assume $P,Q$ are two distributions
defined on the same Borel-measurable space. Assume\footnote{For general distributions $P,Q$ such that $P\ll Q$, $D_{\infty}(P\|Q):=\log\esssup_{P}\frac{\mathrm{d}P}{\mathrm{d}Q}$,
where $\frac{\mathrm{d}P}{\mathrm{d}Q}$ denotes Radon\textendash Nikodym
derivative of $P$ respect to $Q$, and $\esssup_{P}\frac{\mathrm{d}P}{\mathrm{d}Q}$
denotes the essential supremum of $\frac{\mathrm{d}P}{\mathrm{d}Q}$
with respect to $P$. Moreover, if $P\centernot\ll Q$, then $D_{\infty}(P\|Q):=+\infty$. } $D_{\infty}(P\|Q)\leq\epsilon$ for some $\epsilon\in[0,\infty]$.
Then 
\begin{equation}
Q=e^{-\epsilon}P+\left(1-e^{-\epsilon}\right)\widehat{P},\label{eq:-100-1}
\end{equation}
where 
\begin{equation}
\widehat{P}:=\begin{cases}
\textrm{any distribution} & \epsilon=0\\
\frac{e^{\epsilon}Q-P}{e^{\epsilon}-1} & \epsilon\in(0,\infty)\\
Q & \epsilon=\infty
\end{cases}.
\end{equation}
Moreover, if we define
\begin{align}
\Lambda(Q,P) & :=\sup\Bigl\{\alpha:\exists\textrm{ a distribution }\widehat{P}\textrm{ s.t. }\nonumber \\
 & \qquad Q=\alpha P+\left(1-\alpha\right)\widehat{P},\alpha\in[0,1]\Bigr\},
\end{align}
then 
\begin{equation}
\Lambda(Q,P)=e^{-D_{\infty}(P\|Q)}=\begin{cases}
\frac{1}{\esssup_{P}\frac{\mathrm{d}P}{\mathrm{d}Q}} & P\ll Q\\
0 & P\centernot\ll Q
\end{cases}.
\end{equation}
\end{lem}
\begin{rem}
\label{rem:Given-a-set}Given a set of distributions $\left\{ P_{i}:i\in[1:n]\right\} $
and a target distribution $Q$ defined on the same space $\left(\mathcal{Y},\mathbb{B}\right)$,
a natural question is to determine the minimum value $\alpha_{0}\geq0$
such that 
\begin{align}
Q & =\sum_{i=1}^{n}\alpha_{i}P_{i}+\alpha_{0}\widehat{P}
\end{align}
for some distribution $\widehat{P}$ and some values $\alpha_{i}\geq0,i\in[1:n]$
and $\sum_{i=0}^{n}\alpha_{i}=1$. By Lemma \ref{lem:mixturedecomposition}
such a mixture decomposition problem is equivalent to 
\begin{equation}
\min_{\left\{ \hat{\alpha}_{i}\right\} :\hat{\alpha}_{i}\geq0,\sum_{i=1}^{n}\hat{\alpha}_{i}=1}D_{\infty}(\sum_{i=1}^{n}\hat{\alpha}_{i}P_{i}\|Q).\label{eq:-103}
\end{equation}
If we consider $\left\{ P_{i}\right\} $ as a channel $P_{Y|X}$ with
$P_{Y|X=i}=P_{i}$ and denote $Q_{Y}:=Q$, then \eqref{eq:-103} can
be rewritten as 
\begin{equation}
\min_{P_{X}}D_{\infty}(P_{Y}\|Q_{Y})\label{eq:-104}
\end{equation}
where $P_{X}$ is a distribution on $[1:n]$ and $P_{Y}$ is the output
distribution of $P_{Y|X}$ when the input distribution is $P_{X}$.
The problem in \eqref{eq:-104} is just the so-called \emph{$\infty$-R\'enyi
resolvability problem} (or \emph{channel resolvability problem under
$\infty$-R\'enyi divergence measure}).   In \cite{yu2019renyi},
the present authors studied the $\infty$-R\'enyi resolvability problem
in which the channels and target distributions are of product forms
and $P_{X}$ restricted to be a function of a given uniform random
variable.
\end{rem}
Now we consider the other direction of the equivalence $T_{\mathrm{Exact}}(\pi_{XY})=T_{\infty}(\pi_{XY})$.
\begin{lem}
\label{lem:exactlarger}If there exists a sequence of variable-length
synthesis codes with asymptotic rate $R$ that exactly generates $\pi_{XY}^{n}$,
then there must exist a sequence of fixed-length synthesis codes with
rate $R$ that generates $P_{X^{n}Y^{n}}$ such that $D_{\infty}(P_{X^{n}Y^{n}}\|\pi_{XY}^{n})\to0$.
That is, $T_{\mathrm{Exact}}(\pi_{XY})\ge T_{\infty}(\pi_{XY}).$
\end{lem}
\begin{rem}
\label{rem:Note-that-this}Note that by checking our proof, one can
find that this lemma holds not only for sources with finite alphabets,
but also for those with countably infinite or continuous/uncountable
alphabets. 
\end{rem}
\begin{IEEEproof}
Let $\left\{ c_{k}\right\} _{k=1}^{\infty}$ be a sequence of variable-length
codes with rate $R$ that exactly generates $\pi_{XY}^{k}$. Let $W_{k}$
be the common random variable, and $P_{X^{k}|W_{k}}$ and $P_{Y^{k}|W_{k}}$
the two generators that define $c_{k}$. Hence $\sum_{w}P_{W_{k}}(w)P_{X^{k}|W_{k}}(\cdot|w)P_{Y^{k}|W_{k}}(\cdot|w)=\pi_{XY}^{k}$,
and $\frac{1}{k}H(W_{k})\to R$ as $k\to\infty$. Now we consider
a superblock code that consists of $n$ independent $k$-length codes
as defined above. That is, $W_{k}^{n}\sim P_{W_{k}}^{n}$ is the common
random variable and $P_{X^{k}|W_{k}}^{n}$ and $P_{Y^{k}|W_{k}}^{n}$
are the two generators. Observe that $W_{k}^{n}$ is an $n$-length
i.i.d. random sequence with each $W_{k,i}\sim P_{W_{k}}$. Hence we
have 
\begin{equation}
\mathbb{P}\left(W_{k}^{n}\in\mathcal{A}_{\epsilon}^{(n)}(P_{W_{k}})\right)\to1
\end{equation}
as $n\to\infty$ for fixed $k$. Furthermore, $|\mathcal{A}_{\epsilon}^{(n)}|\le e^{n(H(W_{k})+\epsilon)}$.
Define a truncated distribution 
\begin{equation}
Q_{W_{k}^{n}}\left(w_{k}^{n}\right):=\frac{P_{W_{k}}^{n}\left(w_{k}^{n}\right)1\left\{ w_{k}^{n}\in\mathcal{A}_{\epsilon}^{(n)}\right\} }{P_{W_{k}}^{n}\left(\mathcal{A}_{\epsilon}^{(n)}\right)}.
\end{equation}
Now we adopt a simulation scheme $f_{n}$ as used in \cite[Theorem 7]{yu2019simulation}
to simulate the truncated distribution $Q_{W_{k}^{n}}$ from a uniform
random variable $M\sim\mathrm{Unif}[1:e^{nkR'}]$. For each $w_{k}^{n}\in\mathcal{A}_{\epsilon}^{(n)}$,
we map either $\bigl\lfloor e^{nkR'}Q_{W_{k}^{n}}\left(w_{k}^{n}\right)\bigr\rfloor$
or $\bigl\lceil e^{nkR'}Q_{W_{k}^{n}}\left(w_{k}^{n}\right)\bigr\rceil$
number of elements $m\in[1:e^{nkR'}]$ to it. Hence  the output
distribution $\widetilde{P}_{W_{k}^{n}}$ induced by such a mapping
satisfies $\widetilde{P}_{W_{k}^{n}}\left(w_{k}^{n}\right)=e^{-nkR'}\bigl\lfloor e^{nkR'}Q_{W_{k}^{n}}\left(w_{k}^{n}\right)\bigr\rfloor$
or $\widetilde{P}_{W_{k}^{n}}\left(w_{k}^{n}\right)=e^{-nkR'}\bigl\lceil e^{nkR'}Q_{W_{k}^{n}}\left(w_{k}^{n}\right)\bigr\rceil$
for $w_{k}^{n}\in\mathcal{A}_{\epsilon}^{(n)}$. Hence
\begin{align}
 & D_{\infty}(\widetilde{P}_{W_{k}^{n}}\|Q_{W_{k}^{n}})\nonumber \\
 & =\log\max_{w_{k}^{n}\in\mathcal{A}_{\epsilon}^{(n)}}\frac{\widetilde{P}_{W_{k}^{n}}\left(w_{k}^{n}\right)}{Q_{W_{k}^{n}}\left(w_{k}^{n}\right)}\\
 & \leq\log\max_{w_{k}^{n}\in\mathcal{A}_{\epsilon}^{(n)}}\frac{Q_{W_{k}^{n}}\left(w_{k}^{n}\right)+e^{-nkR'}}{Q_{W_{k}^{n}}\left(w_{k}^{n}\right)}\\
 & \leq\log\left(1+\frac{e^{-nkR'}}{e^{-n\left(H(W_{k})+\epsilon\right)}}\right)\\
 & =\log\left(1+e^{-nk\left(R'-\frac{1}{k}\left(H(W_{k})+\epsilon\right)\right)}\right).
\end{align}
Therefore, if $R'>\frac{1}{k}\left(H(W_{k})+\epsilon\right)$, then
\begin{align}
 & D_{\infty}(\widetilde{P}_{W_{k}^{n}}\|Q_{W_{k}^{n}})\nonumber \\
 & =\log\max_{w_{k}^{n}\in\mathcal{A}_{\epsilon}^{(n)}}\frac{\widetilde{P}_{W_{k}^{n}}\left(w_{k}^{n}\right)}{Q_{W_{k}^{n}}\left(w_{k}^{n}\right)}\to0,
\end{align}
as $n\to\infty$ for fixed $k$. Such a simulation code $f_{n}$ is
also valid for simulating $P_{W_{k}}^{n}$. This is because 
\begin{align}
 & D_{\infty}(\widetilde{P}_{W_{k}^{n}}\|P_{W_{k}}^{n})\nonumber \\
 & =\log\max_{w_{k}^{n}\in\mathcal{A}_{\epsilon}^{(n)}}\frac{\widetilde{P}_{W_{k}^{n}}\left(w_{k}^{n}\right)}{P_{W_{k}}^{n}\left(w_{k}^{n}\right)}\\
 & \leq\log\max_{w_{k}^{n}\in\mathcal{A}_{\epsilon}^{(n)}}\frac{\widetilde{P}_{W_{k}^{n}}\left(w_{k}^{n}\right)}{Q_{W_{k}^{n}}\left(w_{k}^{n}\right)}+\log\max_{w_{k}^{n}\in\mathcal{A}_{\epsilon}^{(n)}}\frac{Q_{W_{k}^{n}}\left(w_{k}^{n}\right)}{P_{W_{k}}^{n}\left(w_{k}^{n}\right)}\\
 & =D_{\infty}(\widetilde{P}_{W_{k}^{n}}\|Q_{W_{k}^{n}})-\log P_{W_{k}}^{n}\left(\mathcal{A}_{\epsilon}^{(n)}\right)\\
 & \to0,
\end{align}
as $n\to\infty$ for fixed $k$.

Now we consider a cascaded synthesis code by concatenating the simulation
code $f_{n}$ above with the two generators $P_{X^{k}|W_{k}}^{n}$
and $P_{Y^{k}|W_{k}}^{n}$ of the variable-length synthesis code.
Observe that $P_{X^{kn}Y^{kn}}$ and $\pi_{XY}^{kn}$ are respectively
the outputs of the channel $P_{X^{k}|W_{k}}^{n}P_{Y^{k}|W_{k}}^{n}$
respectively induced by the channel inputs $\widetilde{P}_{W_{k}^{n}}$
and $P_{W_{k}}^{n}$. Hence by the data processing inequality \cite{Erven},
for such a cascaded code, we have 
\begin{align}
 & D_{\infty}(P_{X^{kn}Y^{kn}}\|\pi_{XY}^{kn})\nonumber \\
 & \leq D_{\infty}(\widetilde{P}_{W_{k}^{n}}\|P_{W_{k}}^{n})\\
 & \to0
\end{align}
as $n\to\infty$ for fixed $k$, as long as the code rate $R'>\frac{1}{k}\left(H(W_{k})+\epsilon\right)$.

As for the case where the blocklength $n'$ is not a multiple of $k$,
i.e., $n'=kn+l$ with $l\in[1:k-1]$, we need to construct a code
with blocklength $k(n+1)$ and then truncate the outputs $\left(X^{k(n+1)},Y^{k(n+1)}\right)$
to $\left(X^{n'},Y^{n'}\right)$. Obviously, $D_{\infty}(P_{X^{n'}Y^{n'}}\|\pi_{XY}^{n'})\leq D_{\infty}(P_{X^{k(n+1)}Y^{k(n+1)}}\|\pi_{XY}^{k(n+1)})\to0$
as $n\to\infty$. Furthermore, the code rate for such a code is $\frac{k(n+1)R'}{n'}\leq(1+\frac{1}{n})R'\to R'$
as $n\to\infty$. On the other hand, $\frac{1}{k}H(W_{k})\to R$ as
$k\to\infty$. Therefore, there exists a sequence of fixed-length
synthesis codes with asymptotic rate $R$ that generates $P_{X^{n'}Y^{n'}}$
such that $D_{\infty}(P_{X^{n'}Y^{n'}}\|\pi_{XY}^{n'})\to0$ as $n'\to\infty$. 
\end{IEEEproof}

\subsection{\label{subsec:achievability} Proof of $T_{\infty}(\pi_{XY})\protect\leq\liminf_{n\to\infty}\frac{1}{n}\Gamma(\pi_{XY}^{n})$}

Here we prove the achievability result from the perspective of $\infty$-R\'enyi
common information problem. We borrow an idea from \cite{vellambi2016sufficient}.
The corresponding coding scheme was also independently used by the
present authors in \cite{yu2019renyi,yu2018wyner}.

To show the achievability part, we only need to show that the single-letter
expression $\Gamma(\pi_{XY})$ satisfies $T_{\infty}(\pi_{XY})\le\Gamma(\pi_{XY})$.
This is because we can obtain the upper bound $\Gamma(\pi_{XY}^{k})$
by substituting $\pi_{XY}$ with $\pi_{XY}^{k}$ into the single-letter
expression\footnote{Note that by definition, $T_{\infty}(\pi_{XY}^{k})$ is additive in
$k$, i.e., $T_{\infty}(\pi_{XY}^{k})=kT_{\infty}(\pi_{XY})$. This
is because, on one hand, the superblock code that consists of $k$
independent copies of a $\left(n,R\right)$ code for $\pi_{XY}$ forms
a $\left(n,kR\right)$ code for $\pi_{XY}^{k}$. On the other hand,
a $\left(n,kR\right)$ code for $\pi_{XY}^{k}$ forms a $\left(nk,R\right)$
code for $\pi_{XY}$.}.  For fixed $\epsilon>0$ and a fixed joint distribution $Q_{WXY}=Q_{W}Q_{X|W}Q_{Y|W}$,
define the distributions 
\begin{align}
P_{W^{n}}\left(w^{n}\right) & \propto Q_{W}^{n}\left(w^{n}\right)1\left\{ w^{n}\in\mathcal{T}_{\frac{\epsilon}{2}}^{\left(n\right)}\left(Q_{W}\right)\right\} ,\label{eq:-66}\\
P_{X^{n}|W^{n}}\left(x^{n}|w^{n}\right) & \propto Q_{X|W}^{n}\left(x^{n}|w^{n}\right)\nonumber \\
 & \qquad\times1\left\{ x^{n}\in\mathcal{T}_{\epsilon}^{\left(n\right)}\left(Q_{WX}|w^{n}\right)\right\} ,\\
P_{Y^{n}|W^{n}}\left(y^{n}|w^{n}\right) & \propto Q_{Y|W}^{n}\left(y^{n}|w^{n}\right)\nonumber \\
 & \qquad\times1\left\{ y^{n}\in\mathcal{T}_{\epsilon}^{\left(n\right)}\left(Q_{WY}|w^{n}\right)\right\} .\label{eq:-73}
\end{align}
We set $\mathcal{C}_{n}=\left\{ W^{n}\left(m\right)\right\} _{m\in\calM_{n}}$
with $W^{n}\left(m\right),m\in\calM_{n}$ drawn independently for
different $m$'s and according to the same distribution $P_{W^{n}}$.
Upon receiving $W^{n}\left(M_{n}\right)$, the generators respectively
use random mappings $P_{X^{n}|W^{n}}$ and $P_{Y^{n}|W^{n}}$ to generate
$X^{n}$ and $Y^{n}$. Define $P_{M_{n}}:=\mathrm{Unif}[1:e^{nR}]$.
For random mappings $\left(P_{X^{n}|W^{n}},P_{Y^{n}|W^{n}}\right)$,
we define 
\begin{align}
 & P_{X^{n}Y^{n}|\mathcal{C}_{n}}(x^{n},y^{n}|\left\{ W^{n}\left(m\right)\right\} )\nonumber \\
 & :=\sum_{m}P_{M_{n}}(m)P_{X^{n}|W^{n}}\left(x^{n}|W^{n}\left(m\right)\right)P_{Y^{n}|W^{n}}\left(y^{n}|W^{n}\left(m\right)\right),
\end{align}
which is the output distribution induced by the codebook $\mathcal{C}_{n}$
in a distributed source simulation system with simulators $\left(P_{X^{n}|W^{n}},P_{Y^{n}|W^{n}}\right)$.
For such a code, we have the following distributed R\'enyi-covering
lemma. 
\begin{lem}[Distributed R\'enyi-Covering]
\label{lem:Renyicovering} For the random code described above, if
\begin{align}
R & >\mathcal{I}\left(Q\right):=-H_{Q}(XY|W)\nonumber \\
 & \qquad+\sum_{w}Q(w)\mathcal{H}(Q_{X|W=w},Q_{Y|W=w}\|Q_{XY}),\label{eq:-74}
\end{align}
then there exists some $\alpha,\epsilon>0$ such that 
\begin{align}
 & \mathbb{P}_{\mathcal{C}_{n}}\left(D_{\infty}(P_{X^{n}Y^{n}|\mathcal{C}_{n}}\|Q_{XY}^{n})\leq e^{-n\alpha}\right)\to1
\end{align}
doubly exponentially fast.
\end{lem}
\begin{rem}
The soft-covering problem under the $\infty$-R\'enyi divergence measure
was also studied in \cite{vellambi2016sufficient} as a key step of
proving sufficient conditions for equality of Wyner's common information
and the exact common information. However, no explicit rate bound
(e.g., $\mathcal{I}\left(Q\right)$ as defined in \eqref{eq:-74})
for an arbitrary $\pi_{XY}$ was provided in \cite{vellambi2016sufficient}.
\end{rem}
Setting $Q_{WXY}$ as an optimal distribution attaining $\Gamma(\pi_{XY})$,
we obtain $\mathcal{I}\left(Q\right)=\Gamma(\pi_{XY})$. Hence this
lemma implies that there exists a sequence of codebooks $\left\{ c_{n}\right\} $
with rate $R$ such that $D_{\infty}(P_{X^{n}Y^{n}|\mathcal{C}_{n}=c_{n}}\|Q_{XY}^{n})\leq e^{-n\alpha}$
as long as $R>\Gamma(\pi_{XY})$.  This completes the proof of $T_{\infty}(\pi_{XY})\le\Gamma(\pi_{XY})$.
Hence what we need to do is to prove Lemma \ref{lem:Renyicovering}.
The proof is provided in the following.
\begin{IEEEproof}[Proof of Lemma \ref{lem:Renyicovering}]
 For the fixed $\epsilon>0$ and the fixed joint distribution $Q_{WXY}=Q_{W}Q_{X|W}Q_{Y|W}$,
define 
\begin{align}
 & \calB_{\epsilon}:=\Bigl\{ P_{WXY}\in\mathcal{P}(\mathcal{W}\times\mathcal{X}\times\mathcal{Y}):\nonumber \\
 & \forall w,\left|P_{W}(w)-Q_{W}(w)\right|\leq\frac{\epsilon}{2}Q_{W}(w),\nonumber \\
 & \forall(w,x),\left|P_{WX}(w,x)-Q_{WX}(w,x)\right|\leq\epsilon Q_{WX}(w,x),\nonumber \\
 & \forall(w,y),\left|P_{WY}(w,y)-Q_{WY}(w,y)\right|\leq\epsilon Q_{WY}(w,y)\Bigr\},
\end{align}
and 
\begin{align}
\mathcal{I}_{\epsilon}\left(Q\right) & :=\max_{\widetilde{P}_{WXY}\in\calB_{\epsilon}}\sum_{w,x}\widetilde{P}\left(w,x\right)\log Q\left(x|w\right)\nonumber \\
 & \qquad+\sum_{w,y}\widetilde{P}\left(w,y\right)\log Q\left(y|w\right)\nonumber \\
 & \qquad-\sum_{x,y}\widetilde{P}\left(x,y\right)\log Q\left(x,y\right).\label{eq:-32}
\end{align}
Obviously, $\mathcal{I}_{\epsilon}\left(Q\right)\ge\mathcal{I}\left(Q\right)$,
hence $\lim_{\epsilon\downarrow0}\mathcal{I}_{\epsilon}\left(Q\right)\ge\mathcal{I}\left(Q\right)$,
where $\mathcal{I}\left(Q\right)$ is defined in \eqref{eq:-74}.
Now we prove $\lim_{\epsilon\downarrow0}\mathcal{I}_{\epsilon}\left(Q\right)\le\mathcal{I}\left(Q\right)$.
Let $\left\{ \epsilon_{k}\right\} _{k=1}^{\infty}$ be a sequence
of decreasing positive numbers with $\lim_{k\to\infty}\epsilon_{k}=0$.
Let $\left\{ P_{WXY}^{\left(k\right)}\right\} _{k=1}^{\infty}$ be
a sequence of optimal distributions attaining $\widetilde{\Gamma}_{\epsilon_{k}}(\pi_{XY})$.
Since $\mathcal{P}(\mathcal{W}\times\mathcal{X}\times\mathcal{Y})$
is compact, there exists some subsequence $\left\{ \epsilon_{k_{i}}\right\} _{i=1}^{\infty}$
such that $P_{WXY}^{\left(k_{i}\right)}$ converges to some distribution
$\widehat{P}_{WXY}$ as $i\to\infty$. Since $\lim_{i\to\infty}\epsilon_{k_{i}}=0$,
we must have 
\begin{align}
\widehat{P}_{WX} & =Q_{WX}\\
\widehat{P}_{WY} & =Q_{WY}.
\end{align}
Since the objective function in the right hand side of \eqref{eq:-32}
is continuous in $\widetilde{P}_{WXY}$, we have 
\begin{align}
 & \lim_{\epsilon\downarrow0}\mathcal{I}_{\epsilon}\left(Q\right)\nonumber \\
 & =\sum_{w,x}\widehat{P}\left(w,x\right)\log Q\left(x|w\right)+\sum_{w,y}\widehat{P}\left(w,y\right)\log Q\left(y|w\right)\nonumber \\
 & \qquad-\sum_{x,y}\widehat{P}\left(x,y\right)\log Q\left(x,y\right)\label{eq:-32-1}\\
 & \leq\max_{\widetilde{P}_{WXY}:\widetilde{P}_{WX}=Q_{WX},\widetilde{P}_{WY}=Q_{WY}}\sum_{w,x}\widetilde{P}\left(w,x\right)\log Q\left(x|w\right)\nonumber \\
 & \qquad+\sum_{w,y}\widetilde{P}\left(w,y\right)\log Q\left(y|w\right)-\sum_{x,y}\widetilde{P}\left(x,y\right)\log Q\left(x,y\right)\\
 & =\mathcal{I}\left(Q\right).
\end{align}
Therefore, 
\begin{equation}
\lim_{\epsilon\downarrow0}\mathcal{I}_{\epsilon}\left(Q\right)=\mathcal{I}\left(Q\right).\label{eq:-37}
\end{equation}
By the continuity of $\epsilon\mapsto\mathcal{I}_{\epsilon}\left(Q\right)$
shown in \eqref{eq:-37}, we can choose $\epsilon>0$, used in definitions
\eqref{eq:-66}-\eqref{eq:-73}, so small such that 
\begin{equation}
R>\mathcal{I}_{\epsilon}\left(Q\right)+\epsilon.\label{eq:-75}
\end{equation}
The reason for this choice of $\epsilon$ is to ensure \eqref{eq:-116}
(at the end of this proof) to hold.

For brevity, in the following we denote $\mathsf{M}=e^{nR}$. According
to the definition of the R\'enyi divergence, we first have\footnote{For brevity, we denote $P_{X^{n}Y^{n}|\mathcal{C}_{n}}$ as $P_{X^{n}Y^{n}}$. }
\begin{align}
e^{D_{\infty}(P_{X^{n}Y^{n}}\|\pi_{XY}^{n})} & =\max_{x^{n},y^{n}}\frac{P_{X^{n}Y^{n}}\left(x^{n},y^{n}\right)}{Q_{XY}^{n}\left(x^{n},y^{n}\right)}\\
 & =\max_{x^{n},y^{n}}\widetilde{g}(x^{n},y^{n}|\mathcal{C}_{n}),\label{eq:-151}
\end{align}
where $\widetilde{g}(x^{n},y^{n}|\mathcal{C}_{n}):=\sum_{m\in\calM_{n}}\frac{1}{\mathsf{M}}g(x^{n},y^{n}|W^{n}(m))$
with $g(x^{n},y^{n}|w^{n}):=\frac{1}{Q_{XY}^{n}\left(x^{n},y^{n}\right)}P_{X^{n}|W^{n}}\left(x^{n}|w^{n}\right)P_{Y^{n}|W^{n}}\left(y^{n}|w^{n}\right)$.
Then for $w^{n}\in\mathcal{T}_{\frac{\epsilon}{2}}^{n}\left(Q_{W}\right)$,
\begin{align}
 & g(x^{n},y^{n}|w^{n})\nonumber \\
 & =\frac{1}{Q_{XY}^{n}\left(x^{n},y^{n}\right)}\frac{Q_{X|W}^{n}\left(x^{n}|w^{n}\right)1\left\{ x^{n}\in\mathcal{T}_{\epsilon}^{\left(n\right)}\left(Q_{WX}|w^{n}\right)\right\} }{Q_{X|W}^{n}\left(\mathcal{T}_{\epsilon}^{\left(n\right)}\left(Q_{WX}|w^{n}\right)|w^{n}\right)}\nonumber \\
 & \qquad\times\frac{Q_{Y|W}^{n}\left(y^{n}|w^{n}\right)1\left\{ y^{n}\in\mathcal{T}_{\epsilon}^{\left(n\right)}\left(Q_{WY}|w^{n}\right)\right\} }{Q_{Y|W}^{n}\left(\mathcal{T}_{\epsilon}^{\left(n\right)}\left(Q_{WY}|w^{n}\right)|w^{n}\right)}\\
 & \leq\frac{1\left\{ x^{n}\in\mathcal{T}_{\epsilon}^{\left(n\right)}\left(Q_{WX}|w^{n}\right),y^{n}\in\mathcal{T}_{\epsilon}^{\left(n\right)}\left(Q_{WY}|w^{n}\right)\right\} }{\left(1-\delta_{1,n}\right)\left(1-\delta_{2,n}\right)}\nonumber \\
 & \qquad\times e^{n\sum_{w,x}T_{w^{n}x^{n}}\left(w,x\right)\log Q\left(x|w\right)}\nonumber \\
 & \qquad\times e^{n\sum_{w,y}T_{w^{n}y^{n}}\left(w,y\right)\log Q\left(y|w\right)}\nonumber \\
 & \qquad\times e^{-n\sum_{x,y}T_{x^{n}y^{n}}\left(x,y\right)\log Q\left(x,y\right)}\\
 & \leq\frac{1}{\left(1-\delta_{1,n}\right)\left(1-\delta_{2,n}\right)}e^{n\mathcal{I}_{\epsilon}\left(Q\right)}\\
 & =:\beta_{n},\label{eq:-93}
\end{align}
where by \cite[Lemma 4]{yu2018wyner}, both $\delta_{1,n}:=1-Q_{X|W}^{n}\left(\mathcal{T}_{\epsilon}^{\left(n\right)}\left(Q_{WX}|w^{n}\right)|w^{n}\right)$
and $\delta_{2,n}:=1-Q_{Y|W}^{n}\left(\mathcal{T}_{\epsilon}^{\left(n\right)}\left(Q_{WY}|w^{n}\right)|w^{n}\right)$
converge to zero exponentially fast as $n\to\infty$, and $\mathcal{I}_{\epsilon}\left(Q\right)$
is defined in \eqref{eq:-32}.

Continuing \eqref{eq:-151}, we get for any $\delta>0$, 
\begin{align}
 & \mathbb{P}_{\mathcal{C}_{n}}\left(D_{\infty}(P_{X^{n}Y^{n}}\|\pi_{XY}^{n})\geq\delta\right)\nonumber \\
 & \leq\mathbb{P}_{\mathcal{C}_{n}}\left(e^{D_{\infty}(P_{X^{n}Y^{n}}\|\pi_{XY}^{n})}-1\geq\delta\right)\\
 & =\mathbb{P}_{\mathcal{C}_{n}}\left(\max_{x^{n},y^{n}}\widetilde{g}(x^{n},y^{n}|\mathcal{C}_{n})\geq1+\delta\right)\\
 & \leq\left|\mathcal{X}\right|^{n}\left|\mathcal{Y}\right|^{n}\max_{x^{n},y^{n}}\mathbb{P}_{\mathcal{C}_{n}}\left(\widetilde{g}(x^{n},y^{n}|\mathcal{C}_{n})\geq1+\delta\right),\label{eq:-152}
\end{align}
where \eqref{eq:-152} follows from the union bound. Obviously, $\left|\mathcal{X}\right|^{n}\left|\mathcal{Y}\right|^{n}$
is only exponentially growing. Therefore, if the probability vanishes
doubly exponentially fast, then $\max_{x^{n},y^{n}}\widetilde{g}(x^{n},y^{n}|\mathcal{C}_{n})<1+\delta$
with probability of failure decaying to zero doubly exponentially
fast as $n\to\infty$. To this end, we use the Bernstein inequality
to bound the probability. Observe that $g(x^{n},y^{n}|W^{n}(m)),m\in\calM_{n}$
are i.i.d. random variables with mean 
\begin{align}
\mu_{n} & :=\mathbb{E}_{W^{n}}\left[g(x^{n},y^{n}|W^{n})\right]\\
 & =\frac{1}{Q_{XY}^{n}\left(x^{n},y^{n}\right)}\sum_{w^{n}}\frac{Q_{W}^{n}\left(w^{n}\right)1\left\{ w^{n}\in\mathcal{T}_{\frac{\epsilon}{2}}^{\left(n\right)}\left(Q_{W}\right)\right\} }{Q_{W}^{n}\left(\mathcal{T}_{\frac{\epsilon}{2}}^{\left(n\right)}\left(Q_{W}\right)\right)}\nonumber \\
 & \qquad\times\frac{Q_{X|W}^{n}\left(x^{n}|w^{n}\right)1\left\{ x^{n}\in\mathcal{T}_{\epsilon}^{\left(n\right)}\left(Q_{WX}|w^{n}\right)\right\} }{Q_{X|W}^{n}\left(\mathcal{T}_{\epsilon}^{\left(n\right)}\left(Q_{WX}|w^{n}\right)|w^{n}\right)}\nonumber \\
 & \qquad\times\frac{Q_{Y|W}^{n}\left(y^{n}|w^{n}\right)1\left\{ y^{n}\in\mathcal{T}_{\epsilon}^{\left(n\right)}\left(Q_{WY}|w^{n}\right)\right\} }{Q_{Y|W}^{n}\left(\mathcal{T}_{\epsilon}^{\left(n\right)}\left(Q_{WY}|w^{n}\right)|w^{n}\right)}\\
 & \leq\frac{1}{\left(1-\delta_{0,n}\right)\left(1-\delta_{1,n}\right)\left(1-\delta_{2,n}\right)}\label{eq:-31}\\
 & \to1\textrm{ exponentially fast as }n\to\infty,\label{eq:-30}
\end{align}
and variance 
\begin{align}
\mathrm{Var}_{W^{n}}\left[g(x^{n},y^{n}|W^{n})\right] & \leq\mathbb{E}_{W^{n}}\left[g(x^{n},y^{n}|W^{n})^{2}\right]\\
 & \leq\beta_{n}\mu_{n}.
\end{align}
Here \eqref{eq:-30} follows since $\delta_{0,n}:=1-Q_{W}^{n}\left(\mathcal{T}_{\frac{\epsilon}{2}}^{\left(n\right)}\left(Q_{W}\right)\right)$
converges to zero exponentially fast as $n\to\infty$. Then we bound
the probability in \eqref{eq:-152} as follows:
\begin{align}
 & \mathbb{P}_{\mathcal{C}_{n}}\left(\widetilde{g}(x^{n},y^{n}|\mathcal{C}_{n})\geq1+\delta\right)\nonumber \\
 & =\mathbb{P}_{\mathcal{C}_{n}}\Biggl(\sum_{m\in\calM_{n}}g(x^{n},y^{n}|W^{n}(m))-\mu_{n}\mathsf{M}\nonumber \\
 & \qquad\geq\left(1+\delta-\mu_{n}\right)\mathsf{M}\Biggr)\label{eq:-94}\\
 & \leq\exp\left(-\frac{\frac{1}{2}\left(1+\delta-\mu_{n}\right)^{2}\mathsf{M}^{2}}{\mathsf{M}\beta_{n}\mu_{n}+\frac{1}{3}\left(1+\delta-\mu_{n}\right)\mathsf{M}\beta_{n}}\right)\label{eq:-117}\\
 & =\exp\left(-\frac{3\left(1+\delta-\mu_{n}\right)^{2}\mathsf{M}}{2\left(1+\delta+2\mu_{n}\right)\beta_{n}}\right),\label{eq:-153}
\end{align}
where \eqref{eq:-117} follows from Bernstein's inequality, stated
here for the readers' convenience.
\begin{lem}
\cite{boucheron2013concentration} Let $X_{1},\ldots,X_{n}$ be independent
zero-mean random variables such that $|X_{i}|\leq M$ almost surely,
for all $i$. Then, for any $t>0$,
\begin{equation}
\mathbb{P}\left(\sum_{i=1}^{n}X_{i}>t\right)\leq\exp\left(-\frac{\tfrac{1}{2}t^{2}}{\sum_{i=1}^{n}\mathbb{E}\left[X_{i}^{2}\right]+\tfrac{1}{3}Mt}\right).
\end{equation}
\end{lem}
Observe that
\begin{equation}
\frac{\mathsf{M}}{\beta_{n}}=\left(1-\delta_{1,n}\right)\left(1-\delta_{2,n}\right)e^{n\left(R-\mathcal{I}_{\epsilon}\left(Q\right)\right)}.
\end{equation}
Denote $\alpha_{0}$ as the exponent of $\frac{1}{\left(1-\delta_{0,n}\right)\left(1-\delta_{1,n}\right)\left(1-\delta_{2,n}\right)}-1$.
By \cite[Lemma 4]{yu2018wyner}, 
\begin{equation}
\alpha_{0}\geq\min\left\{ \frac{1}{3}\epsilon^{2}Q_{W}^{(\mathsf{min})},\frac{1}{3}\left(\frac{\epsilon}{2+\epsilon}\right)^{2}\min\left\{ Q_{X|W}^{(\mathsf{min})},Q_{Y|W}^{(\mathsf{min})}\right\} \right\} ,\label{eq:-96}
\end{equation}
where $Q_{W}^{(\mathsf{min})}:=\min_{w:Q_{W}(w)>0}Q_{W}(w)$, $Q_{X|W}^{(\mathsf{min})}:=\min_{(x,w):Q_{X|W}(x|w)>0}Q_{X|W}(x|w)$,
and similarly for $Q_{Y|W}^{(\mathsf{min})}$. By \eqref{eq:-31},
$\mu_{n}-1\leq e^{-n\frac{\alpha_{0}}{2}}$ for all sufficiently large
$n$.

Set $\delta=e^{-n\alpha_{1}}$ with $\alpha_{1}:=\min\left\{ \frac{\alpha_{0}}{4},\frac{\epsilon}{4}\right\} >0$,
then 
\begin{align}
 & \liminf_{n\to\infty}\frac{1}{n}\log\frac{3\left(1+\delta-\mu_{n}\right)^{2}\mathsf{M}}{2\left(1+\delta+2\mu_{n}\right)\beta_{n}}\nonumber \\
 & \geq R-\mathcal{I}_{\epsilon}\left(Q\right)\nonumber \\
 & \qquad+\liminf_{n\to\infty}\frac{1}{n}\log\frac{3\left(e^{-n\alpha_{1}}-e^{-n\frac{\alpha_{0}}{2}}\right)^{2}}{2\left(1+e^{-n\alpha_{1}}+2\left(1+e^{-n\frac{\alpha_{0}}{2}}\right)\right)}\\
 & =R-\mathcal{I}_{\epsilon}\left(Q\right)-2\alpha_{1},\label{eq:-118}
\end{align}
where \eqref{eq:-118} follows since $\alpha_{1}<\frac{\alpha_{0}}{2}$.
Hence the exponent of $\frac{3\left(1+\delta-\mu_{n}\right)^{2}\mathsf{M}}{2\left(1+\delta+2\mu_{n}\right)\beta_{n}}$
is lower bounded by 
\begin{equation}
R-\mathcal{I}_{\epsilon}\left(Q\right)-2\alpha_{1}\geq\frac{\epsilon}{2},\label{eq:-116}
\end{equation}
where \eqref{eq:-116} holds due to \eqref{eq:-75} and the choice
of $\alpha_{1}$. Hence \eqref{eq:-153} converges to zero doubly
exponentially fast in $n$. Combined this with \eqref{eq:-152} yields
\begin{equation}
\mathbb{P}_{\mathcal{C}_{n}}\left(D_{\infty}(P_{X^{n}Y^{n}|\mathcal{C}_{n}}\|\pi_{XY}^{n})\geq e^{-n\alpha_{1}}\right)\rightarrow0\label{eq:-76}
\end{equation}
doubly exponentially fast as $n\to\infty$.
\end{IEEEproof}

\subsection{\label{subsec:Converse-Part} Proof of $T_{\mathrm{Exact}}(\pi_{XY})\protect\geq\limsup_{n\to\infty}\frac{1}{n}\Gamma(\pi_{XY}^{n})$}

We prove the converse result from the perspective of exact common
information, i.e., 
\begin{equation}
T_{\mathrm{Exact}}(\pi_{XY})\geq\limsup_{n\to\infty}\frac{1}{n}\Gamma(\pi_{XY}^{n}).
\end{equation}

Similar to the idea used in Appendix \ref{subsec:Proof-of}, we first
independently replicate a $k$-length optimal exact common information
code $\left(P_{W_{k}},P_{X^{k}|W_{k}},P_{Y^{k}|W_{k}}\right)$ $n$
times. Then the resulting superblock code is also an exact common
information code, i.e., $\sum_{w^{n}}P_{W_{k}}^{n}(w^{n})P_{X^{k}|W_{k}}^{n}(\cdot|w^{n})P_{Y^{k}|W_{k}}^{n}(\cdot|w^{n})=\pi_{XY}^{kn}$.
Observe that $W_{k}^{n}=(W_{k,1},W_{k,2},...,W_{k,n})$ is an $n$-length
i.i.d. random sequence with each $W_{k,i}\sim P_{W_{k}}$. Hence we
have for $\epsilon>0$, 
\begin{equation}
\mathbb{P}\left(W_{k}^{n}\in\mathcal{A}_{\epsilon}^{(n)}\left(P_{W_{k}}\right)\right)\to1
\end{equation}
as $n\to\infty$ for fixed $k$. Furthermore, $|\mathcal{A}_{\epsilon}^{(n)}|\le e^{n(H(W_{k})+\epsilon)}$.
Consider
\begin{align}
 & D_{\infty}(P_{X^{kn}Y^{kn}}\|\pi_{XY}^{kn})\nonumber \\
 & =\log\Biggl(\max_{x^{kn},y^{kn}}\nonumber \\
 & \qquad\frac{\sum_{w^{n}}P_{W_{k}}^{n}(w^{n})P_{X^{k}|W_{k}}^{n}(x^{kn}|w^{n})P_{Y^{k}|W_{k}}^{n}(y^{kn}|w^{n})}{\pi_{XY}^{kn}\left(x^{kn},y^{kn}\right)}\Biggr)\\
 & \geq\log\Biggl(\max_{x^{kn},y^{kn}}\max_{w^{n}\in\mathcal{A}_{\epsilon}^{(n)}}\nonumber \\
 & \qquad\frac{P_{W_{k}}^{n}(w^{n})P_{X^{k}|W_{k}}^{n}(x^{kn}|w^{n})P_{Y^{k}|W_{k}}^{n}(y^{kn}|w^{n})}{\pi_{XY}^{kn}\left(x^{kn},y^{kn}\right)}\Biggr)\\
 & \geq\log\Biggl(\max_{x^{kn},y^{kn}}\max_{w^{n}\in\mathcal{A}_{\epsilon}^{(n)}}\nonumber \\
 & \qquad\frac{e^{-n\left(H(W_{k})+\epsilon\right)}P_{X^{k}|W_{k}}^{n}(x^{kn}|w^{n})P_{Y^{k}|W_{k}}^{n}(y^{kn}|w^{n})}{\pi_{XY}^{kn}\left(x^{kn},y^{kn}\right)}\Biggr)\\
 & =-n\left(H(W_{k})+\epsilon\right)+\log\Biggl(\max_{x^{kn},y^{kn}}\max_{w^{n}\in\mathcal{A}_{\epsilon}^{(n)}}\nonumber \\
 & \qquad\frac{P_{X^{k}|W_{k}}^{n}(x^{kn}|w^{n})P_{Y^{k}|W_{k}}^{n}(y^{kn}|w^{n})}{\pi_{XY}^{kn}\left(x^{kn},y^{kn}\right)}\Biggr).
\end{align}
Since for the exact common information superblock code, $D_{\infty}(P_{X^{kn}Y^{kn}}\|\pi_{XY}^{kn})=0$,
we have
\begin{align}
 & \frac{1}{k}\left(H(W_{k})+\epsilon\right)\nonumber \\
 & \geq\limsup_{n\to\infty}\frac{1}{kn}\log\nonumber \\
 & \qquad\left(\max_{x^{kn},y^{kn}}\max_{w^{n}\in\mathcal{A}_{\epsilon}^{(n)}}\frac{P_{X^{k}|W_{k}}^{n}(x^{kn}|w^{n})P_{Y^{k}|W_{k}}^{n}(y^{kn}|w^{n})}{\pi_{XY}^{kn}\left(x^{kn},y^{kn}\right)}\right)\\
 & =\limsup_{n\to\infty}\frac{1}{kn}\max_{w^{n}\in\mathcal{A}_{\epsilon}^{(n)}}\nonumber \\
 & \qquad\sum_{i=1}^{n}\max_{x^{k},y^{k}}\log\frac{P_{X^{k}|W_{k}}(x^{k}|w_{i})P_{Y^{k}|W_{k}}(y^{k}|w_{i})}{\pi_{XY}^{k}\left(x^{k},y^{k}\right)}.\label{eq:-1}
\end{align}
Continuing \eqref{eq:-1}, we obtain
\begin{align}
 & \max_{w^{n}\in\mathcal{A}_{\epsilon}^{(n)}}\sum_{i=1}^{n}\max_{x^{k},y^{k}}\log\frac{P_{X^{k}|W_{k}}(x^{k}|w_{i})P_{Y^{k}|W_{k}}(y^{k}|w_{i})}{\pi_{XY}^{k}\left(x^{k},y^{k}\right)}\nonumber \\
 & \geq\sum_{w^{n}\in\mathcal{A}_{\epsilon}^{(n)}}\frac{P_{W_{k}}^{n}(w^{n})}{P_{W_{k}}^{n}(\mathcal{A}_{\epsilon}^{(n)})}\sum_{i=1}^{n}\max_{\substack{Q_{X^{k}Y^{k}|W_{k}}\in\\
C(P_{X^{k}|W_{k}},P_{Y^{k}|W_{k}})
}
}\nonumber \\
 & \qquad\sum_{x^{k},y^{k}}Q_{X^{k}Y^{k}|W_{k}}\left(x^{k},y^{k}|w_{i}\right)\nonumber \\
 & \qquad\times\log\frac{P_{X^{k}|W_{k}}(x^{k}|w_{i})P_{Y^{k}|W_{k}}(y^{k}|w_{i})}{\pi_{XY}^{k}\left(x^{k},y^{k}\right)}\label{eq:-11}\\
 & =\sum_{w^{n}\in\mathcal{A}_{\epsilon}^{(n)}}\frac{P_{W_{k}}^{n}(w^{n})}{P_{W_{k}}^{n}(\mathcal{A}_{\epsilon}^{(n)})}\sum_{i=1}^{n}g\left(w_{i},P_{X^{k}|W_{k}},P_{Y^{k}|W_{k}}\right)\label{eq:-7}\\
 & =\sum_{w^{n}}\frac{P_{W_{k}}^{n}(w^{n})}{P_{W_{k}}^{n}(\mathcal{A}_{\epsilon}^{(n)})}\sum_{i=1}^{n}g\left(w_{i},P_{X^{k}|W_{k}},P_{Y^{k}|W_{k}}\right)\nonumber \\
 & \qquad-\sum_{w^{n}\notin\mathcal{A}_{\epsilon}^{(n)}}\frac{P_{W_{k}}^{n}(w^{n})}{P_{W_{k}}^{n}(\mathcal{A}_{\epsilon}^{(n)})}\sum_{i=1}^{n}g\left(w_{i},P_{X^{k}|W_{k}},P_{Y^{k}|W_{k}}\right)\\
 & \geq\frac{n}{P_{W_{k}}^{n}(\mathcal{A}_{\epsilon}^{(n)})}\mathbb{E}_{W\sim P_{W_{k}}}g\left(W,P_{X^{k}|W_{k}},P_{Y^{k}|W_{k}}\right)\nonumber \\
 & \qquad-\frac{1-P_{W_{k}}^{n}(\mathcal{A}_{\epsilon}^{(n)})}{P_{W_{k}}^{n}(\mathcal{A}_{\epsilon}^{(n)})}nk\log\min_{x,y:\pi_{XY}\left(x,y\right)>0}\pi_{XY}\left(x,y\right),\label{eq:-25}
\end{align}
where \eqref{eq:-11} follows since the maximum is no smaller than
the average; in \eqref{eq:-7},

\begin{align}
 & g\left(w,P_{X^{k}|W_{k}},P_{Y^{k}|W_{k}}\right)\nonumber \\
 & :=-H(X^{k}|W_{k}=w)-H(Y^{k}|W_{k}=w)\\
 & \qquad+\mathcal{H}(P_{X^{k}|W_{k}=w},P_{Y^{k}|W_{k}=w}\|\pi_{XY}^{k});
\end{align}
and \eqref{eq:-25} follows since 
\begin{align}
 & g\left(w,P_{X^{k}|W_{k}},P_{Y^{k}|W_{k}}\right)\nonumber \\
 & \leq\mathcal{H}(P_{X^{k}|W_{k}=w},P_{Y^{k}|W_{k}=w}\|\pi_{XY}^{k})\\
 & \leq-k\log\min_{x,y:\pi_{XY}\left(x,y\right)>0}\pi_{XY}\left(x,y\right).
\end{align}

Since $P_{W_{k}}^{n}(\mathcal{A}_{\epsilon}^{(n)})\to1$, combining
this fact with \eqref{eq:-1} and \eqref{eq:-7}, we have 
\begin{align}
 & \frac{1}{k}\left(H(W_{k})+\epsilon\right)\nonumber \\
 & \geq\frac{1}{k}\biggl(-H(X^{k}|W_{k})-H(Y^{k}|W_{k})\nonumber \\
 & \qquad+\sum_{w}P_{W_{k}}(w)\mathcal{H}(P_{X^{k}|W_{k}=w},P_{Y^{k}|W_{k}=w}\|\pi_{XY}^{k})\biggr)\\
 & \geq\frac{1}{k}\biggl(\inf_{\substack{P_{W_{k}}P_{X^{k}|W_{k}}P_{Y^{k}|W_{k}}:\\
P_{X^{k}Y^{k}}=\pi_{XY}^{k}
}
}-H(X^{k}|W_{k})-H(Y^{k}|W_{k})\nonumber \\
 & \qquad+\sum_{w}P_{W_{k}}(w)\mathcal{H}(P_{X^{k}|W_{k}=w},P_{Y^{k}|W_{k}=w}\|\pi_{XY}^{k})\biggr)\\
 & =\frac{1}{k}\Gamma(\pi_{XY}^{k}).
\end{align}

Furthermore, since $\frac{1}{k}H(W_{k})\to R$ as $k\to\infty$, we
have 
\begin{align}
 & R\geq\limsup_{k\to\infty}\frac{1}{k}\Gamma(\pi_{XY}^{k}).
\end{align}

\section{\label{sec:singleletter}Proof of Theorem \ref{thm:singleletter}}

The inequality $\widetilde{T}_{\infty}(\pi_{XY})\ge C_{\mathsf{Wyner}}(\pi_{XY})$
follows since $\widetilde{T}_{\infty}(\pi_{XY})\ge\widetilde{T}_{1}(\pi_{XY})\ge C_{\mathsf{Wyner}}(\pi_{XY})$,
where the last inequality is the converse result for Wyner's common
information \cite{Wyner}. On the other hand, the upper bound $\Gamma^{\mathrm{UB}}(\pi_{XY})$
(i.e., $\Gamma(\pi_{XY})$) has been proved in Appendix \ref{subsec:achievability}.
Hence we only need to prove the lower bound, i.e., $\widetilde{T}_{\infty}(\pi_{XY})\ge\Gamma^{\mathrm{LB}}(\pi_{XY})$.
The proof for this inequality is divided into two parts: single-letterization
and simplifying constraints.

\subsection{Single-letterization}

Observe by Remark \ref{rem:By-a-similar}, since $\widetilde{T}_{\infty}(\pi_{XY})\geq\lim_{\epsilon\downarrow0}\lim_{n\to\infty}\frac{1}{n}\Gamma_{\epsilon}(\pi_{XY}^{n}),$
in order to lower bound $\widetilde{T}_{\infty}(\pi_{XY})$, it suffices
to lower bound $\frac{1}{n}\Gamma_{\epsilon}(\pi_{XY}^{n})$. According
to the definition of $\Gamma_{\epsilon}(\pi_{XY}^{n})$ in \eqref{eq:GammaEpsilon},
we have \eqref{eq:-54} (given on page \pageref{eq:-54}). 
\begin{figure*}
\begin{align}
\frac{1}{n}\Gamma_{\epsilon}(\pi_{XY}^{n}) & \geq\frac{1}{n}\inf_{\substack{\substack{P_{W}P_{X^{n}|W}P_{Y^{n}|W}:}
\\
\frac{1}{n}D_{\infty}\left(P_{X^{n}Y^{n}}\|\pi_{XY}^{n}\right)\le\epsilon
}
}\max_{\substack{\substack{Q_{X^{n}Y^{n}|W}\in}
\\
C(P_{X^{n}|W},P_{Y^{n}|W})
}
}-\frac{1}{n}\sum_{i=1}^{n}H(X_{i}|X^{i-1}W)-\frac{1}{n}\sum_{i=1}^{n}H(Y_{i}|Y^{i-1}W)\nonumber \\
 & \qquad-\sum_{w}P(w)\frac{1}{n}\sum_{i=1}^{n}\left(\sum_{x^{i-1},y^{i-1}}Q(x^{i-1},y^{i-1}|w)\sum_{x_{i},y_{i}}Q(x_{i},y_{i}|x^{i-1},y^{i-1},w)\log\pi\left(x_{i},y_{i}\right)\right).\label{eq:-54}
\end{align}

\hrulefill{}
\end{figure*}
  Denote $J\sim P_{J}:=\mathrm{Unif}[1:n]$ as a time index which
is independent of $(W,X^{n},Y^{n})\sim P_{W}P_{X^{n}|W}P_{Y^{n}|W}$.
Then 
\begin{align}
 & -\frac{1}{n}\sum_{i=1}^{n}H(X_{i}|X^{i-1}W)-\frac{1}{n}\sum_{i=1}^{n}H(Y_{i}|Y^{i-1}W)\nonumber \\
 & =-H(X_{J}|X^{J-1}WJ)-H(Y_{J}|Y^{J-1}WJ).\label{eq:-114}
\end{align}

Next we single-letterize the last term in \eqref{eq:-54}. On one
hand,
\begin{align}
 & \sum_{x^{i-1},y^{i-1}}Q(x^{i-1},y^{i-1}|w)\nonumber \\
 & \qquad\times\sum_{x_{i},y_{i}}Q(x_{i},y_{i}|x^{i-1},y^{i-1},w)\log\frac{1}{\pi\left(x_{i},y_{i}\right)}\nonumber \\
 & \geq\min_{\substack{\widetilde{Q}_{X^{i-1}Y^{i-1}|W}\in\\
C(P_{X^{i-1}|W},P_{Y^{i-1}|W})
}
}\sum_{x^{i-1},y^{i-1}}\widetilde{Q}(x^{i-1},y^{i-1}|w)\nonumber \\
 & \qquad\times\sum_{x_{i},y_{i}}Q(x_{i},y_{i}|x^{i-1},y^{i-1},w)\log\frac{1}{\pi\left(x_{i},y_{i}\right)}.\label{eq:-56}
\end{align}
On the other hand, in order to get a further lower bound on \eqref{eq:-56},
we need the following ``chain rule'' for coupling sets. 
\begin{lem}[``Chain Rule'' for Coupling Sets]
\label{lem:coupling} For a pair of conditional distributions $(P_{X^{n}|W},P_{Y^{n}|W})$,
we have
\begin{equation}
\prod_{i=1}^{n}C(P_{X_{i}|X^{i-1}W},P_{Y_{i}|Y^{i-1}W})\subseteq C(P_{X^{n}|W},P_{Y^{n}|W}),
\end{equation}
where 
\begin{align}
 & C(P_{X_{i}|X^{i-1}W},P_{Y_{i}|Y^{i-1}W})\nonumber \\
 & :=\Bigl\{ Q_{X_{i}Y_{i}|X^{i-1}Y^{i-1}W}:\,Q_{X_{i}|X^{i-1}Y^{i-1}W}=P_{X_{i}|X^{i-1}W},\nonumber \\
 & \qquad Q_{Y_{i}|X^{i-1}Y^{i-1}W}=P_{Y_{i}|Y^{i-1}W}\Bigr\},i\in[1:n]
\end{align}
and
\begin{align}
 & \prod_{i=1}^{n}C(P_{X_{i}|X^{i-1}W},P_{Y_{i}|Y^{i-1}W})\nonumber \\
 & :=\Bigl\{\prod_{i=1}^{n}Q_{X_{i}Y_{i}|X^{i-1}Y^{i-1}W}:Q_{X_{i}Y_{i}|X^{i-1}Y^{i-1}W}\in\nonumber \\
 & \qquad C(P_{X_{i}|X^{i-1}W},P_{Y_{i}|Y^{i-1}W}),i\in[1:n]\Bigr\}.
\end{align}
 
\end{lem}
\begin{IEEEproof}
If $\left\{ Q_{X_{i}Y_{i}|X^{i-1}Y^{i-1}W}\right\} _{i\in[1:n]}$
is a set of distributions such that $Q_{X_{i}Y_{i}|X^{i-1}Y^{i-1}W}\in C(P_{X_{i}|X^{i-1}W},P_{Y_{i}|Y^{i-1}W})$
for all $i\in[1:n]$, then we have that for any $\left(w,x^{n}\right)$,
\begin{align}
 & \sum_{y^{n}}\prod_{i=1}^{n}Q\left(x_{i},y_{i}|x^{i-1},y^{i-1},w\right)\nonumber \\
 & =\sum_{y^{n-1}}\prod_{i=1}^{n-1}Q\left(x_{i},y_{i}|x^{i-1},y^{i-1},w\right)\nonumber \\
 & \qquad\times\sum_{y_{n}}Q\left(x_{n},y_{n}|x^{n-1},y^{n-1},w\right)\\
 & =\sum_{y^{n-1}}\prod_{i=1}^{n-1}Q\left(x_{i},y_{i}|x^{i-1},y^{i-1},w\right)Q\left(x_{n}|x^{n-1},y^{n-1},w\right)\\
 & =P\left(x_{n}|x^{n-1},w\right)\sum_{y^{n-1}}\prod_{i=1}^{n-1}Q\left(x_{i},y_{i}|x^{i-1},y^{i-1},w\right)\label{eq:-99}\\
 & =P\left(x_{n}|x^{n-1},w\right)P\left(x_{n-1}|x^{n-2},w\right)\nonumber \\
 & \qquad\times\sum_{y^{n-2}}\prod_{i=1}^{n-2}Q\left(x_{i},y_{i}|x^{i-1},y^{i-1},w\right)\\
 & \vdots\nonumber \\
 & =\prod_{i=1}^{n}P\left(x_{i}|x^{i-1},w\right)\\
 & =P\left(x^{n}|w\right),
\end{align}
where  \eqref{eq:-99} follows since $Q_{X_{i}Y_{i}|X^{i-1}Y^{i-1}W}\in C(P_{X_{i}|X^{i-1}W},P_{Y_{i}|Y^{i-1}W})$. 

Hence $\prod_{i=1}^{n}Q_{X_{i}Y_{i}|X^{i-1}Y^{i-1}W}$ has marginal
conditional distributions $P_{X^{n}|W}$ and $P_{Y^{n}|W}$, i.e.,
$\prod_{i=1}^{n}Q_{X_{i}Y_{i}|X^{i-1}Y^{i-1}W}\in C(P_{X^{n}|W},P_{Y^{n}|W})$.
Since for any $i\in[1:n]$, $Q_{X_{i}Y_{i}|X^{i-1}Y^{i-1}W}$ is an
arbitrary distribution in $C(P_{X_{i}|X^{i-1}W},P_{Y_{i}|Y^{i-1}W})$,
we have that $\prod_{i=1}^{n}C(P_{X_{i}|X^{i-1}W},P_{Y_{i}|Y^{i-1}W})\subseteq C(P_{X^{n}|W},P_{Y^{n}|W})$.
\end{IEEEproof}
By Lemma \ref{lem:coupling}, we have that for any function $f:\mathcal{P}\left(\mathcal{X}^{n}\times\mathcal{Y}^{n}\right)\to\mathbb{R}$,
\begin{align}
 & \max_{\substack{\substack{Q_{X^{n}Y^{n}|W}\in}
\\
C(P_{X^{n}|W},P_{Y^{n}|W})
}
}f\left(Q_{X^{n}Y^{n}|W}\right)\nonumber \\
 & \geq\max_{\substack{\substack{Q_{X^{n}Y^{n}|W}\in}
\\
\prod_{i=1}^{n}C(P_{X_{i}|X^{i-1}W},P_{Y_{i}|Y^{i-1}W})
}
}f\left(\prod_{i=1}^{n}Q_{X_{i}Y_{i}|X^{i-1}Y^{i-1}W}\right).\label{eq:-57}
\end{align}
Therefore, substituting \eqref{eq:-56} into the last term in \eqref{eq:-54}
and utilizing \eqref{eq:-57}, we obtain \eqref{eq:-15}-\eqref{eq:-115}
(given on page \pageref{eq:-15}), where the swapping of min and max
in \eqref{eq:-14} follows since on one hand, maximin is no larger
than minimax, and on the other hand, 
\begin{align}
\eqref{eq:-15} & \geq\sum_{w}P(w)\frac{1}{n}\sum_{i=1}^{n}\min_{\substack{\widetilde{Q}_{X^{i-1}Y^{i-1}|W}\in\\
C(P_{X^{i-1}|W},P_{Y^{i-1}|W})
}
}\nonumber \\
 & \qquad\sum_{x^{i-1},y^{i-1}}\widetilde{Q}(x^{i-1},y^{i-1}|w)\nonumber \\
 & \qquad\times\sum_{x_{i},y_{i}}Q^{*}(x_{i},y_{i}|x^{i-1},y^{i-1},w)\log\frac{1}{\pi\left(x_{i},y_{i}\right)}\label{eq:-15-2}\\
 & =\eqref{eq:-14}
\end{align}
with 
\begin{align}
 & Q_{X_{i}Y_{i}|X^{i-1}Y^{i-1}W}^{*}\nonumber \\
 & :=\arg\max_{\substack{\substack{Q_{X_{i}Y_{i}|X^{i-1}Y^{i-1}W}\in}
\\
C(P_{X_{i}|X^{i-1}W},P_{Y_{i}|Y^{i-1}W})
}
}\sum_{x_{i},y_{i}}Q(x_{i},y_{i}|x^{i-1},y^{i-1},w)\nonumber \\
 & \qquad\times\log\frac{1}{\pi\left(x_{i},y_{i}\right)}.
\end{align}

\begin{figure*}
\begin{align}
 & \max_{\substack{\substack{Q_{X^{n}Y^{n}|W}\in}
\\
C(P_{X^{n}|W},P_{Y^{n}|W})
}
}\sum_{w}P(w)\frac{1}{n}\sum_{i=1}^{n}\sum_{x^{i-1},y^{i-1}}Q(x^{i-1},y^{i-1}|w)\sum_{x_{i},y_{i}}Q(x_{i},y_{i}|x^{i-1},y^{i-1},w)\log\frac{1}{\pi\left(x_{i},y_{i}\right)}\nonumber \\
 & \geq\sum_{w}P(w)\frac{1}{n}\sum_{i=1}^{n}\max_{\substack{\substack{Q_{X_{i}Y_{i}|X^{i-1}Y^{i-1}W}\in}
\\
C(P_{X_{i}|X^{i-1}W},P_{Y_{i}|Y^{i-1}W})
}
}\min_{\substack{\widetilde{Q}_{X^{i-1}Y^{i-1}|W}\in\\
C(P_{X^{i-1}|W},P_{Y^{i-1}|W})
}
}\nonumber \\
 & \qquad\sum_{x^{i-1},y^{i-1}}\widetilde{Q}(x^{i-1},y^{i-1}|w)\sum_{x_{i},y_{i}}Q(x_{i},y_{i}|x^{i-1},y^{i-1},w)\log\frac{1}{\pi\left(x_{i},y_{i}\right)}\label{eq:-15}\\
 & =\sum_{w}P(w)\frac{1}{n}\sum_{i=1}^{n}\min_{\substack{\widetilde{Q}_{X^{i-1}Y^{i-1}|W}\in\\
C(P_{X^{i-1}|W},P_{Y^{i-1}|W})
}
}\sum_{x^{i-1},y^{i-1}}\widetilde{Q}(x^{i-1},y^{i-1}|w)\nonumber \\
 & \qquad\times\max_{\substack{\substack{Q_{X_{i}Y_{i}|X^{i-1}Y^{i-1}W}\in}
\\
C(P_{X_{i}|X^{i-1}W},P_{Y_{i}|Y^{i-1}W})
}
}\sum_{x_{i},y_{i}}Q(x_{i},y_{i}|x^{i-1},y^{i-1},w)\log\frac{1}{\pi\left(x_{i},y_{i}\right)}\label{eq:-14}\\
 & =\sum_{w}P(w)\sum_{j=1}^{n}P_{J}(j)\min_{\substack{\substack{\widetilde{Q}_{X^{J-1}Y^{J-1}|WJ}\in}
\\
C(P_{X^{J-1}|WJ},P_{Y^{J-1}|WJ})
}
}\sum_{x^{j-1},y^{j-1}}\widetilde{Q}(x^{j-1},y^{j-1}|w,j)\nonumber \\
 & \qquad\times\max_{\substack{\substack{Q_{X_{J}Y_{J}|X^{J-1}Y^{J-1}WJ}\in}
\\
C(P_{X_{J}|X^{J-1}WJ},P_{Y_{J}|Y^{J-1}WJ})
}
}\sum_{x_{j},y_{j}}Q(x_{j},y_{j}|x^{j-1},y^{j-1},w,j)\log\frac{1}{\pi\left(x_{j},y_{j}\right)},\label{eq:-115}
\end{align}

\hrulefill{}
\end{figure*}

For brevity, we set 
\begin{equation}
W\leftarrow WJ,U\leftarrow X^{J-1},V\leftarrow Y^{J-1},X\leftarrow X_{J},Y\leftarrow Y_{J}.\label{eq:-37-1}
\end{equation}
  Then $\frac{1}{n}D_{\infty}\left(P_{X^{n}Y^{n}}\|\pi_{XY}^{n}\right)\le\epsilon$
implies that $D\left(P_{XY}\|\pi_{XY}\right)\le\epsilon$. Since $\pi_{XY}$
has finite support, $D\left(P_{XY}\|\pi_{XY}\right)\to0$ if and only
if $D_{\infty}\left(P_{XY}\|\pi_{XY}\right)\to0$. Therefore, substituting
\eqref{eq:-114} and \eqref{eq:-115} into \eqref{eq:-54} and utilizing
the identification of the random variables in \eqref{eq:-37-1}, we
obtain \eqref{eq:-40-2} (given on page \pageref{eq:-40-2}).
\begin{figure*}
\begin{align}
\widetilde{T}_{\infty}(\pi_{XY}) & \geq\lim_{\epsilon\downarrow0}\inf_{\substack{P_{W}P_{U|W}P_{V|W}P_{X|U}P_{Y|V}:\\
D_{\infty}\left(P_{XY}\|\pi_{XY}\right)\le\epsilon
}
}-H(X|UW)-H(Y|VW)\nonumber \\
 & \qquad+\sum_{w}P(w)\inf_{\substack{\widetilde{Q}_{UV|W}\in\\
C(P_{U|W},P_{V|W})
}
}\sum_{u,v}\widetilde{Q}(u,v|w)\max_{\substack{Q_{XY|UVW}\in\\
C(P_{X|UW},P_{Y|VW})
}
}\sum_{x,y}Q(x,y|u,v,w)\log\frac{1}{\pi\left(x,y\right)}.\label{eq:-40-2}
\end{align}

\hrulefill{}
\end{figure*}
 For $\widetilde{Q}_{UV|W}\in C(P_{U|W},P_{V|W})$, define a joint
distribution induced by $\widetilde{Q}_{UV|W}$ as 
\begin{align}
 & \widehat{Q}_{\left(U,V',W\right),\left(U',V,W'\right)}(u,v',w,u',v,w')\nonumber \\
 & :=P_{W}(w)\widetilde{Q}_{UV|W}(u,v|w)1\left\{ w'=w\right\} \nonumber \\
 & \qquad\times P_{V|W}(v'|w)P_{U|W}(u'|w').
\end{align}
Then this joint distribution satisfies the following marginal constraints:
\begin{align}
\widehat{Q}_{UVW}(u,v,w) & =P_{W}(w)\widetilde{Q}_{UV|W}(u,v|w)\label{eq:-47-2}\\
\widehat{Q}_{UV'W}(u,v',w) & =P_{UVW}(u,v',w)\\
\widehat{Q}_{U'VW'}(u',v,w') & =P_{UVW}(u',v,w').\label{eq:-48-1}
\end{align}
Utilizing this induced distribution, its properties in \eqref{eq:-47-2}-\eqref{eq:-48-1},
and the lower bound in \eqref{eq:-40-2}, we obtain 
\begin{align}
 & \widetilde{T}_{1+s}(\pi_{XY})\nonumber \\
 & \geq\lim_{\epsilon\downarrow0}\inf_{\substack{P_{W}P_{U|W}P_{V|W}P_{X|U}P_{Y|V}:\\
D_{\infty}\left(P_{XY}\|\pi_{XY}\right)\le\epsilon
}
}-\left(H(X|UW)+H(Y|VW)\right)\nonumber \\
 & \qquad+\inf_{\substack{\widehat{Q}_{\left(U,V',W\right),\left(U',V,W'\right)}\in\\
C(P_{UVW},P_{UVW})
}
}\sum_{u,u',v,v',w,w'}\widehat{Q}(u,v',w,u',v,w')\nonumber \\
 & \qquad\times\max_{\substack{Q_{XY}\in\\
C(P_{X|UW=u,w},P_{Y|VW=v,w'})
}
}\sum_{x,y}Q(x,y)\log\frac{1}{\pi\left(x,y\right)}.\label{eq:-49-1}
\end{align}
Observe that $P_{X|UW=u,w}=P_{X|\left(U,V,W\right)=\left(u,v',w\right)}$
and $P_{Y|VW=v,w'}=P_{Y|\left(U,V,W\right)=\left(u',v,w'\right)}$
(since $X\to UW\to V$ and $Y\to VW\to U$ form Markov chains under
$P$). Hence the coupling set $C(P_{X|UW=u,w},P_{Y|VW=v,w'})$ in
the last term in \eqref{eq:-49-1} can be replaced by $C(P_{X|\left(U,V,W\right)=\left(u,v',w\right)},P_{Y|\left(U,V,W\right)=\left(u',v,w'\right)})$.

Substituting $W\leftarrow\left(U,V,W\right)$, we can simplify \eqref{eq:-49-1}
as follows:
\begin{align}
 & \widetilde{T}_{1+s}(\pi_{XY})\nonumber \\
 & \geq\lim_{\epsilon\downarrow0}\inf_{\substack{P_{W}P_{X|W}P_{Y|W}:\\
D_{\infty}\left(P_{XY}\|\pi_{XY}\right)\le\epsilon
}
}-\left(H(X|W)+H(Y|W)\right)\nonumber \\
 & \qquad+\inf_{\substack{Q_{WW'}\in\\
C(P_{W},P_{W})
}
}\sum_{w,w'}Q(w,w')\mathcal{H}(P_{X|W=w},P_{Y|W=w'}\|\pi_{XY}).\label{eq:-23-1}
\end{align}

\subsection{Simplifying Constraints}

 Next we prove that the constraint $D_{\infty}\left(P_{XY}\|\pi_{XY}\right)\le\epsilon$
in \eqref{eq:-23-1} can be replaced with $P_{XY}=\pi_{XY}$. For
$D_{\infty}\left(P_{XY}\|\pi_{XY}\right)\le\epsilon$, using the splitting
technique, we can write
\begin{equation}
\pi_{XY}\left(x,y\right)=e^{-\epsilon}P_{XY}\left(x,y\right)+\left(1-e^{-\epsilon}\right)\widehat{P}_{XY}\left(x,y\right)\label{eq:-100-2}
\end{equation}
where 
\begin{equation}
\widehat{P}_{XY}\left(x,y\right):=\frac{e^{\epsilon}\pi_{XY}\left(x,y\right)-P_{XY}\left(x,y\right)}{e^{\epsilon}-1}.
\end{equation}
Define 
\begin{align}
 & \widetilde{P}_{XYWU}(x,y,w,u)\nonumber \\
 & =\begin{cases}
e^{-\epsilon}P_{W}P_{X|W}P_{Y|W} & \textrm{if }u=1\\
\left(1-e^{-\epsilon}\right)\widehat{P}_{XY}\left(x,y\right)1\left\{ w=(x,y)\right\}  & \textrm{if }u=0
\end{cases}.
\end{align}
Then consider \eqref{eq:-154}-\eqref{eq:-156} (given on page \pageref{eq:-154}),
where \eqref{eq:-154} follows since $\widetilde{P}_{U}(u)\widetilde{P}_{U}(u')Q_{WW'}(w,w')$
with $Q_{WW'}\in C(\widetilde{P}_{W|U=u},\widetilde{P}_{W|U=u'})$
forms a coupling of $\left(\widetilde{P}_{WU},\widetilde{P}_{WU}\right)$,
and \eqref{eq:-155} follows since $\mathcal{H}(Q_{X},Q_{Y}\|\pi_{XY})\leq\max_{\left(x,y\right)\in\supp(\pi_{XY})}\log\frac{1}{\pi\left(x,y\right)}$
for any $\left(Q_{X},Q_{Y}\right)$. 
\begin{figure*}
\begin{align}
 & -H_{\widetilde{P}}(XY|WU)+\inf_{\substack{Q_{WUW'U'}\in\\
C(\widetilde{P}_{WU},\widetilde{P}_{WU})
}
}\sum_{w,u,w',u'}Q(w,u,w',u')\mathcal{H}(\widetilde{P}_{X|(W,U)=(w,u)},\widetilde{P}_{Y|(W,U)=(w',u')}\|\pi_{XY})\nonumber \\
 & \leq-e^{-\epsilon}H(XY|W)+\sum_{u,u'}\widetilde{P}_{U}(u)\widetilde{P}_{U}(u')\nonumber \\
 & \qquad\times\inf_{\substack{Q_{WW'}\in\\
C(\widetilde{P}_{W|U=u},\widetilde{P}_{W|U=u'})
}
}\sum_{w,w'}Q(w,w')\mathcal{H}(\widetilde{P}_{X|(W,U)=(w,u)},\widetilde{P}_{Y|(W,U)=(w',u')}\|\pi_{XY})\label{eq:-154}\\
 & \leq-e^{-\epsilon}H(XY|W)+e^{-2\epsilon}\inf_{Q_{WW'}\in C(P_{W},P_{W})}\sum_{w,w'}Q(w,w')\mathcal{H}(P_{X|W=w},P_{Y|W=w'}\|\pi_{XY})\nonumber \\
 & \qquad+\left(1-e^{-2\epsilon}\right)\max_{\left(x,y\right)\in\supp(\pi_{XY})}\log\frac{1}{\pi\left(x,y\right)}\label{eq:-155}\\
 & \leq e^{-\epsilon}\left(-H(XY|W)+\inf_{Q_{WW'}\in C(P_{W},P_{W})}\sum_{w,w'}Q(w,w')\mathcal{H}(P_{X|W=w},P_{Y|W=w'}\|\pi_{XY})\right)+O(\epsilon).\label{eq:-156}
\end{align}

\hrulefill{}
\end{figure*}

Hence substituting \eqref{eq:-156} into \eqref{eq:-23-1}, we obtain
\eqref{eq:-174}-\eqref{eq:-175} (given on page \pageref{eq:-174}).
\begin{figure*}
\begin{align}
\widetilde{T}_{\infty}(\pi_{XY}) & \geq\lim_{\epsilon\downarrow0}\inf_{\substack{P_{W}P_{X|W}P_{Y|W}:\\
D_{\infty}\left(P_{XY}\|\pi_{XY}\right)\le\epsilon
}
}e^{\epsilon}\biggl\{-H_{\widetilde{P}}(XY|WU)\nonumber \\
 & \qquad+\inf_{\substack{Q_{WUW'U'}\in\\
C(\widetilde{P}_{WU},\widetilde{P}_{WU})
}
}\sum_{w,u,w',u'}Q(w,u,w',u')\mathcal{H}(\widetilde{P}_{X|(W,U)=(w,u)},\widetilde{P}_{Y|(W,U)=(w',u')}\|\pi_{XY})+O(\epsilon)\biggr\}\label{eq:-174}\\
 & =\lim_{\epsilon\downarrow0}\inf_{\substack{P_{W}P_{X|W}P_{Y|W}:\\
D_{\infty}\left(P_{XY}\|\pi_{XY}\right)\le\epsilon
}
}-H_{\widetilde{P}}(XY|WU)\nonumber \\
 & \qquad+\inf_{\substack{Q_{WUW'U'}\in\\
C(\widetilde{P}_{WU},\widetilde{P}_{WU})
}
}\sum_{w,u,w',u'}Q(w,u,w',u')\mathcal{H}(\widetilde{P}_{X|(W,U)=(w,u)},\widetilde{P}_{Y|(W,U)=(w',u')}\|\pi_{XY})\\
 & \geq\inf_{\substack{\widetilde{P}_{WU}\widetilde{P}_{X|WU}\widetilde{P}_{Y|WU}:\\
\widetilde{P}_{XY}=\pi_{XY}
}
}-H_{\widetilde{P}}(XY|WU)\nonumber \\
 & \qquad+\inf_{\substack{Q_{WUW'U'}\in\\
C(\widetilde{P}_{WU},\widetilde{P}_{WU})
}
}\sum_{w,u,w',u'}Q(w,u,w',u')\mathcal{H}(\widetilde{P}_{X|(W,U)=(w,u)},\widetilde{P}_{Y|(W,U)=(w',u')}\|\pi_{XY})\\
 & =\Gamma^{\mathrm{LB}}(\pi_{XY}).\label{eq:-175}
\end{align}

\hrulefill{}
\end{figure*}

\section{\label{sec:DSBS}Proof of Theorem \ref{thm:DSBS}}

\emph{Upper Bound: }Set $X=W\oplus A$ and $Y=W\oplus B$ with $W\sim\mathrm{Bern}(\frac{1}{2})$,
$A\sim\mathrm{Bern}(a)$, and $B\sim\mathrm{Bern}(a)$ mutually independent,
where $a:=\frac{1-\sqrt{1-2p}}{2}\in(0,\frac{1}{2})$. 
\begin{align}
 & \mathcal{H}(P_{X|W=w},P_{Y|W=w}\|\pi_{XY})\nonumber \\
 & =\max_{Q_{XY}\in C(P_{X|W=w},P_{Y|W=w})}\sum_{x,y}Q_{XY}(x,y)\log\frac{1}{\pi\left(x,y\right)}\\
 & =\log\frac{1}{\alpha_{0}}+2\min\{a,\overline{a}\}\log\frac{\alpha_{0}}{\beta_{0}}\\
 & =\log\frac{1}{\alpha_{0}}+2a\log\frac{\alpha_{0}}{\beta_{0}}
\end{align}
Hence we have 
\begin{align}
 & \Gamma^{\mathrm{UB}}(\pi_{XY})\nonumber \\
 & \leq-H_{2}(a)-H_{2}(a)+\log\frac{1}{\alpha_{0}}+2a\log\frac{\alpha_{0}}{\beta_{0}}\\
 & =-2H_{2}(a)+\log\frac{1}{\alpha_{0}}+2a\log\frac{\alpha_{0}}{\beta_{0}}.\label{eq:-60}
\end{align}
Substituting $\alpha_{0},\beta_{0}$ into \eqref{eq:-60}, we get
the right hand side of \eqref{eq:-61}.

\emph{Lower Bound: }We adopt similar techniques as ones used by Wyner
\cite{Wyner}. Denote 
\begin{align}
\alpha(w) & :=\mathbb{P}\left(X=0|W=w\right)\\
\beta(w) & :=\mathbb{P}\left(Y=0|W=w\right).
\end{align}
Hence $P_{XY}=\pi_{XY}$ implies 
\begin{align}
\mathbb{E}\alpha(W) & =\mathbb{P}\left(X=0\right)=\frac{1}{2}\\
\mathbb{E}\beta(W) & =\mathbb{P}\left(Y=0\right)=\frac{1}{2}\\
\mathbb{E}\alpha(W)\beta(W) & =\mathbb{P}\left(X=0,Y=0\right)=\alpha_{0}.
\end{align}
Observe that 
\begin{align}
 & \mathcal{H}(P_{X|W=w},P_{Y|W=w'}\|\pi_{XY})\nonumber \\
 & =\max_{Q_{XY}\in C(P_{X|W=w},P_{Y|W=w'})}\sum_{x,y}Q_{XY}(x,y)\log\frac{1}{\pi\left(x,y\right)}\\
 & =\log\frac{1}{\alpha_{0}}\nonumber \\
 & \qquad+\left(\min\{\alpha(w),\overline{\beta(w')}\}+\min\{\overline{\alpha(w)},\beta(w')\right)\log\frac{\alpha_{0}}{\beta_{0}}\\
 & =\log\frac{1}{\alpha_{0}}+\min\{\alpha(w)+\beta(w'),\overline{\alpha(w)}+\overline{\beta(w')}\}\log\frac{\alpha_{0}}{\beta_{0}}\\
 & \geq\log\frac{1}{\alpha_{0}}\nonumber \\
 & \qquad+\left(\min\{\alpha(w),\overline{\alpha(w)}\}+\min\{\beta(w'),\overline{\beta(w')}\right)\log\frac{\alpha_{0}}{\beta_{0}}.
\end{align}
Here $\overline{a}=1-a$.

Define $\alpha'(W):=\left|\alpha(W)-\frac{1}{2}\right|,\beta'(W):=\left|\beta(W)-\frac{1}{2}\right|$,
$\gamma(W):=\frac{\alpha'(W)+\beta'(W)}{2}$, $\delta(W):=\gamma^{2}(W)$,
and $\theta:=\sqrt{\mathbb{E}\delta(W)}$. Then we can lower bound
$\Gamma^{\mathrm{LB}}(\pi_{XY})$ as \eqref{eq:-176}-\eqref{eq:-5}
(given on page \pageref{eq:-176}), where \eqref{eq:-8} follows from
\cite[Prop. 3.2]{Wyner}; \eqref{eq:-9} follows since $-H_{2}(t)$
is convex in $t$; \eqref{eq:-3} follows from \cite[Prop. 3.3]{Wyner}
and the fact $x\mapsto\sqrt{x}$ is a concave function; \eqref{eq:-5}
follows since the objective function in \eqref{eq:-4} is non-decreasing
in $\theta$ (this can be seen from the facts that the stationary
point $\theta^{*}=\nicefrac{\frac{1}{2}\left(\frac{\alpha_{0}}{\beta_{0}}-1\right)}{\frac{\alpha_{0}}{\beta_{0}}+1}$
of the objective function is not larger than $\sqrt{\alpha_{0}-\frac{1}{4}}$,
the objective function is convex, and the derivative of the objective
function is continuous). 
\begin{figure*}
\begin{align}
 & \Gamma^{\mathrm{LB}}(\pi_{XY})\nonumber \\
 & \geq\inf_{\substack{P_{W},\alpha(\cdot),\beta(\cdot):\\
\mathbb{E}\alpha(W)=\frac{1}{2}\\
\mathbb{E}\beta(W)=\frac{1}{2}\\
\mathbb{E}\alpha(W)\beta(W)=\alpha_{0}
}
}-\mathbb{E}H_{2}(\alpha(W))-\mathbb{E}H_{2}(\beta(W))+\log\frac{1}{\alpha_{0}}+\left(\mathbb{E}\min\{\alpha(W),\overline{\alpha(W)}\}+\mathbb{E}\min\{\beta(W),\overline{\beta(W)}\}\right)\log\frac{\alpha_{0}}{\beta_{0}}\label{eq:-176}\\
 & \geq\inf_{\substack{P_{W},\alpha(\cdot),\beta(\cdot):\\
\mathbb{E}\alpha(W)=\frac{1}{2}\\
\mathbb{E}\beta(W)=\frac{1}{2}\\
\mathbb{E}\alpha(W)\beta(W)\geq\alpha_{0}
}
}-\mathbb{E}H_{2}(\alpha(W))-\mathbb{E}H_{2}(\beta(W))+\log\frac{1}{\alpha_{0}}+\left(\mathbb{E}\min\{\alpha(W),\overline{\alpha(W)}\}+\mathbb{E}\min\{\beta(W),\overline{\beta(W)}\}\right)\log\frac{\alpha_{0}}{\beta_{0}}\\
 & \geq\inf_{\substack{P_{W},\alpha'(\cdot),\beta'(\cdot):\\
0\leq\alpha'(W),\beta'(W)\leq\frac{1}{2}\\
\mathbb{E}\alpha'(W)\beta'(W)\geq\alpha_{0}-\frac{1}{4}
}
}-\mathbb{E}H_{2}\left(\frac{1}{2}+\alpha'(W)\right)-\mathbb{E}H_{2}\left(\frac{1}{2}+\beta'(W)\right)+\log\frac{1}{\alpha_{0}}\nonumber \\
 & \qquad+\left(\mathbb{E}\left(\frac{1}{2}-\alpha'(W)\right)+\mathbb{E}\left(\frac{1}{2}-\beta'(W)\right)\right)\log\frac{\alpha_{0}}{\beta_{0}}\label{eq:-8}\\
 & \geq\inf_{\substack{P_{W},\gamma(\cdot):\\
0\leq\gamma(W)\leq\frac{1}{2}\\
\mathbb{E}\gamma^{2}(W)\geq\alpha_{0}-\frac{1}{4}
}
}-2\mathbb{E}H_{2}\left(\frac{1}{2}+\gamma(W)\right)+\log\frac{1}{\alpha_{0}}+\left(1-2\mathbb{E}\gamma(W)\right)\log\frac{\alpha_{0}}{\beta_{0}}\label{eq:-9}\\
 & =\inf_{\substack{P_{W},\delta(\cdot):\\
0\leq\delta(W)\leq\frac{1}{4}\\
\mathbb{E}\delta(W)\geq\sqrt{\alpha_{0}-\frac{1}{4}}
}
}-2\mathbb{E}H_{2}\left(\frac{1}{2}+\sqrt{\delta(W)}\right)+\log\frac{1}{\alpha_{0}}+\left(1-2\mathbb{E}\sqrt{\delta(W)}\right)\log\frac{\alpha_{0}}{\beta_{0}}\label{eq:-2}\\
 & \geq\inf_{\substack{P_{W},\delta(\cdot):\\
0\leq\delta(W)\leq\frac{1}{4}\\
\mathbb{E}\delta(W)\geq\sqrt{\alpha_{0}-\frac{1}{4}}
}
}-2H_{2}\left(\frac{1}{2}+\sqrt{\mathbb{E}\delta(W)}\right)+\log\frac{1}{\alpha_{0}}+\left(1-2\sqrt{\mathbb{E}\delta(W)}\right)\log\frac{\alpha_{0}}{\beta_{0}}\label{eq:-3}\\
 & =\inf_{\theta\geq\sqrt{\alpha_{0}-\frac{1}{4}}}-2H_{2}\left(\frac{1}{2}+\theta\right)+\log\frac{1}{\alpha_{0}}+\left(1-2\theta\right)\log\frac{\alpha_{0}}{\beta_{0}}\label{eq:-4}\\
 & =-2H_{2}\left(\frac{1}{2}+\sqrt{\alpha_{0}-\frac{1}{4}}\right)+\log\frac{1}{\alpha_{0}}+\left(1-2\sqrt{\alpha_{0}-\frac{1}{4}}\right)\log\frac{\alpha_{0}}{\beta_{0}}.\label{eq:-5}
\end{align}

\hrulefill{}
\end{figure*}

Substituting $a=\frac{1}{2}+\sqrt{\alpha_{0}-\frac{1}{4}}$ into \eqref{eq:-5}
, we obtain the desired result.

\section{\label{sec:GeneralWyner}Proof of Theorem \ref{thm:GeneralWyner}}

\emph{Achievability Part: }The achievability part is obtained by the
following lemma.
\begin{lem}[One-Shot Soft-Covering]
\cite{yu2019renyi}\label{lem:oneshotach} Assume $P_{W}$ and $P_{X|W}$
are unconditional and conditional distributions respectively (which
can be defined on any countable or uncountable alphabets). Consider
 a random codebook $\mathcal{C}=\{W(i)\}_{i\in\calM}$ with $W(i)\sim P_{W},i\in\calM$,
where $\calM=\{1,\ldots,e^{R}\}$. We define 
\begin{equation}
P_{X|\mathcal{C}}(\cdot|\left\{ w(i)\right\} _{i\in\calM}):=\frac{1}{|{\cal M}|}\sum_{m\in{\cal M}}P_{X|W}(\cdot|w(m))\label{eq:-113-1-1}
\end{equation}
Assume $\pi_{X}$ is a distribution such that for some $s\in(0,1]$,
$D_{1+s}\left(P_{X|W}\|\pi_{X}|P_{W}\right)$ and $D_{1+s}(P_{X}\|\pi_{X})$
exist (and hence are finite). Then we have
\begin{align}
 & e^{sD_{1+s}(P_{X|\mathcal{C}}\|\pi_{X}|P_{\mathcal{C}})}\nonumber \\
 & \leq e^{sD_{1+s}\left(P_{X|W}\|\pi_{X}|P_{W}\right)-sR}+e^{sD_{1+s}(P_{X}\|\pi_{X})}.\label{eq:-123-1}
\end{align}
\end{lem}
Now we set $\pi_{X},P_{X|W},P_{W},R$ to $\pi_{XY}^{n},P_{X|W}^{n}P_{Y|W}^{n},P_{W}^{n},nR$
respectively\footnote{The pair $(X^{n},Y^{n})$ plays the role of $X$ in Lemma \ref{lem:oneshotach}.}
for some distribution $P_{W}P_{X|W}P_{Y|W}$ such that  the marginal
distribution of $P_{W}P_{X|W}P_{Y|W}$ on $(X,Y)$ is equal to $\pi_{XY}$.
Then Lemma \ref{lem:oneshotach} implies that if 
\begin{equation}
R>D_{1+s}(P_{X|W}P_{Y|W}\|\pi_{XY}|P_{W}),
\end{equation}
then $D_{1+s}(P_{X^{n}Y^{n}|\mathcal{C}_{n}}\|\pi_{XY}^{n}|P_{\mathcal{C}_{n}})\to0$.
That is, there exists at least one sequence of codebooks indexed by
$\{c_{n}\}_{n=1}^{\infty}$ such that $D(P_{X^{n}Y^{n}|\mathcal{C}_{n}=c_{n}}\|\pi_{XY}^{n})\leq D_{1+s}(P_{X^{n}Y^{n}|\mathcal{C}_{n}=c_{n}}\|\pi_{XY}^{n})\rightarrow0$.
 This completes the achievability proof.

\emph{Converse Part: }Observe that 
\begin{align}
R & =\frac{1}{n}H\left(M\right)\\
 & \geq\frac{1}{n}I\left(X^{n}Y^{n};M\right)\\
 & =\frac{1}{n}D\left(P_{X^{n}Y^{n}M}\|P_{X^{n}Y^{n}}P_{M}\right)\\
 & =\frac{1}{n}D\left(P_{X^{n}Y^{n}M}\|\pi_{XY}^{n}P_{M}\right)-\frac{1}{n}D\left(P_{X^{n}Y^{n}}\|\pi_{XY}^{n}\right).\label{eq:-125}
\end{align}

We lower bound the first term in \eqref{eq:-125} as follows:
\begin{align}
 & \frac{1}{n}D\left(P_{X^{n}Y^{n}M}\|\pi_{XY}^{n}P_{M}\right)\nonumber \\
 & =\frac{1}{n}\sum_{i=1}^{n}D\left(P_{X_{i}Y_{i}|MX^{i-1}Y^{i-1}}\|\pi_{XY}|P_{MX^{i-1}Y^{i-1}}\right)\label{eq:-126}\\
 & \geq\frac{1}{n}\sum_{i=1}^{n}D\left(P_{X_{i}Y_{i}|M}\|\pi_{XY}|P_{M}\right)\label{eq:-127}\\
 & =D\left(P_{X_{J}Y_{J}|MJ}\|\pi_{XY}|P_{MJ}\right)\label{eq:-128}\\
 & =D\left(P_{XY|W}\|\pi_{XY}|P_{W}\right),\label{eq:-129}
\end{align}
where \eqref{eq:-126} follows by chain rule, \eqref{eq:-127} follows
by the convexity of relative entropy \cite[Theorem 2.7.2]{Cover},
in \eqref{eq:-128}, $J\sim P_{J}:=\mathrm{Unif}[1:n]$ is a time
index independent of $(M,X^{n},Y^{n})$, and in \eqref{eq:-129},
$X:=X_{J},Y:=Y_{J},W:=MJ$.

On the other hand, by assumption, the second term in \eqref{eq:-125}
satisfies 
\begin{equation}
\frac{1}{n}D\left(P_{X^{n}Y^{n}}\|\pi_{XY}^{n}\right)\to0\label{eq:-44}
\end{equation}
 as $n\to\infty$. Moreover, by similar derivation as \eqref{eq:-126}-\eqref{eq:-127},
we can lower bound it as follows:
\begin{align}
 & \frac{1}{n}D\left(P_{X^{n}Y^{n}}\|\pi_{XY}^{n}\right)\geq D\left(P_{XY}\|\pi_{XY}\right).\label{eq:-45}
\end{align}
Hence combining \eqref{eq:-125}, \eqref{eq:-129}, \eqref{eq:-44},
and \eqref{eq:-45} yields that 
\begin{align}
R & \geq\lim_{\epsilon\downarrow0}\inf_{P_{W}P_{X|W}P_{Y|W}:D\left(P_{XY}\|\pi_{XY}\right)\le\epsilon}D\left(P_{XY|W}\|\pi_{XY}|P_{W}\right)\\
 & =\lim_{\epsilon\downarrow0}\inf_{P_{W}P_{X|W}P_{Y|W}:D\left(P_{XY}\|\pi_{XY}\right)\le\epsilon}I(XY;W).
\end{align}

\section{\label{sec:countable}Proof of Corollary \ref{cor:countable}}

First we introduce the following lemma, which upper bounds R\'enyi divergences
in terms of R\'enyi entropies.
\begin{lem}
For a distribution $P_{UV}$ with $\mathcal{U}$ countable, we have
for $s\in[-1,\infty]$, 
\begin{align}
D_{1+s}(P_{UV}\|P_{U}P_{V}) & \leq H_{1-s}(P_{U})
\end{align}
\end{lem}
\begin{IEEEproof}
Consider,
\begin{align}
D_{1+s}(P_{UV}\|P_{U}P_{V}) & =\frac{1}{s}\log\mathbb{E}_{P_{UV}}\left(\frac{P_{U|V}(U|V)}{P_{U}(U)}\right)^{s}\\
 & \leq\frac{1}{s}\log\mathbb{E}_{P_{UV}}\left(\frac{1}{P_{U}(U)}\right)^{s}\\
 & =H_{1-s}(P_{U}).
\end{align}
\end{IEEEproof}
By the lemma above, we obtain $D_{1+s}(P_{X|W}P_{Y|W}\|P_{XY}|P_{W})\le H_{1-s}(\pi_{XY})<\infty$
for all $P_{W}P_{X|W}P_{Y|W}$ such that $P_{XY}=\pi_{XY}$. Then
by Proposition \ref{prop:GeneralWyner-1}, we have that 
\begin{equation}
\widehat{C}_{\mathrm{Wyner}}(\pi_{XY})=C_{\mathrm{Wyner}}(\pi_{XY}).\label{eq:-138}
\end{equation}
Furthermore, by Theorem \ref{thm:GeneralWyner}, to prove Corollary
\ref{cor:countable}, we only need prove 
\begin{equation}
\widetilde{C}_{\mathrm{Wyner}}(\pi_{XY})\geq\widehat{C}_{\mathrm{Wyner}}(\pi_{XY}).\label{eq:-139}
\end{equation}
Hence we only need prove $\widetilde{C}_{\mathrm{Wyner}}(\pi_{XY})\geq C_{\mathrm{Wyner}}(\pi_{XY})$.
 In this appendix, we combine the distribution truncation technique
and the mixture decomposition to prove this. 

Without loss of generality, we assume $X,Y$ are integer-valued. Define
the $n$-truncation operator $\left[\cdot\right]_{n}$ as follows:
$\left[z\right]_{n}:=z$ if $\left|z\right|\le n$, and $\left[z\right]_{n}:=n+1$
if $\left|z\right|>n$. We introduce a random variable (in fact, a
function of $\left(X,Y\right)$ or $(\left[X\right]_{n},\left[Y\right]_{n})$)
\begin{equation}
V:=1\left\{ (X,Y)\in[-n:n]^{2}\right\} =1\left\{ (\left[X\right]_{n},\left[Y\right]_{n})\in[-n:n]^{2}\right\} .\label{eq:-86}
\end{equation}
Hence $P_{V|W\left[X\right]_{n}\left[Y\right]_{n}}(v|w,x,y)=1\left\{ (x,y)\in[-n:n]^{2}\right\} $,
and $q_{n}:=P_{V}(1)=P_{\left[X\right]_{n}\left[Y\right]_{n}}\left([-n:n]^{2}\right)$.
Then 
\begin{align}
 & \widetilde{C}_{\mathrm{Wyner}}(\pi_{XY})\nonumber \\
 & =\lim_{\epsilon\downarrow0}\inf_{\substack{P_{W}P_{X|W}P_{Y|W}:\\
D\left(P_{XY}\|\pi_{XY}\right)\le\epsilon
}
}I\left(XY;W\right)\\
 & \geq\limsup_{n\to\infty}\lim_{\epsilon\downarrow0}\inf_{\substack{P_{W}P_{X|W}P_{Y|W}:\\
D\left(P_{XY}\|\pi_{XY}\right)\le\epsilon
}
}I\left(\left[X\right]_{n}\left[Y\right]_{n};W\right)\label{eq:-171}\\
 & \geq\limsup_{n\to\infty}\lim_{\epsilon\downarrow0}\inf_{\substack{P_{W}P_{X|W}P_{Y|W}:\\
D\left(P_{\left[X\right]_{n}\left[Y\right]_{n}}\|\pi_{\left[X\right]_{n}\left[Y\right]_{n}}\right)\le\epsilon
}
}I\left(\left[X\right]_{n}\left[Y\right]_{n};W\right)\label{eq:-22}\\
 & =\limsup_{n\to\infty}\lim_{\epsilon\downarrow0}\inf_{\substack{P_{W}P_{\left[X\right]_{n}|W}P_{\left[Y\right]_{n}|W}:\\
D\left(P_{\left[X\right]_{n}\left[Y\right]_{n}}\|\pi_{\left[X\right]_{n}\left[Y\right]_{n}}\right)\le\epsilon
}
}I\left(\left[X\right]_{n}\left[Y\right]_{n};W\right)\label{eq:-172}\\
 & =\limsup_{n\to\infty}\min_{\substack{P_{W}P_{\left[X\right]_{n}|W}P_{\left[Y\right]_{n}|W}:\\
P_{\left[X\right]_{n}\left[Y\right]_{n}}=\pi_{\left[X\right]_{n}\left[Y\right]_{n}}
}
}I\left(\left[X\right]_{n}\left[Y\right]_{n};W\right),\label{eq:-62}
\end{align}
where \eqref{eq:-171} follows by the data processing inequality $I\left(\left[X\right]_{n}\left[Y\right]_{n};W\right)\le I\left(XY;W\right)$;
\eqref{eq:-22} follows by the data processing inequality $D\left(P_{\left[X\right]_{n}\left[Y\right]_{n}}\|\pi_{\left[X\right]_{n}\left[Y\right]_{n}}\right)\le D\left(P_{XY}\|\pi_{XY}\right)$;
\eqref{eq:-172} follows since the objective function and the constraint
depend on $(W,X,Y)$ through their truncated version $(W,\left[X\right]_{n},\left[Y\right]_{n})$;
and \eqref{eq:-62} follows since the alphabet size of $W$ can be
restricted to be no larger than $(2n+1)^{2}$ (by standard cardinality
bounding techniques) and hence for such discrete $W$, the probability
simplex defined on the alphabet of $\left(W,X,Y\right)$ is compact.

By basic information-theoretic inequalities, we obtain that 
\begin{align}
I\left(\left[X\right]_{n}\left[Y\right]_{n};W\right) & =I\left(\left[X\right]_{n}\left[Y\right]_{n}V;W\right)\label{eq:-120}\\
 & =I\left(\left[X\right]_{n}\left[Y\right]_{n};W|V\right)+I\left(V;W\right)\\
 & \geq I\left(\left[X\right]_{n}\left[Y\right]_{n};W|V\right)\\
 & \geq q_{n}I\left(\left[X\right]_{n}\left[Y\right]_{n};W|V=1\right).\label{eq:-121}
\end{align}

Observe that under the condition $P_{\left[X\right]_{n}\left[Y\right]_{n}}=\pi_{\left[X\right]_{n}\left[Y\right]_{n}}$,
it holds that $q_{n}=P_{\left[X\right]_{n}\left[Y\right]_{n}}\left([-n:n]^{2}\right)=\pi_{\left[X\right]_{n}\left[Y\right]_{n}}\left([-n:n]^{2}\right)=\pi_{XY}\left([-n:n]^{2}\right)\to1$
as $n\to\infty$. Hence 
\begin{align}
I\left(\left[X\right]_{n}\left[Y\right]_{n};W\right) & \geq I\left(\left[X\right]_{n}\left[Y\right]_{n};W|V=1\right).\label{eq:-119}
\end{align}

Combining \eqref{eq:-62} and \eqref{eq:-119}, we obtain 
\begin{align}
 & \widetilde{C}_{\mathrm{Wyner}}(\pi_{XY})\nonumber \\
 & \geq\limsup_{n\to\infty}\min_{\substack{P_{W}P_{\left[X\right]_{n}|W}P_{\left[Y\right]_{n}|W}:\\
P_{\left[X\right]_{n}\left[Y\right]_{n}}=\pi_{\left[X\right]_{n}\left[Y\right]_{n}}
}
}I\left(\left[X\right]_{n}\left[Y\right]_{n};W|V=1\right).\label{eq:-18}
\end{align}
To simplify the RHS of \eqref{eq:-18}, we need the following lemma.
\begin{lem}[Conditional Markov Chain]
\label{lem:conditionalMarkov} If $X\to W\to Y$ form a Markov chain,
then $X\to W\to Y$ also form a Markov chain conditioned on $\left\{ X\in A,Y\in B\right\} $
for any $A\subseteq\mathcal{X},B\subseteq\mathcal{Y}$ such that $\mathbb{P}\left(X\in A\right),\mathbb{P}\left(Y\in B\right)>0$.
\end{lem}
\begin{IEEEproof}
Consider,
\begin{align}
 & \mathbb{P}\left(\left(W,X,Y\right)=\left(w,x,y\right)|X\in A,Y\in B\right)\nonumber \\
 & =\frac{P_{W}(w)P_{X|W}(x|w)P_{Y|W}(y|w)1\left\{ \left(x,y\right)\in A\times B\right\} }{P_{XY}(A\times B)}\\
 & =\frac{P_{W}(w)P_{X|W}(A|w)P_{Y|W}(B|w)}{P_{XY}(A\times B)}\nonumber \\
 & \qquad\times\frac{P_{X|W}(x|w)1\left\{ x\in A\right\} }{P_{X|W}(A|w)}\frac{P_{Y|W}(y|w)1\left\{ y\in B\right\} }{P_{Y|W}(B|w)}\\
 & =:\widetilde{P}_{W}(w)\widetilde{P}_{X|W}(x|w)\widetilde{P}_{Y|W}(y|w),\label{eq:-87-2}
\end{align}
i.e., $X\to W\to Y$ forms a Markov chain under $\widetilde{P}$. 
\end{IEEEproof}
By Lemma \ref{lem:conditionalMarkov}, for $(x,y)\in[-n:n]^{2}$,
$P_{W\left[X\right]_{n}\left[Y\right]_{n}|V}(w,x,y|1)$ can be factorized
as 
\begin{align}
P_{W\left[X\right]_{n}\left[Y\right]_{n}|V}(w,x,y|1) & =\widetilde{P}_{W}(w)\widetilde{P}_{X|W}(x|w)\widetilde{P}_{Y|W}(y|w)\label{eq:-87}
\end{align}
i.e., $X\to W\to Y$ forms a Markov chain under $\widetilde{P}$.
Hence 
\begin{equation}
I\left(\left[X\right]_{n}\left[Y\right]_{n};W|V=1\right)=I_{\widetilde{P}}\left(XY;W\right).\label{eq:-16}
\end{equation}
On the other hand, $P_{\left[X\right]_{n}\left[Y\right]_{n}}=\pi_{\left[X\right]_{n}\left[Y\right]_{n}}$
implies 
\begin{align}
 & \sum_{w}\widetilde{P}_{W}(w)\widetilde{P}_{X|W}(x|w)\widetilde{P}_{Y|W}(y|w)\nonumber \\
 & =P_{\left[X\right]_{n}\left[Y\right]_{n}|V}(x,y|1)\\
 & =\frac{\pi_{\left[X\right]_{n}\left[Y\right]_{n}}(x,y)P_{V|\left[X\right]_{n}\left[Y\right]_{n}}(1|x,y)}{P_{V}(1)}\\
 & =\frac{\pi_{\left[X\right]_{n}\left[Y\right]_{n}}(x,y)1\left\{ (x,y)\in[-n,n]^{2}\right\} }{\pi_{\left[X\right]_{n}\left[Y\right]_{n}}\left([-n,n]^{2}\right)}\\
 & =:\pi_{XY}^{(n)}(x,y).\label{eq:-17}
\end{align}
Hence \eqref{eq:-18} implies that 
\begin{align}
 & \widetilde{C}_{\mathrm{Wyner}}(\pi_{XY})\nonumber \\
 & \geq\limsup_{n\to\infty}\min_{\widetilde{P}_{W}\widetilde{P}_{X|W}\widetilde{P}_{Y|W}:\widetilde{P}_{XY}=\pi_{XY}^{(n)}}I_{\widetilde{P}}\left(XY;W\right)\\
 & =\limsup_{n\to\infty}C_{\mathrm{Wyner}}(\pi_{XY}^{(n)}).\label{eq:-85}
\end{align}

Next we prove $C_{\mathrm{Wyner}}(\pi_{XY})\le\liminf_{n\to\infty}C_{\mathrm{Wyner}}(\pi_{XY}^{(n)}).$
Obviously, $p_{n}:=\pi_{XY}([-n,n]^{2})\to1$ as $n\to\infty$. Then
for $(x,y)\in\supp\left(\pi_{XY}\right)$, 
\begin{align}
\frac{\pi_{XY}^{(n)}(x,y)}{\pi_{XY}(x,y)} & =\frac{1\left\{ (x,y)\in[-n,n]^{2}\right\} }{p_{n}}\\
 & \leq\frac{1}{p_{n}},\label{eq:-13-3}
\end{align}
and 
\begin{align}
 & H\left(\pi_{XY}^{(n)}\right)\nonumber \\
 & =-\sum_{(x,y)\in[-n,n]^{2}}\frac{\pi_{XY}(x,y)}{p_{n}}\log\frac{\pi_{XY}(x,y)}{p_{n}}\\
 & =\log p_{n}-\frac{1}{p_{n}}\sum_{(x,y)\in[-n,n]^{2}}\pi_{XY}(x,y)\log\pi_{XY}(x,y).
\end{align}
According to the definition of entropy, 
\begin{equation}
-\sum_{(x,y)\in[-n,n]^{2}}\pi_{XY}(x,y)\log\pi_{XY}(x,y)\to H(\pi_{XY})
\end{equation}
as $n\to\infty$. Hence 
\begin{equation}
\lim_{n\to\infty}H\left(\pi_{XY}^{(n)}\right)\to H(\pi_{XY}).\label{eq:-12-2}
\end{equation}

We construct a new distribution 
\begin{align}
\widehat{\pi}_{XY}^{(n)}\left(x,y\right) & :=\frac{\frac{1}{p_{n}}\pi_{XY}\left(x,y\right)-\pi_{XY}^{(n)}(x,y)}{\frac{1}{p_{n}}-1}\\
 & =\frac{\pi_{XY}\left(x,y\right)1\left\{ (x,y)\notin[-n:n]^{2}\right\} }{1-p_{n}}.
\end{align}
Hence $\pi_{XY}$ can be written as a mixture distribution $\pi_{XY}(x,y)=p_{n}\pi_{XY}^{(n)}(x,y)+\left(1-p_{n}\right)\widehat{\pi}_{XY}^{(n)}\left(x,y\right)$.
Define $U$ as a Bernoulli random variable $U$ with $P_{U}(1)=p_{n}$.
Define 
\begin{align}
 & Q_{XYWU}^{(n)}(x,y,w,u)\nonumber \\
 & =\begin{cases}
p_{n}\pi_{XY}^{(n)}(x,y)P_{W|XY}^{(n)}(w|x,y) & \textrm{if }u=1\\
\left(1-p_{n}\right)\widehat{\pi}_{XY}^{(n)}\left(x,y\right)1\left\{ w=(x,y)\right\}  & \textrm{if }u=0
\end{cases},
\end{align}
where $P_{W|XY}^{(n)}$ is induced by an optimal joint distribution
$P_{W}^{(n)}P_{X|W}^{(n)}P_{Y|W}^{(n)}$ (with $P_{XY}^{(n)}=\pi_{XY}^{(n)}$
and $W$ having a finite support) attaining $C_{\mathrm{Wyner}}(\pi_{XY}^{(n)})$.
Obviously, $Q_{XY}^{(n)}=\pi_{XY}$, and $X\to\left(W,U\right)\to Y$
under $Q^{(n)}$. Therefore, we have 
\begin{align}
 & C_{\mathrm{Wyner}}(\pi_{XY})\nonumber \\
 & =\inf_{P_{W}P_{X|W}P_{Y|W}:P_{XY}=\pi_{XY}}I\left(XY;W\right)\label{eq:-65-3}\\
 & \leq I_{Q^{(n)}}\left(XY;WU\right)\\
 & =H(\pi_{XY})-H_{Q^{(n)}}(XY|WU)\\
 & =H(\pi_{XY})-p_{n}H_{Q^{(n)}}(XY|W,U=1)\nonumber \\
 & \qquad-\left(1-p_{n}\right)H_{Q^{(n)}}(XY|W,U=0)\\
 & =H(\pi_{XY})-p_{n}H_{P^{(n)}}(XY|W)\\
 & =H(\pi_{XY})-p_{n}H\left(\pi_{XY}^{(n)}\right)+p_{n}I_{P^{(n)}}(XY;W).\label{eq:-64-3}
\end{align}
Taking limits and using \eqref{eq:-12-2} and the fact that $p_{n}\to1$
as $n\to\infty$, we have 
\begin{align}
C_{\mathrm{Wyner}}(\pi_{XY}) & \leq\liminf_{n\to\infty}I_{P^{(n)}}(XY;W)\\
 & =\liminf_{n\to\infty}C_{\mathrm{Wyner}}(\pi_{XY}^{(n)}).\label{eq:-21-3}
\end{align}

Combining \eqref{eq:-85} and \eqref{eq:-21-3} gives us the desired
result.

\section{\label{sec:continuous}Proof of Corollary \ref{cor:continuous}}

In this section, we extend the proof in Appendix \ref{sec:countable}
to the continuous distribution case by combining it with the discretization
technique and dyadic decomposition results in \cite{li2017distributed}.

By assumption, $\widehat{C}_{\mathrm{Wyner}}(\pi_{XY})=C_{\mathrm{Wyner}}(\pi_{XY})$.
Hence to prove Corollary \ref{cor:continuous}, we only need to prove
$\widetilde{C}_{\mathrm{Wyner}}(\pi_{XY})\geq C_{\mathrm{Wyner}}(\pi_{XY})$.
To this end, similar to \eqref{eq:-86}, we introduce a random variable
\begin{equation}
V_{d}:=1\left\{ (X,Y)\in[-d,d)^{2}\right\} .\label{eq:-42-1-2}
\end{equation}
Denote $P_{WXY}=P_{W}P_{X|W}P_{Y|W}$. Similarly to \eqref{eq:-87},
we define $\widetilde{P}_{WXY}(\cdot):=P_{WXY|V_{d}}(\cdot|1)$. Then
$\widetilde{P}_{WXY}=\widetilde{P}_{W}\widetilde{P}_{X|W}\widetilde{P}_{Y|W}$,
i.e., $X\to W\to Y$ forms a Markov chain under $\widetilde{P}$.
Define $q_{d}:=P_{XY}\left([-d,d)^{2}\right)$. The conclusions similar
to \eqref{eq:-16} and \eqref{eq:-17} hold.

\subsection{Proof of $\widetilde{C}_{\mathrm{Wyner}}(\pi_{XY})\protect\geq\limsup_{d\to\infty}\widetilde{C}_{\mathrm{Wyner}}(\pi_{XY|V_{d}=1})$}

Consider that 
\begin{align}
 & \widetilde{C}_{\mathrm{Wyner}}(\pi_{XY})\nonumber \\
 & =\lim_{\epsilon\downarrow0}\inf_{\substack{P_{W}P_{X|W}P_{Y|W}:\\
D\left(P_{XY}\|\pi_{XY}\right)\le\epsilon
}
}I\left(XY;W\right)\\
 & \geq\limsup_{d\to\infty}\lim_{\epsilon\downarrow0}\inf_{\substack{P_{W}P_{X|W}P_{Y|W}:\\
D\left(P_{XY}\|\pi_{XY}\right)\le\epsilon
}
}q_{d}I\left(XY;W|V_{d}=1\right)\label{eq:-122}\\
 & =\limsup_{d\to\infty}\lim_{\epsilon\downarrow0}\inf_{\substack{P_{W}P_{X|W}P_{Y|W}:\\
D\left(P_{XY}\|\pi_{XY}\right)\le\epsilon
}
}\pi_{XY}\left([-d,d)^{2}\right)\nonumber \\
 & \qquad\times I\left(XY;W|V_{d}=1\right)\label{eq:-67}\\
 & \geq\limsup_{d\to\infty}\lim_{\epsilon\downarrow0}\inf_{\substack{P_{W}P_{X|W}P_{Y|W}:\\
\left(\pi_{XY}\left([-d,d)^{2}\right)-\sqrt{2\epsilon}\right)D\left(P_{XY|V_{d}=1}\|\pi_{XY|V_{d}=1}\right)\le\epsilon
}
}\nonumber \\
 & \qquad\pi_{XY}\left([-d,d)^{2}\right)I\left(XY;W|V_{d}=1\right)\label{eq:-82}\\
 & =\limsup_{d\to\infty}\lim_{\epsilon\downarrow0}\inf_{\substack{P_{W}P_{X|W}P_{Y|W}:\\
D\left(P_{XY|V_{d}=1}\|\pi_{XY|V_{d}=1}\right)\le\epsilon
}
}I\left(XY;W|V_{d}=1\right)\label{eq:-130}\\
 & \geq\limsup_{d\to\infty}\lim_{\epsilon\downarrow0}\inf_{\substack{\widetilde{P}_{W}\widetilde{P}_{X|W}\widetilde{P}_{Y|W}:\\
D\left(\widetilde{P}_{XY}\|\pi_{XY|V_{d}=1}\right)\le\epsilon
}
}I_{\widetilde{P}}\left(XY;W\right)\\
 & =\limsup_{d\to\infty}\widetilde{C}_{\mathrm{Wyner}}(\pi_{XY|V_{d}=1}),
\end{align}
where \eqref{eq:-122} follows similarly as \eqref{eq:-120}-\eqref{eq:-121};
\eqref{eq:-67} follows from that by Pinsker's inequality $|P_{XY}-\pi_{XY}|\leq\sqrt{2D\left(P_{XY}\|\pi_{XY}\right)}\le\sqrt{2\epsilon}$,
we have 
\begin{equation}
q_{d}\in\pi_{XY}\left([-d,d)^{2}\right)+[-\sqrt{2\epsilon},\sqrt{2\epsilon}];\label{eq:-83}
\end{equation}
 \eqref{eq:-82} follows from \eqref{eq:-83} and the fact that 
\begin{align}
 & q_{d}D\left(P_{XY|V_{d}=1}\|\pi_{XY|V_{d}=1}\right)\nonumber \\
 & \le D\left(P_{XY|V_{d}}\|\pi_{XY|V_{d}}|P_{V_{d}}\right)\\
 & \le D\left(P_{XY|V_{d}}\|\pi_{XY|V_{d}}|P_{V_{d}}\right)+D\left(P_{V_{d}}\|\pi_{V_{d}}\right)\\
 & =D\left(P_{XYV_{d}}\|\pi_{XYV_{d}}\right)\\
 & =D\left(P_{XY}\|\pi_{XY}\right),\label{eq:-84}
\end{align}
(\eqref{eq:-84} follows since $V_{d}$ is a function of $\left(X,Y\right)$);
and \eqref{eq:-130} follows since $\pi_{XY}\left([-d,d)^{2}\right)\to1$
as $d\to\infty$.

\subsection{Proof of $C_{\mathrm{Wyner}}(\pi_{XY})\protect\leq\liminf_{d\to\infty}\widetilde{C}_{\mathrm{Wyner}}(\pi_{XY|V_{d}=1})$}

Next we prove $C_{\mathrm{Wyner}}(\pi_{XY})\leq\liminf_{d\to\infty}\widetilde{C}_{\mathrm{Wyner}}(\pi_{XY|V_{d}=1}).$
Since in definition of $\widetilde{C}_{\mathrm{Wyner}}(\pi_{XY|V_{d}=1})$,
a joint distribution $\widetilde{P}_{W}\widetilde{P}_{X|W}\widetilde{P}_{Y|W}$
generates a distribution $\widetilde{P}_{XY}$, which is an approximate
version of $\pi_{XY|V_{d}=1}$ and hence is also an approximation
of $\pi_{XY}$. In this subsection, we combine mixture decomposition
technique with dyadic decomposition schemes \cite{li2017distributed}
to make the joint distribution $\widetilde{P}_{XY}$ exactly equal
to $\pi_{XY}$ (by constructing a modified version of $\widetilde{P}_{W}\widetilde{P}_{X|W}\widetilde{P}_{Y|W}$).

Define 
\begin{align}
\pi_{XY}^{(d)}(x,y) & :=\frac{1}{p_{d}}\pi_{XY}(x,y)1\left\{ (x,y)\in[-d,d)^{2}\right\} \\
 & =\pi_{XY|V_{d}=1},
\end{align}
where $p_{d}:=\pi_{XY}\left([-d,d)^{2}\right)\to1$ as $d\to\infty$.
Then given an integer $n>0$, we define $\Delta:=\frac{d}{n}$, and
we quantize $X,Y$ as $A:=\left\lfloor \frac{X}{\Delta}\right\rfloor ,B:=\left\lfloor \frac{Y}{\Delta}\right\rfloor $.
The induced distribution of $\left(A,B\right)$ is $\pi_{AB}^{(n)}(a,b)=\frac{1}{p_{d}}\int_{\Delta\left(a,b\right)+\left[0,\Delta\right)^{2}}\pi_{XY}(x,y)\mathrm{d}x\mathrm{d}y1\left\{ (a,b)\in[-n,n-1]^{2}\right\} $.
By adding an independent uniform vector $(U,V)\sim\mathrm{Unif}([0,\Delta)^{2})$
to $\Delta\left(A,B\right)$ with $\left(A,B\right)\sim\pi_{AB}^{(n)}$,
we get a continuous distribution 
\begin{equation}
\Delta\left(A,B\right)+(U,V)\sim\pi_{XY}^{(n)}(x,y):=\frac{1}{\Delta^{2}}\pi_{AB}^{(n)}\left(\left\lfloor \frac{x}{\Delta}\right\rfloor ,\left\lfloor \frac{y}{\Delta}\right\rfloor \right).
\end{equation}
 Then for $(x,y)\in\supp\left(\pi_{XY}\right)$, 
\begin{align}
\frac{\pi_{XY}^{(n)}(x,y)}{\pi_{XY}(x,y)} & =\frac{\frac{1}{\Delta^{2}}\pi_{AB}^{(n)}(\left\lfloor \frac{x}{\Delta}\right\rfloor ,\left\lfloor \frac{y}{\Delta}\right\rfloor )}{\pi_{XY}(x,y)}\\
 & =\frac{\frac{1}{\Delta^{2}}\int_{\Delta\left(\left\lfloor \frac{x}{\Delta}\right\rfloor ,\left\lfloor \frac{y}{\Delta}\right\rfloor \right)+\left[0,\Delta\right)^{2}}\pi_{XY}(x,y)\mathrm{d}x\mathrm{d}y}{\pi_{XY}(x,y)p_{d}}\nonumber \\
 & \qquad\times1\left\{ (x,y)\in[-d,d)^{2}\right\} \\
 & =\frac{\pi_{XY}(\widehat{x},\widehat{y})1\left\{ (x,y)\in[-d,d)^{2}\right\} }{\pi_{XY}(x,y)p_{d}}\label{eq:-63}\\
 & \leq\sup_{(x,y)\in[-d,d)^{2}}\frac{\pi_{XY}(\widehat{x},\widehat{y})}{\pi_{XY}(x,y)p_{d}}\label{eq:-13-1}
\end{align}
where \eqref{eq:-63} follows by the mean value theorem, and it holds
for some $\left(\widehat{x},\widehat{y}\right)\in\Delta\left(\left\lfloor \frac{x}{\Delta}\right\rfloor ,\left\lfloor \frac{y}{\Delta}\right\rfloor \right)+\left[0,\Delta\right)^{2}$. 
\begin{lem}
\label{lem:RatioBounds} Assume $\pi_{XY}$ is differentiable. Then
for any $\left(x,y\right),\left(\hat{x},\hat{y}\right)\in[-d,d]^{2}$
satisfying $\left|x-\hat{x}\right|,\left|y-\hat{y}\right|\leq\Delta$,
we have 
\begin{equation}
\exp\left(-\Delta L_{d}\right)\leq\frac{\pi_{XY}\left(x,y\right)}{\pi_{XY}\left(\hat{x},\hat{y}\right)}\leq\exp\left(\Delta L_{d}\right),\label{eq:-27}
\end{equation}
where $L_{d}$ is defined in \eqref{eq:-78}.
\end{lem}
\begin{IEEEproof}[Proof of Lemma \ref{lem:RatioBounds}]
By Taylor's theorem, 
\begin{align}
 & \log\pi_{XY}\left(x,y\right)\nonumber \\
 & =\log\pi_{XY}\left(\hat{x},\hat{y}\right)+\frac{\partial}{\partial x}\log\pi_{XY}\left(\widetilde{x},\widetilde{y}\right)\left(x-\hat{x}\right)\nonumber \\
 & \qquad+\frac{\partial}{\partial y}\log\pi_{XY}\left(\widetilde{x},\widetilde{y}\right)\left(y-\hat{y}\right)\label{eq:-173}\\
 & \leq\log\pi_{XY}\left(\hat{x},\hat{y}\right)+\frac{\left(\left|\frac{\partial\pi_{XY}}{\partial x}\left(\widetilde{x},\widetilde{y}\right)\right|+\left|\frac{\partial\pi_{XY}}{\partial y}\left(\widetilde{x},\widetilde{y}\right)\right|\right)\Delta}{\pi_{XY}\left(\widetilde{x},\widetilde{y}\right)}\label{eq:-46-2}\\
 & \leq\log\pi_{XY}\left(\hat{x},\hat{y}\right)+\Delta L_{d},
\end{align}
where \eqref{eq:-173} holds for some $\left(\widetilde{x},\widetilde{y}\right)$
on the line segment joining $\left(\hat{x},\hat{y}\right)$ and $\left(x,y\right)$.
By symmetry, $\log\pi_{XY}\left(\hat{x},\hat{y}\right)\leq\log\pi_{XY}\left(x,y\right)+\Delta L_{d}$
also holds.
\end{IEEEproof}
Using Lemma \ref{lem:RatioBounds}, we obtain 
\begin{align}
\frac{\pi_{XY}^{(n)}(x,y)}{\pi_{XY}(x,y)} & \leq\frac{1}{p_{d}}\exp\left(\Delta L_{d}\right)=\exp\left(\Delta L_{d}-\log p_{d}\right).\label{eq:-131}
\end{align}
Define 
\begin{equation}
\epsilon_{n}':=\Delta L_{d}-\log p_{d}+\delta_{n}\label{eq:-52-1}
\end{equation}
for some positive sequence $\delta_{n}\to0$ as $n\to\infty$, which
will be specified later. Then \eqref{eq:-131} implies 
\begin{align}
e^{D_{\infty}\left(\pi_{XY}^{(n)}\|\pi_{XY}\right)}=\sup_{x,y} & \frac{\pi_{XY}^{(n)}(x,y)}{\pi_{XY}(x,y)}\leq e^{\epsilon_{n}'-\delta_{n}},\label{eq:-137}
\end{align}
i.e., 
\begin{align}
\frac{e^{\epsilon_{n}'}\pi_{XY}(x,y)}{\pi_{XY}^{(n)}(x,y)} & \geq e^{\delta_{n}}\label{eq:-51-1}
\end{align}
for all $\left(x,y\right)\in[-d,d)^{2}$.

We construct a new distribution 
\begin{equation}
\widehat{\pi}_{XY}^{(n)}\left(x,y\right):=\frac{e^{\epsilon_{n}'}\pi_{XY}(x,y)-\pi_{XY}^{(n)}(x,y)}{e^{\epsilon_{n}'}-1}.\label{eq:-157}
\end{equation}
Hence $\pi_{XY}$ can be written as a mixture distribution $\pi_{XY}(x,y)=e^{-\epsilon_{n}'}\pi_{XY}^{(n)}(x,y)+\left(1-e^{-\epsilon_{n}'}\right)\widehat{\pi}_{XY}^{(n)}\left(x,y\right)$.
Furthermore, by \eqref{eq:-51-1}, we have 
\begin{align}
\widehat{\pi}_{XY}^{(n)}\left(x,y\right) & \geq\frac{e^{\delta_{n}}-1}{e^{\epsilon_{n}'}-1}\pi_{XY}^{(n)}(x,y)\\
 & =\frac{e^{\delta_{n}}-1}{e^{\epsilon_{n}'}-1}\frac{1}{\Delta^{2}}\pi_{AB}^{(n)}\left(\left\lfloor \frac{x}{\Delta}\right\rfloor ,\left\lfloor \frac{y}{\Delta}\right\rfloor \right).\label{eq:-133}
\end{align}
Define $U$ as a Bernoulli random variable $U$ with $P_{U}(1)=e^{-\epsilon_{n}'}$.
Let $\left[z\right]_{n}:=z$, if $z\in[-n,n-1]$; $n$, if $z\ge n$;
and $-(n+1)$, otherwise, denote the truncation operation on integers.
Define 
\begin{align}
 & \widehat{\pi}_{XY|W_{1}}^{(n)}(x',y'|w_{1})\nonumber \\
 & :=\frac{\widehat{\pi}_{XY}^{(n)}\left(x',y'\right)1\left\{ \left(\left[\left\lfloor \frac{x'}{\Delta}\right\rfloor \right]_{n},\left[\left\lfloor \frac{y'}{\Delta}\right\rfloor \right]_{n}\right)=w_{1}\right\} }{\widehat{\pi}_{XY}^{(n)}\left\{ \left(x',y'\right):\left(\left[\left\lfloor \frac{x'}{\Delta}\right\rfloor \right]_{n},\left[\left\lfloor \frac{y'}{\Delta}\right\rfloor \right]_{n}\right)=w_{1}\right\} }
\end{align}
for $w_{1}\in[-(n+1),n]^{2}$. Define 
\begin{align}
 & Q_{XYWU}^{(n)}(x,y,w,u)\nonumber \\
 & :=\begin{cases}
e^{-\epsilon_{n}'}\pi_{XY}^{(n)}(x,y)P_{W|AB}^{(n)}\left(w|\left\lfloor \frac{x}{\Delta}\right\rfloor ,\left\lfloor \frac{y}{\Delta}\right\rfloor \right) & \textrm{if }u=1\\
\left(1-e^{-\epsilon_{n}'}\right)\widehat{\pi}_{XY}^{(n)}\left(x,y\right)\widehat{P}_{W|XY}^{(n)}(w|x,y) & \textrm{if }u=0
\end{cases},
\end{align}
where $P_{W|AB}^{(n)}$ is induced by an optimal joint distribution
$P_{W}^{(n)}P_{A|W}^{(n)}P_{B|W}^{(n)}$ (with $P_{AB}^{(n)}=\pi_{AB}^{(n)}$
and $W$ having a finite support) attaining $C_{\mathrm{Wyner}}(\pi_{AB}^{(n)})$;
\begin{align}
 & \widehat{P}_{W|XY}^{(n)}\left((w_{1},w_{2})|x,y\right)\nonumber \\
 & :=\widehat{P}_{W_{1}|XY}^{(n)}(w_{1}|x,y)\widehat{P}_{W_{2}|XYW_{1}}^{(n)}(w_{2}|x,y,w_{1})
\end{align}
with $\widehat{P}_{W_{1}|XY}^{(n)}(w_{1}|x,y)=1\left\{ w_{1}=\left(\left[\left\lfloor \frac{x}{\Delta}\right\rfloor \right]_{n},\left[\left\lfloor \frac{y}{\Delta}\right\rfloor \right]_{n}\right)\right\} $
(i.e., $W=(W_{1},W_{2})$ and $W_{1}=\left(\left[\left\lfloor \frac{X}{\Delta}\right\rfloor \right]_{n},\left[\left\lfloor \frac{Y}{\Delta}\right\rfloor \right]_{n}\right)$
under $\widehat{P}^{(n)}$); and $\widehat{P}_{W_{2}|XYW_{1}}^{(n)}$
is induced by an optimal joint distribution $\widehat{P}_{W_{1}W_{2}}^{(n)}\widehat{P}_{X|W_{1}W_{2}}^{(n)}\widehat{P}_{Y|W_{1}W_{2}}^{(n)}$
(with $\widehat{P}_{XY|W_{1}}^{(n)}=\widehat{\pi}_{XY|W_{1}}^{(n)}$)
such that $\widehat{P}_{W_{2}|W_{1}=w_{1}}^{(n)}\widehat{P}_{X|W_{2},W_{1}=w_{1}}^{(n)}\widehat{P}_{Y|W_{2},W_{1}=w_{1}}^{(n)}$
attains  the common entropy $G(\widehat{\pi}_{XY|W_{1}=w_{1}}^{(n)})$
defined in \eqref{eq:commonentropy} (or $G(\widehat{\pi}_{XY|W_{1}=w_{1}}^{(n)})+\delta_{n}'$
for a sequence $\delta_{n}'>0$ satisfying $\delta_{n}'\to0$ as $n\to\infty$
if the infimization in $G(\widehat{\pi}_{XY|W_{1}=w_{1}}^{(n)})$
is not attained) for $w_{1}\in[-(n+1),n]^{2}$.

Partition $\mathbb{R}^{2}$ into $9$ subregions by the lines $x=\pm d$
and $y=\pm d$. Denote them as $\mathrm{R}_{0},\mathrm{R}_{1},...,\mathrm{R}_{8}$,
where $\mathrm{R}_{0}:=[-d,d)^{2}$ and $\mathrm{R}_{1},\mathrm{R}_{2},...,\mathrm{R}_{8}$
denote others. Obviously,  $\mathrm{R}_{k},0\le k\le8$ can be expressed
as $\mathrm{R}_{k}=I_{1}^{(k)}\times I_{2}^{(k)}$ with $I_{i}^{(k)}\in\left\{ \mathcal{L}_{d}^{-},\mathcal{L}_{d},\mathcal{L}_{d}^{+}\right\} $,
where $\mathcal{L}_{d}^{-}:=(-\infty,-d)$, $\mathcal{L}_{d}:=[-d,d)$,
and $\mathcal{L}_{d}^{+}:=[d,+\infty)$. Note that $(X,Y)\in\mathrm{R}_{0}$
corresponds to $W_{1}\in[-n,n-1]^{2}$; and $(X,Y)\in\bigcup_{k=1}^{8}\mathrm{R}_{k}$
corresponds to the case that the first or the second component of
$W_{1}$ is $-(n+1)$ or $n$. According to the definition of $\widehat{\pi}_{XY|W_{1}}^{(n)}$,
for the subregion $\mathrm{R}_{0}$, we have $\widehat{\pi}_{XY|W_{1}}^{(n)}(\cdot|(a,b))=\widehat{\pi}_{XY}^{(n)}(\cdot|I_{\Delta}^{2})$
with $I_{\Delta}^{2}:=\Delta(a,b)+\left[0,\Delta\right)^{2}$ for
$(a,b)\in[-n,n-1]^{2}$; and for the subregion $\mathrm{R}_{k},1\le k\le8$,
we have $\widehat{\pi}_{XY|W_{1}}^{(n)}(\cdot|w_{1})=\widehat{\pi}_{XY}^{(n)}(\cdot|\mathrm{R}_{k})=\pi_{XY}^{(n)}(\cdot|\mathrm{R}_{k})$
for some $1\le k\le8$, where the first or the second component of
$w_{1}$ is $-(n+1)$ or $n$.

By the following lemma, we know that $\widehat{\pi}_{XY}^{(n)}$ is
log-concave.
\begin{lem}[Invariance of Log-Concavity]
\cite[Exercise 3.48]{boyd2004convex} \label{lem:logconcave} If
a pdf $P_{Z^{n}}$ is log-concave, then for any $0\le a<\inf_{z^{n}}P_{Z^{n}}(z^{n})$,
$P_{Z^{n}}-a$ is also log-concave.
\end{lem}
Since $\widehat{\pi}_{XY}^{(n)}$ is log-concave and so is $\widehat{\pi}_{XY|W_{1}}^{(n)}(\cdot|w_{1})$
for each $w_{1}$, the dyadic decomposition scheme in \cite{li2017distributed}
can be applied to $\widehat{\pi}_{XY|W_{1}}^{(n)}(\cdot|w_{1})$.
Hence \cite{li2017distributed} implies that the common entropy $G(\widehat{\pi}_{XY|W_{1}}^{(n)}(\cdot|w_{1}))$
defined in \eqref{eq:commonentropy} satisfies 
\begin{align}
 & H_{\widehat{P}^{(n)}}(W_{2}|W_{1}=w_{1})\nonumber \\
 & =G(\widehat{\pi}_{XY|W_{1}}^{(n)}(\cdot|w_{1}))\\
 & \le I_{\widehat{\pi}_{XY|W_{1}}^{(n)}(\cdot|w_{1})}(X;Y)+24\log2\label{eq:-140}
\end{align}
nats/symbol for $w_{1}\in[-(n+1),n]^{2}$.  We first consider the
case of $w_{1}\in[-n,n-1]^{2}$. For any square $I_{\Delta}^{2}=\Delta(a,b)+\left[0,\Delta\right)^{2}$
in $\mathrm{R}_{0}$ with $(a,b)\in[-n,n-1]^{2}$, we have that 
\begin{align}
 & I_{\widehat{\pi}_{XY}^{(n)}}\left(X;Y|\left(X,Y\right)\in I_{\Delta}^{2}\right)\nonumber \\
 & =\int_{I_{\Delta}^{2}}\widehat{\pi}^{(n)}(x,y|I_{\Delta}^{2})\log\frac{\widehat{\pi}^{(n)}(x,y|I_{\Delta}^{2})}{\widehat{\pi}^{(n)}(x|I_{\Delta})\widehat{\pi}^{(n)}(y|I_{\Delta})}\mathrm{d}x\mathrm{d}y\label{eq:-47}\\
 & =\int_{I_{\Delta}^{2}}\widehat{\pi}^{(n)}(x,y|I_{\Delta}^{2})\log\frac{\frac{\widehat{\pi}^{(n)}(x,y)}{\widehat{\pi}^{(n)}(I_{\Delta}^{2})}}{\frac{\widehat{\pi}^{(n)}(x,I_{\Delta})}{\widehat{\pi}^{(n)}(I_{\Delta}^{2})}\frac{\widehat{\pi}^{(n)}(I_{\Delta},y)}{\widehat{\pi}^{(n)}(I_{\Delta}^{2})}}\mathrm{d}x\mathrm{d}y\\
 & \leq\sup_{\left(x,y\right)\in I_{\Delta}^{2}}\log\frac{\frac{\widehat{\pi}^{(n)}(x,y)}{\widehat{\pi}^{(n)}(I_{\Delta}^{2})}}{\frac{\widehat{\pi}^{(n)}(x,I_{\Delta})}{\widehat{\pi}^{(n)}(I_{\Delta}^{2})}\frac{\widehat{\pi}^{(n)}(I_{\Delta},y)}{\widehat{\pi}^{(n)}(I_{\Delta}^{2})}}\label{eq:-132}\\
 & =\sup_{\left(x,y\right)\in I_{\Delta}^{2}}\log\frac{\widehat{\pi}^{(n)}(x,y)\widehat{\pi}^{(n)}(I_{\Delta}^{2})}{\widehat{\pi}^{(n)}(x,I_{\Delta})\widehat{\pi}^{(n)}(I_{\Delta},y)}\\
 & =\sup_{\left(x,y\right)\in I_{\Delta}^{2}}\log\frac{\widehat{\pi}^{(n)}(x,y)\widehat{\pi}^{(n)}(x',y')}{\widehat{\pi}^{(n)}(x,\widehat{y})\widehat{\pi}^{(n)}(\widehat{x},y)}\label{eq:-53-1}\\
 & \leq\sup_{\left(x,y\right)\in I_{\Delta}^{2}}\log\Biggl\{\frac{1}{\left(\frac{e^{\delta_{n}}-1}{e^{\epsilon_{n}'}-1}\frac{1}{\Delta^{2}}\pi_{AB}^{(n)}\left(a,b\right)\right)^{2}}\nonumber \\
 & \qquad\times\left(\frac{e^{\epsilon_{n}'}\pi_{XY}(x,y)-\frac{1}{\Delta^{2}}\pi_{AB}^{(n)}\left(a,b\right)}{e^{\epsilon_{n}'}-1}\right)\nonumber \\
 & \qquad\times\left(\frac{e^{\epsilon_{n}'}\pi_{XY}(x',y')-\frac{1}{\Delta^{2}}\pi_{AB}^{(n)}\left(a,b\right)}{e^{\epsilon_{n}'}-1}\right)\Biggr\}\label{eq:-57-1}\\
 & \leq2\sup_{\left(x,y\right)\in I_{\Delta}^{2}}\log\frac{e^{\epsilon_{n}'}\pi_{XY}(x,y)-\frac{1}{\Delta^{2}}\pi_{AB}^{(n)}\left(a,b\right)}{\left(e^{\delta_{n}}-1\right)\frac{1}{\Delta^{2}}\pi_{AB}^{(n)}\left(a,b\right)}\\
 & \leq2\log\frac{e^{\epsilon_{n}'+\Delta L_{d}}\frac{1}{\Delta^{2}}\pi_{AB}^{(n)}\left(a,b\right)-\frac{1}{\Delta^{2}}\pi_{AB}^{(n)}\left(a,b\right)}{\left(e^{\delta_{n}}-1\right)\frac{1}{\Delta^{2}}\pi_{AB}^{(n)}\left(a,b\right)}\label{eq:-54-1}\\
 & =2\log\frac{e^{\epsilon_{n}'+\Delta L_{d}}-1}{e^{\delta_{n}}-1}\label{eq:-56-1}\\
 & =2\log\frac{\left(\epsilon_{n}'+\Delta L_{d}\right)\left(1+o\left(1\right)\right)}{\delta_{n}\left(1+o\left(1\right)\right)}\label{eq:-89}\\
 & =2\log\left(\frac{\epsilon_{n}'+\Delta L_{d}}{\delta_{n}}\right)+o\left(1\right)\\
 & \leq\frac{4\Delta L_{d}-2\log p_{d}}{\delta_{n}}+o\left(1\right),\label{eq:-90}
\end{align}
where \eqref{eq:-132} follows since the average is no greater than
the supremum; \eqref{eq:-53-1} holds for some $\left(x',y'\right),\left(\widehat{x},\widehat{y}\right)\in I_{\Delta}^{2}$,
since by the mean value theorem, $\widehat{\pi}^{(n)}(I_{\Delta}^{2})=\Delta^{2}\widehat{\pi}^{(n)}(x',y'),\widehat{\pi}^{(n)}(x,I_{\Delta})=\Delta\widehat{\pi}^{(n)}(x,\widehat{y}),\widehat{\pi}^{(n)}(I_{\Delta},y)=\Delta\widehat{\pi}^{(n)}(\widehat{x},y)$
for some $\left(x',y'\right),\left(\widehat{x},\widehat{y}\right)\in I_{\Delta}^{2}$;
\eqref{eq:-57-1} follows from \eqref{eq:-157} and \eqref{eq:-133};
\eqref{eq:-54-1} follows from \eqref{eq:-55}; \eqref{eq:-56-1}
follows from \eqref{eq:-48}; in \eqref{eq:-89}, $o\left(1\right)$
denotes a term tending to zero as $\epsilon_{n}',\Delta L_{d},\delta_{n}\to0$;
and \eqref{eq:-90} follows from \eqref{eq:-52-1}. By introducing
the positive sequence $\delta_{n}$, the denominators in equations
after \eqref{eq:-57-1} are ensured to be positive. This is the reason
why we introduce $\delta_{n}$ in \eqref{eq:-52-1}.

On the other hand, for the case of $w_{1}\notin[-n,n-1]^{2}$, i.e.,
for the subregions $\mathrm{R}_{k}=I_{1}^{(k)}\times I_{2}^{(k)},1\le k\le8$,
we have $\widehat{\pi}_{XY}^{(n)}(\cdot|\mathrm{R}_{k})=\pi_{XY}(\cdot|I_{1}^{(k)}\times I_{2}^{(k)})$.
Hence 
\begin{align}
 & I_{\widehat{\pi}_{XY}^{(n)}}\left(X;Y|\left(X,Y\right)\in\mathrm{R}_{k}\right)\nonumber \\
 & =I_{\pi}\left(X;Y|\left(X,Y\right)\in I_{1}^{(k)}\times I_{2}^{(k)}\right).\label{eq:-88}
\end{align}
Now we bound the RHS of \eqref{eq:-88} by using the following lemma. 
\begin{lem}[Estimation of Conditional Mutual Information]
\label{lem:conditionalMI} Assume $\pi_{XY}$ is an absolutely continuous
distribution such that $\lim_{x\to+\infty}\pi_{X}(x)=\lim_{x\to-\infty}\pi_{X}(x)=\lim_{y\to+\infty}\pi_{Y}(y)=\lim_{y\to-\infty}\pi_{Y}(y)=0$.
For $A,B\in\left\{ \mathcal{L}_{d}^{-},\mathcal{L}_{d},\mathcal{L}_{d}^{+}\right\} $,
we have 
\begin{equation}
I_{\pi}\left(X;Y|\left(X,Y\right)\in A\times B\right)\leq\Upsilon_{\pi}\left(A,B\right),
\end{equation}
where 
\begin{align}
 & \Upsilon_{\pi}\left(A,B\right)\nonumber \\
 & :=\begin{cases}
\frac{1}{\pi_{XY}(A\times B)}\left(I_{\pi}\left(X;Y\right)+o(1)\right) & A=B=\mathcal{L}_{d}\\
\frac{1}{\pi_{XY}(A\times B)}o(1) & \textrm{otherwise}
\end{cases}
\end{align}
and $o(1)$ denotes a term tending to zero as $d\to\infty$. 
\end{lem}
The proof of Lemma \ref{lem:conditionalMI} is deferred to Appendix
\ref{subsec:conditionalMI}.

It is easy to verify that a absolutely continuous log-concave pdf
satisfies the conditions prescribed in Lemma \ref{lem:conditionalMI}.
Hence by Lemma \ref{lem:conditionalMI}, we have 
\begin{equation}
I_{\widehat{\pi}_{XY}^{(n)}}\left(X;Y|\left(X,Y\right)\in\mathrm{R}_{k}\right)\leq\Upsilon_{\pi}\left(I_{1}^{(k)},I_{2}^{(k)}\right).
\end{equation}
Substituting this into \eqref{eq:-140}, we have 
\begin{align}
 & G(\widehat{\pi}_{XY|\left(X,Y\right)\in\mathrm{R}_{k}}^{(n)})\nonumber \\
 & \leq I_{\widehat{\pi}_{XY}^{(n)}}\left(X;Y|\left(X,Y\right)\in\mathrm{R}_{k}\right)+24\log2\\
 & \leq\Upsilon_{\pi}\left(I_{1}^{(k)},I_{2}^{(k)}\right)+24\log2.\label{eq:-141}
\end{align}

According to the definition of $Q_{XY}^{(n)}$, we have $Q_{XY}^{(n)}=\pi_{XY}$,
and $X\to\left(W,U\right)\to Y$ under $Q^{(n)}$. Similarly to the
countable case, we obtain that 

\begin{align}
 & C_{\mathrm{Wyner}}(\pi_{XY})\nonumber \\
 & =\inf_{P_{W}P_{X|W}P_{Y|W}:P_{XY}=\pi_{XY}}I\left(XY;W\right)\\
 & \leq I_{Q^{(n)}}\left(XY;WU\right)\\
 & =I_{Q^{(n)}}\left(XY;U\right)+e^{-\epsilon_{n}'}I_{Q^{(n)}}\left(XY;W|U=1\right)\nonumber \\
 & \qquad+\left(1-e^{-\epsilon_{n}'}\right)I_{Q^{(n)}}\left(XY;W|U=0\right)\\
 & \leq H\left(U\right)+e^{-\epsilon_{n}'}I_{P^{(n)}}(XY;W)\nonumber \\
 & \qquad+\left(1-e^{-\epsilon_{n}'}\right)H_{Q^{(n)}}\left(W_{1}W_{2}|U=0\right).\label{eq:-80}
\end{align}
Since $\epsilon_{n}'\to0$ as $n\to\infty$, the first term in \eqref{eq:-80}
is bounded as $H\left(U\right)=H\left(e^{-\epsilon_{n}'}\right)\to0$
as $n\to\infty$. For the second term in \eqref{eq:-80}, 
\begin{align}
e^{-\epsilon_{n}'}I_{P^{(n)}}(XY;W) & =e^{-\epsilon_{n}'}I_{P^{(n)}}(AB;W)\\
 & \leq C_{\mathrm{Wyner}}(\pi_{AB}^{(n)})\\
 & =\widetilde{C}_{\mathrm{Wyner}}(\pi_{AB}^{(n)})\label{eq:-82-1-1}\\
 & \leq\widetilde{C}_{\mathrm{Wyner}}(\pi_{XY}^{(d)}),\label{eq:-67-3-1}
\end{align}
where \eqref{eq:-82-1-1} follows by Corollary \ref{cor:countable}
since $\pi_{AB}^{(n)}$ is supported on a finite alphabet, and \eqref{eq:-67-3-1}
follows by the data processing inequality.

We bound the last term in \eqref{eq:-80} as 
\begin{align}
 & H_{Q^{(n)}}\left(W_{1}W_{2}|U=0\right)\nonumber \\
 & =H_{\widehat{P}^{(n)}}\left(W_{1}W_{2}\right)\\
 & =H_{\widehat{P}^{(n)}}(W_{1})+H_{\widehat{P}^{(n)}}\left(W_{2}|W_{1}\right)\\
 & \leq H_{\widehat{P}^{(n)}}(W_{1})+\sum_{w_{1}}\widehat{P}^{(n)}(w_{1})\nonumber \\
 & \qquad\times\left(I_{\widehat{\pi}_{XY|W_{1}}^{(n)}(\cdot|w_{1})}(X;Y)+24\log2\right)\\
 & \leq2\log(2n+2)+I_{\widehat{P}^{(n)}}(X;Y|W_{1})+24\log2,\label{eq:-143}
\end{align}
where \eqref{eq:-143} follows from $H_{\widehat{P}^{(n)}}(W_{1})\le2\log(2n+2)$
since $W_{1}$ is defined on $[-(n+1),n]^{2}$.

On the other hand, by applying \eqref{eq:-90} and \eqref{eq:-141},
we obtain that 
\begin{align}
 & \left(1-e^{-\epsilon_{n}'}\right)I_{\widehat{P}^{(n)}}(X;Y|W_{1})\nonumber \\
 & \leq\left(1-e^{-\epsilon_{n}'}\right)\widehat{\pi}_{XY}^{(n)}\left([-d,d)^{2}\right)\left(\frac{4\Delta L_{d}-2\log p_{d}}{\delta_{n}}+o\left(1\right)\right)\nonumber \\
 & \qquad+\left(1-e^{-\epsilon_{n}'}\right)\sum_{k=1}^{8}\widehat{\pi}_{XY}^{(n)}(\mathrm{R}_{k})\Upsilon_{\pi}\left(I_{1}^{(k)},I_{2}^{(k)}\right)\\
 & \leq\left(1-e^{-\epsilon_{n}'}\right)\left(\frac{4\Delta L_{d}-2\log p_{d}}{\delta_{n}}+o\left(1\right)\right)\nonumber \\
 & \qquad+\sum_{k=1}^{8}\pi_{XY}(\mathrm{R}_{k})\Upsilon_{\pi}\left(I_{1}^{(k)},I_{2}^{(k)}\right)\label{eq:-158}\\
 & =\left(1-e^{-\epsilon_{n}'}\right)\left(\frac{4\Delta L_{d}-2\log p_{d}}{\delta_{n}}+o\left(1\right)\right)+o\left(1\right)\label{eq:-142}
\end{align}
where \eqref{eq:-158} follows since $\widehat{\pi}_{XY}^{(n)}\left([-d,d)^{2}\right)\le1$
and $\widehat{\pi}_{XY}^{(n)}(\mathrm{R}_{k})=\frac{\pi_{XY}(\mathrm{R}_{k})}{1-e^{-\epsilon_{n}'}}$
(the latter follows by \eqref{eq:-157} and the fact that $\pi_{XY}^{(n)}(x,y)$
is defined on $[-d,d)^{2}$); and \eqref{eq:-142} follow by Lemma
\ref{lem:conditionalMI}.

Combining \eqref{eq:-80}, \eqref{eq:-67-3-1}, \eqref{eq:-143},
and \eqref{eq:-142} yields \eqref{eq:-79}-\eqref{eq:-81} (given
on page \pageref{eq:-79}). 
\begin{figure*}
\begin{align}
C_{\mathrm{Wyner}}(\pi_{XY}) & \leq\liminf_{n\to\infty}\left\{ \widetilde{C}_{\mathrm{Wyner}}(\pi_{XY}^{(d)})+\left(1-e^{-\epsilon_{n}'}\right)\left(2\log(2n+2)+24\log2+\frac{4\Delta L_{d}-2\log p_{d}}{\delta_{n}}+o\left(1\right)\right)+o\left(1\right)\right\} \label{eq:-79}\\
 & =\liminf_{n\to\infty}\left\{ \widetilde{C}_{\mathrm{Wyner}}(\pi_{XY}^{(d)})+\left(1-e^{-\left(\Delta L_{d}-\log p_{d}+\delta_{n}\right)}\right)\left(2\log(2n+2)+\frac{4\Delta L_{d}-2\log p_{d}}{\delta_{n}}\right)\right\} .\label{eq:-81}
\end{align}

\hrulefill{}
\end{figure*}

Choose $\delta_{n}=2\Delta L_{d}-\log p_{d}$, then to ensure that
the RHS of \eqref{eq:-81} is no larger than $\liminf_{d\to\infty}\widetilde{C}_{\mathrm{Wyner}}(\pi_{XY}^{(d)})$,
we only require 
\begin{align}
 & \left(1-e^{-\left(3\Delta L_{d}-2\log p_{d}\right)}\right)\log n\to0,\label{eq:-58-1-1-2}
\end{align}
i.e., 
\begin{align}
 & \left(3\Delta L_{d}-2\log p_{d}\right)\log n\to0.\label{eq:-58-1-1}
\end{align}

Set $\Delta$ to $\Delta_{d}=\left(dL_{d}\right)^{-\alpha}$ for $\alpha>1$.
Recall $n=\frac{d}{\Delta}$. Then we have
\begin{align}
 & \Delta L_{d}\log n\to0.\label{eq:-58-1-1-1-1}
\end{align}

Recall $p_{d}=1-\epsilon_{d}$. By the hypothesis that $\epsilon_{d}\log\left(dL_{d}\right)\to0$
as $d\to+\infty$, we have
\begin{align}
 & \left(\log p_{d}\right)\left(\log n\right)\to0.\label{eq:-58-1-1-1}
\end{align}

Hence for such a choice of $\Delta$, \eqref{eq:-58-1-1} is satisfied,
which implies that 
\begin{align}
C_{\mathrm{Wyner}}(\pi_{XY}) & \leq\liminf_{d\to\infty}\widetilde{C}_{\mathrm{Wyner}}(\pi_{XY}^{(d)}).\label{eq:-21-1}
\end{align}

\subsubsection{\label{subsec:conditionalMI}Proof of Lemma \ref{lem:conditionalMI} }

Consider that 
\begin{align}
 & I_{\pi}\left(X;Y|\left(X,Y\right)\in A\times B\right)\nonumber \\
 & =\int_{A\times B}\frac{\pi_{XY}(x,y)}{\pi_{XY}(A\times B)}\nonumber \\
 & \qquad\times\log\frac{\pi_{XY}(x,y)\pi_{XY}(A\times B)}{\pi_{X}(x)\pi_{Y|X}(B|x)\pi_{Y}(y)\pi_{X|Y}(A|y)}\mathrm{d}x\mathrm{d}y\\
 & =\frac{1}{\pi_{XY}(A\times B)}\biggl\{\int_{A\times B}\pi_{XY}(x,y)\log\frac{\pi_{XY}(x,y)}{\pi_{X}(x)\pi_{Y}(y)}\mathrm{d}x\mathrm{d}y\nonumber \\
 & \qquad+\log\pi_{XY}(A\times B)\nonumber \\
 & \qquad-\int_{A}\pi_{X}(x)\pi_{Y|X}(B|x)\log\pi_{Y|X}(B|x)\mathrm{d}x\nonumber \\
 & \qquad-\int_{B}\pi_{Y}(y)\pi_{X|Y}(A|y)\log\pi_{X|Y}(A|y)\mathrm{d}y\biggr\}\\
 & \leq\Upsilon_{\pi}\left(A,B\right)\label{eq:-39}
\end{align}
where \eqref{eq:-39} follows from the facts that $\log\pi_{XY}(A\times B)\le0$
and 
\begin{align}
 & \lim_{n\to\infty}\int_{A\times B}\pi_{XY}(x,y)\log\frac{\pi_{XY}(x,y)}{\pi_{X}(x)\pi_{Y}(y)}\mathrm{d}x\mathrm{d}y\nonumber \\
 & =\begin{cases}
I_{\pi}\left(X;Y\right) & A=B=\mathcal{L}_{d}\\
0 & \textrm{otherwise}
\end{cases},
\end{align}
as well as the following arguments. For all $B\in\left\{ \mathcal{L}_{d}^{-},\mathcal{L}_{d},\mathcal{L}_{d}^{+}\right\} $,
\begin{equation}
-\pi_{X}(x)\pi_{Y|X}(B|x)\log\pi_{Y|X}(B|x)\to0
\end{equation}
pointwise, 
\begin{equation}
\left|-\pi_{X}(x)\pi_{Y|X}(B|x)\log\pi_{Y|X}(B|x)\right|\leq e^{-1}\pi_{X}(x)
\end{equation}
and $e^{-1}\pi_{X}(x)$ is integrable. Hence by Lebesgue's dominated
convergence theorem, we have 
\begin{equation}
\lim_{n\to\infty}-\int_{A}\pi_{X}(x)\pi_{Y|X}(B|x)\log\pi_{Y|X}(B|x)\mathrm{d}x=0.
\end{equation}
Similarly, 
\begin{equation}
\lim_{n\to\infty}-\int_{B}\pi_{Y}(y)\pi_{X|Y}(A|y)\log\pi_{X|Y}(A|y)\mathrm{d}y=0.
\end{equation}

\section{\label{sec:countable-1}Proof of Lemma \ref{lem:countable}}

The proof techniques used in this section are similar to those used
in Appendix \ref{sec:countable}.

Assume $\left(P_{M},P_{X^{n}|M},P_{Y^{n}|M}\right)$ is a sequence
of fixed-length codes with rate $R$ that generates $P_{X^{n}Y^{n}}$
such that $D_{\infty}(P_{X^{n}Y^{n}}\|\pi_{XY}^{n})\to0$, where $P_{M}$
is the uniform distribution on $[1:e^{nR}]$. Similarly to \eqref{eq:-86},
we introduce a random variable 
\begin{equation}
V:=1\left\{ (X^{n},Y^{n})\in\mathcal{A}_{\epsilon}^{(n)}\left(\pi_{X}\right)\times\mathcal{A}_{\epsilon}^{(n)}\left(\pi_{Y}\right)\right\} .\label{eq:-42}
\end{equation}
Similarly to \eqref{eq:-87}, we define $\widetilde{P}_{MX^{n}Y^{n}}:=P_{MX^{n}Y^{n}|V}(m,x^{n},y^{n}|1)$.
Then $\widetilde{P}_{MX^{n}Y^{n}}=\widetilde{P}_{M}\widetilde{P}_{X^{n}|M}\widetilde{P}_{Y^{n}|M}$,
i.e., $X^{n}\to M\to Y^{n}$ forms a Markov chain under $\widetilde{P}$.
On the other hand, 
\begin{equation}
H(\widetilde{P}_{M})\leq R\label{eq:-43}
\end{equation}
(since $\widetilde{P}_{M}$ is defined on an alphabet with size $e^{nR}$)
and 
\begin{align}
 & D_{\infty}(\widetilde{P}_{X^{n}Y^{n}}\|\pi_{XY}^{n})\nonumber \\
 & =D_{\infty}(P_{X^{n}Y^{n}|V=1}\|\pi_{XY}^{n})\\
 & =\log\sup_{\substack{(x^{n},y^{n})\in\\
\mathcal{A}_{\epsilon}^{(n)}\left(\pi_{X}\right)\times\mathcal{A}_{\epsilon}^{(n)}\left(\pi_{Y}\right)
}
}\frac{P_{X^{n}Y^{n}}\left(x^{n},y^{n}\right)}{\pi_{XY}^{n}\left(x^{n},y^{n}\right)}-\log P_{V}(1)\\
 & \leq D_{\infty}(P_{X^{n}Y^{n}}\|\pi_{XY}^{n})-\log P_{V}(1).\label{eq:-38}
\end{align}

We now prove Lemma \ref{lem:countable} by a argument similar as that
in Appendix \ref{sec:equivalence}. According to the definition of
$D_{\infty}$, $D_{\infty}(P_{X^{n}Y^{n}}\|\pi_{XY}^{n})\leq\epsilon_{n}$
implies $D_{\infty}(\widetilde{P}_{X^{n}Y^{n}}\|\pi_{XY}^{n})\leq\epsilon_{n}-\log P_{V}(1)$,
i.e., 
\begin{equation}
\sup_{x^{n},y^{n}}\frac{\widetilde{P}_{X^{n}Y^{n}}\left(x^{n},y^{n}\right)}{\pi_{XY}^{n}\left(x^{n},y^{n}\right)}\leq e^{\epsilon_{n}-\log P_{V}(1)}=:e^{\epsilon_{n}'}.
\end{equation}
 Define $\widehat{P}_{X^{n}Y^{n}}\left(x^{n},y^{n}\right):=\frac{e^{\epsilon_{n}'}\pi_{XY}^{n}\left(x^{n},y^{n}\right)-\widetilde{P}_{X^{n}Y^{n}}\left(x^{n},y^{n}\right)}{e^{\epsilon_{n}'}-1}$,
then obviously $\widehat{P}_{X^{n}Y^{n}}\left(x^{n},y^{n}\right)$
is a distribution. Hence $\pi_{XY}^{n}$ can be written as a mixture
distribution $\pi_{XY}^{n}\left(x^{n},y^{n}\right)=e^{-\epsilon_{n}'}\widetilde{P}_{X^{n}Y^{n}}\left(x^{n},y^{n}\right)+\left(1-e^{-\epsilon_{n}'}\right)\widehat{P}_{X^{n}Y^{n}}\left(x^{n},y^{n}\right)$.
The encoder first generates a Bernoulli random variable $U$ with
$P_{U}(1)=e^{-\epsilon_{n}'}$, compresses it with $1$ bit, and transmits
it to the two generators. If $U=1$, then the encoder and two generators
use the synthesis codes $\left(\widetilde{P}_{M},\widetilde{P}_{X^{n}|M},\widetilde{P}_{Y^{n}|M}\right)$
with rate $R$ (by fixed-length codes) to generate $\widetilde{P}_{X^{n}Y^{n}}$.
If $U=0$, then the encoder generates $\left(X^{n},Y^{n}\right)\sim\widehat{P}_{X^{n}Y^{n}}$,
and uses a variable-length compression code with rate 
\begin{align}
 & \frac{1}{n}\left(H\left(\widehat{P}_{X^{n}Y^{n}}\right)+1\right)\nonumber \\
 & \leq\frac{1}{n}\Bigl(H_{\widehat{P}}(V)+\widehat{P}_{V}(1)\log\left|\mathcal{A}_{\epsilon}^{(n)}\left(\pi_{X}\right)\times\mathcal{A}_{\epsilon}^{(n)}\left(\pi_{Y}\right)\right|\nonumber \\
 & \qquad+\widehat{P}_{V}(0)H\left(\widehat{P}_{X^{n}Y^{n}|V=0}\right)+1\Bigr)
\end{align}
to generate $\widehat{P}_{X^{n}Y^{n}}$. The distribution generated
by such a mixed code is $e^{-\epsilon_{n}'}\widetilde{P}_{X^{n}Y^{n}}\left(x^{n},y^{n}\right)+\left(1-e^{-\epsilon_{n}'}\right)\widehat{P}_{X^{n}Y^{n}}\left(x^{n},y^{n}\right)$,
i.e., $\pi_{XY}^{n}\left(x^{n},y^{n}\right)$. The total code rate
is no larger than 
\begin{align}
 & \frac{1}{n}+e^{-\epsilon_{n}'}R+\left(1-e^{-\epsilon_{n}'}\right)\frac{1}{n}\biggl(H_{\widehat{P}}(V)\nonumber \\
 & \quad+\widehat{P}_{V}(1)\log\left|\mathcal{A}_{\epsilon}^{(n)}\left(\pi_{X}\right)\times\mathcal{A}_{\epsilon}^{(n)}\left(\pi_{Y}\right)\right|\nonumber \\
 & \quad+\widehat{P}_{V}(0)H\left(\widehat{P}_{X^{n}Y^{n}|V=0}\right)+1\biggr).\label{eq:-35}
\end{align}

Observe that $\pi_{V}(0)\to0,$ and by the data processing inequality,
$P_{V}(0)\leq\pi_{V}(0)e^{\epsilon_{n}}\to0$. Hence $\epsilon_{n}'=\epsilon_{n}-\log P_{V}(1)\to0$
as $n\to\infty$. On the other hand, we have 
\begin{align}
H_{\widehat{P}}(V) & \leq\log2\\
\frac{1}{n}\log\left|\mathcal{A}_{\epsilon}^{(n)}\left(\pi_{X}\right)\times\mathcal{A}_{\epsilon}^{(n)}\left(\pi_{Y}\right)\right| & \to H\left(\pi_{X}\right)+H\left(\pi_{Y}\right),
\end{align}
 and 
\begin{align}
 & H\left(\widehat{P}_{X^{n}Y^{n}|V=0}\right)\nonumber \\
 & =H\left(\pi_{X^{n}Y^{n}|V=0}\right)\\
 & =\log\pi_{V}(0)-\frac{1}{\pi_{V}(0)}\nonumber \\
 & \qquad\times\sum_{\substack{(x^{n},y^{n})\notin\\
\mathcal{A}_{\epsilon}^{(n)}\left(\pi_{X}\right)\times\mathcal{A}_{\epsilon}^{(n)}\left(\pi_{Y}\right)
}
}\pi_{XY}^{n}\left(x^{n},y^{n}\right)\log\pi_{XY}^{n}\left(x^{n},y^{n}\right)\\
 & =\log\pi_{V}(0)+\frac{1}{\pi_{V}(0)}\Biggl(nH(\pi_{XY})\nonumber \\
 & \qquad+\sum_{\substack{(x^{n},y^{n})\in\\
\mathcal{A}_{\epsilon}^{(n)}\left(\pi_{X}\right)\times\mathcal{A}_{\epsilon}^{(n)}\left(\pi_{Y}\right)
}
}\pi_{XY}^{n}\left(x^{n},y^{n}\right)\log\pi_{XY}^{n}\left(x^{n},y^{n}\right)\Biggr)\\
 & \leq\log\pi_{V}(0)+\frac{1}{\pi_{V}(0)}\left(nH(\pi_{XY})-n\left(1-\epsilon\right)\left(H(\pi_{XY})-\epsilon\right)\right)\label{eq:-28}\\
 & =\frac{n}{\pi_{V}(0)}\left(\epsilon\left(H(\pi_{XY})+1-\epsilon\right)+\frac{\pi_{V}(0)\log\pi_{V}(0)}{n}\right)\\
 & =\frac{n\left(\epsilon\left(H(\pi_{XY})+1-\epsilon\right)+o(1)\right)}{\pi_{V}(0)},\label{eq:-29}
\end{align}
where \eqref{eq:-28} follows since  $\mathcal{A}_{\epsilon}^{(n)}\left(\pi_{XY}\right)\subseteq\mathcal{A}_{\epsilon}^{(n)}\left(\pi_{X}\right)\times\mathcal{A}_{\epsilon}^{(n)}\left(\pi_{Y}\right)$
and 
\begin{align}
 & \sum_{(x^{n},y^{n})\in\mathcal{A}_{\epsilon}^{(n)}\left(\pi_{XY}\right)}\pi_{XY}^{n}\left(x^{n},y^{n}\right)\log\pi_{XY}^{n}\left(x^{n},y^{n}\right)\nonumber \\
 & \leq-n\sum_{(x^{n},y^{n})\in\mathcal{A}_{\epsilon}^{(n)}\left(\pi_{XY}\right)}\pi_{XY}^{n}\left(x^{n},y^{n}\right)\left(H(\pi_{XY})-\epsilon\right)\label{eq:-145}\\
 & =-n\pi_{XY}^{n}\left(\mathcal{A}_{\epsilon}^{(n)}\left(\pi_{XY}\right)\right)\left(H(\pi_{XY})-\epsilon\right)\\
 & \leq-n\left(1-\epsilon\right)\left(H(\pi_{XY})-\epsilon\right).\label{eq:-144}
\end{align}
Here \eqref{eq:-145} follows by the definition of the $\epsilon$-weakly
jointly typical set $\mathcal{A}_{\epsilon}^{(n)}\left(\pi_{XY}\right)$,
and \eqref{eq:-144} follows by \cite[Theorem 3.1.2]{Cover}.

Hence to ensure \eqref{eq:-35} converges to $R$, we only require
\begin{equation}
\left(1-e^{-\epsilon_{n}'}\right)\widehat{P}_{V}(0)\frac{\epsilon\left(H(\pi_{XY})+1-\epsilon\right)+o(1)}{\pi_{V}(0)}\to0.
\end{equation}
According to the definitions of $\widehat{P}_{X^{n}Y^{n}}$ and $V$,
we know $\widehat{P}_{V}(0)=\frac{e^{\epsilon_{n}'}\pi_{V}(0)}{e^{\epsilon_{n}'}-1}$.
Hence 
\begin{align}
 & \left(1-e^{-\epsilon_{n}'}\right)\widehat{P}_{V}(0)\frac{\epsilon\left(H(\pi_{XY})+1-\epsilon\right)+o(1)}{\pi_{V}(0)}\nonumber \\
 & =\pi_{V}(0)\frac{\epsilon\left(H(\pi_{XY})+1-\epsilon\right)+o(1)}{\pi_{V}(0)}\\
 & =\epsilon\left(H(\pi_{XY})+1-\epsilon\right)+o(1)\to0
\end{align}
by letting $n\to\infty$ first and letting $\epsilon\to\infty$ then.
This completes the proof.

\section{\label{sec:continuous-1}Proof of Lemma \ref{lem:continuous}}

Some proof techniques used in this section are similar to those used
in Appendix \ref{sec:continuous}.

\subsection{\label{subsec:A-Modified-Version}A Modified Version of $\infty$-R\'enyi
Code}

By assumption, there exists a sequence of fixed-length $\infty$-R\'enyi
codes with rate $R$ such that $D_{\infty}(P_{X^{n}Y^{n}}\|\pi_{XY}^{n})\to0$.
In this subsection, we construct another sequence of fixed-length
$\infty$-R\'enyi codes by cascading the original $\infty$-R\'enyi codes
with truncation, discretization, and adding noise. These new $\infty$-R\'enyi
codes will be used to construct the final exact synthesis scheme in
Appendix \ref{subsec:Exact-Synthesis-Scheme}. The original $\infty$-R\'enyi
codes cannot be applied directly, since in the final exact synthesis
scheme, we mix the $\infty$-R\'enyi codes and dyadic decomposition
schemes \cite{li2017distributed}. The $\infty$-R\'enyi codes are used
to generate an approximate distribution $\widetilde{P}_{X^{n}Y^{n}}$
of $\pi_{XY}^{n}$. The dyadic decomposition schemes are used to generate
the residual distribution after subtracting (a scaled version of)
$\widetilde{P}_{X^{n}Y^{n}}$ from $\pi_{XY}^{n}$. The dyadic decomposition
schemes require the residual distribution to be log-concave. The original
$\infty$-R\'enyi codes cannot generate a log-concave residual distribution.
Hence it is necessary to construct new $\infty$-R\'enyi codes to ensure
the residual distribution to be log-concave.

By respectively scaling $X,Y$, we can obtain a bivariate source with
$\mathbb{E}\left[X^{2}\right]=\mathbb{E}\left[Y^{2}\right]=1$. Hence
without loss of generality, we assume $\pi_{XY}$ satisfying $\mathbb{E}\left[X^{2}\right]=\mathbb{E}\left[Y^{2}\right]=1$.
Define an $n$-ball with radius $\sqrt{n\left(1+\epsilon\right)}$
as 
\begin{equation}
\mathcal{B}_{\epsilon}^{(n)}:=\left\{ x^{n}\in\mathbb{R}^{n}:\left\Vert x^{n}\right\Vert \leq\sqrt{n\left(1+\epsilon\right)}\right\} .
\end{equation}
Note that $\mathcal{B}_{\epsilon}^{(n)}$ is a high probability set
for any memoryless source with unit second moment, i.e., $\pi_{X}^{n}(\mathcal{B}_{\epsilon}^{(n)}),\pi_{Y}^{n}(\mathcal{B}_{\epsilon}^{(n)})\to1$.
Hence $\pi_{XY}^{n}(\mathcal{B}_{\epsilon}^{(n)}\times\mathcal{B}_{\epsilon}^{(n)})\to1$.
Obviously, $\mathcal{B}_{\epsilon}^{(n)}$ is contained in the $n$-cube
$\mathcal{L}_{\epsilon,n}^{n}$ with $\mathcal{L}_{\epsilon,n}$ defined
in \eqref{eq:-36}. Hence $\pi_{XY}^{n}(\mathcal{L}_{\epsilon,n}^{2n})\to1$.

Assume $\Delta_{n}$ is a decreasing positive sequence such that $\Delta_{n}\to0$
and $n\Delta_{n}L_{\epsilon,n}\to0$. By Lemma \ref{lem:RatioBounds},
we have that for any $\left(x,y\right),\left(\hat{x},\hat{y}\right)\in\mathcal{L}_{\epsilon,n}^{2}$
satisfying $\left|x-\hat{x}\right|,\left|y-\hat{y}\right|\leq\Delta_{n}$,
\begin{align}
\frac{\pi_{XY}\left(x,y\right)}{\pi_{XY}\left(\hat{x},\hat{y}\right)} & \leq\exp\left(\Delta_{n}L_{\epsilon,n}\right).
\end{align}
Hence for $\left(x^{n},y^{n}\right),\left(\hat{x}^{n},\hat{y}^{n}\right)\in\mathcal{L}_{\epsilon,n}^{n}\times\mathcal{L}_{\epsilon,n}^{n}$,
satisfying $\left|x_{i}-\hat{x}_{i}\right|,\left|y_{i}-\hat{y}_{i}\right|\leq\Delta_{n},\forall i$,
we have 
\begin{align}
\frac{\pi_{XY}^{n}\left(x^{n},y^{n}\right)}{\pi_{XY}^{n}\left(\hat{x}^{n},\hat{y}^{n}\right)} & \leq\exp\left(n\Delta_{n}L_{\epsilon,n}\right).\label{eq:-55}
\end{align}

Assume $\left(P_{M},P_{X^{n}|M},P_{Y^{n}|M}\right)$ is a sequence
of fixed-length $\infty$-R\'enyi codes with rate $R$. That is, $P_{M}$
is the uniform distribution on $[1:e^{nR}]$, and this sequence of
codes generates distributions $P_{X^{n}Y^{n}}$ such that $\epsilon_{n}:=D_{\infty}(P_{X^{n}Y^{n}}\|\pi_{XY}^{n})\to0$.
Similar to \eqref{eq:-42}, we introduce a random variable 
\begin{equation}
V:=1\left\{ (X^{n},Y^{n})\in\mathcal{L}_{\epsilon,n}^{2n}\right\} .\label{eq:-42-1}
\end{equation}
We define $\widetilde{P}_{MX^{n}Y^{n}}:=P_{MX^{n}Y^{n}|V}(m,x^{n},y^{n}|1)$.
Then by Lemma \ref{lem:conditionalMarkov}, $\widetilde{P}_{MX^{n}Y^{n}}=\widetilde{P}_{M}\widetilde{P}_{X^{n}|M}\widetilde{P}_{Y^{n}|M}$,
i.e., $X^{n}\to M\to Y^{n}$ forms a Markov chain under $\widetilde{P}$.
\eqref{eq:-43} and \eqref{eq:-38} still hold. Define $\left[z\right]^{n}:=\Delta_{n}\left\lfloor \frac{z^{n}}{\Delta_{n}}\right\rfloor $
as componentwise quantization operation of a vector $z^{n}$ with
step $\Delta_{n}$ (for simplicity, we choose $\Delta_{n}$ such that
$\sqrt{n\left(1+\epsilon\right)}$ is a multiple of $\Delta_{n}$).
Define $U^{n},V^{n}\sim\mathrm{Unif}\left([0,\Delta_{n}]^{n}\right)$
are mutually independent, and also independent of $\left[X\right]^{n},\left[Y\right]^{n}$.
Then 
\begin{align}
 & \sup_{x^{n},y^{n}}\frac{\widetilde{P}_{\left[X\right]^{n}+U^{n},\left[Y\right]^{n}+V^{n}}\left(x^{n},y^{n}\right)}{\pi_{XY}^{n}\left(x^{n},y^{n}\right)}\nonumber \\
 & \leq\exp\left(n\Delta_{n}L_{\epsilon,n}\right)\sup_{x^{n},y^{n}}\frac{\widetilde{P}_{\left[X\right]^{n}\left[Y\right]^{n}}\left(\left[x\right]^{n},\left[y\right]^{n}\right)/\Delta_{n}^{n}}{\pi_{XY}^{n}\left(\hat{x}^{n},\hat{y}^{n}\right)}\label{eq:-101}\\
 & =\exp\left(n\Delta_{n}L_{\epsilon,n}\right)\sup_{\left[x\right]^{n},\left[y\right]^{n}}\frac{\widetilde{P}_{\left[X\right]^{n}\left[Y\right]^{n}}\left(\left[x\right]^{n},\left[y\right]^{n}\right)}{\pi_{\left[X\right]\left[Y\right]}^{n}\left(\left[x\right]^{n},\left[y\right]^{n}\right)}\label{eq:-102}\\
 & \leq\exp\left(n\Delta_{n}L_{\epsilon,n}\right)\sup_{x^{n},y^{n}}\frac{\widetilde{P}_{X^{n}Y^{n}}\left(x^{n},y^{n}\right)}{\pi_{XY}^{n}\left(x^{n},y^{n}\right)}\label{eq:-46}\\
 & \leq\exp\left(n\Delta_{n}L_{\epsilon,n}+D_{\infty}(P_{X^{n}Y^{n}}\|\pi_{XY}^{n})-\log P_{V}(1)\right)\label{eq:-49}\\
 & =\exp\left(n\Delta_{n}L_{\epsilon,n}-\log P_{V}(1)+\epsilon_{n}\right),\label{eq:-48}
\end{align}
where $\left(\hat{x}^{n},\hat{y}^{n}\right)$ in \eqref{eq:-101}
is a point in $\left(\left[x\right]^{n},\left[y\right]^{n}\right)+[0,\Delta_{n}]^{2n}$
such that $\pi_{XY}^{n}\left(\hat{x}^{n},\hat{y}^{n}\right)=\pi_{\left[X\right]\left[Y\right]}^{n}\left(\left[x\right]^{n},\left[y\right]^{n}\right)/\Delta_{n}^{n}$
(the existence of such a point follows from the mean value theorem),
\eqref{eq:-101} follows from \eqref{eq:-55}, \eqref{eq:-46} follows
from the data processing inequality, and \eqref{eq:-49} follows from
\eqref{eq:-38}. Define 
\begin{equation}
\epsilon_{n}':=n\Delta_{n}L_{\epsilon,n}-\log P_{V}(1)+\epsilon_{n}+\delta_{n}\label{eq:-52}
\end{equation}
for some positive sequence $\delta_{n}\to0$ as $n\to\infty$, which
will be specified later. Then \eqref{eq:-48} implies for all $\left(x^{n},y^{n}\right)\in\mathcal{L}_{\epsilon,n}^{2n}$,
\begin{align}
\frac{e^{\epsilon_{n}'}\pi_{XY}^{n}\left(x^{n},y^{n}\right)}{\widetilde{P}_{\left[X\right]^{n}+U^{n},\left[Y\right]^{n}+V^{n}}\left(x^{n},y^{n}\right)} & \geq e^{\delta_{n}}.\label{eq:-51}
\end{align}

Define 
\begin{align}
 & \widehat{P}_{X^{n}Y^{n}}\left(x^{n},y^{n}\right)\nonumber \\
 & :=\frac{e^{\epsilon_{n}'}\pi_{XY}^{n}\left(x^{n},y^{n}\right)-\widetilde{P}_{\left[X\right]^{n}+U^{n},\left[Y\right]^{n}+V^{n}}\left(x^{n},y^{n}\right)}{e^{\epsilon_{n}'}-1}.\label{eq:Phat}
\end{align}
 Obviously $\widehat{P}_{X^{n}Y^{n}}\left(x^{n},y^{n}\right)$ is
a distribution. Then $\pi_{XY}^{n}$ can be written as a mixture distribution
\begin{align}
\pi_{XY}^{n}\left(x^{n},y^{n}\right) & =e^{-\epsilon_{n}'}\widetilde{P}_{\left[X\right]^{n}+U^{n},\left[Y\right]^{n}+V^{n}}\left(x^{n},y^{n}\right)\nonumber \\
 & \qquad+\left(1-e^{-\epsilon_{n}'}\right)\widehat{P}_{X^{n}Y^{n}}\left(x^{n},y^{n}\right).
\end{align}
Furthermore, by \eqref{eq:-51}, we have 
\begin{align}
\widehat{P}_{X^{n}Y^{n}}\left(x^{n},y^{n}\right) & \geq\frac{e^{\delta_{n}}-1}{e^{\epsilon_{n}'}-1}\widetilde{P}_{\left[X\right]^{n}+U^{n},\left[Y\right]^{n}+V^{n}}\left(x^{n},y^{n}\right)\\
 & =\frac{e^{\delta_{n}}-1}{e^{\epsilon_{n}'}-1}\frac{\widetilde{P}_{\left[X\right]^{n}\left[Y\right]^{n}}\left(\left[x\right]^{n},\left[y\right]^{n}\right)}{\Delta_{n}^{2}}.
\end{align}

Now we partition the space $\mathbb{R}^{2n}$ into a finite number
of subregions so that we can apply dyadic decomposition schemes to
each subregion.  Specifically, partition the whole space $\mathbb{R}^{2n}$
into $3^{2n}$ subregions by $2n$ hyperplanes $x_{i}=\pm\sqrt{n\left(1+\epsilon\right)}$
and $y_{i}=\pm\sqrt{n\left(1+\epsilon\right)},1\le i\le n$. These
subregions can be expressed as $I_{1}\times I_{2}\times...\times I_{2n}$,
where $I_{i}\in\left\{ \mathcal{L}_{\epsilon,n}^{-},\mathcal{L}_{\epsilon,n},\mathcal{L}_{\epsilon,n}^{+}\right\} ,1\le i\le2n$
with $\mathcal{L}_{\epsilon,n}^{-}:=(-\infty,-\sqrt{n\left(1+\epsilon\right)})$
and $\mathcal{L}_{\epsilon,n}^{+}:=(\sqrt{n\left(1+\epsilon\right)},+\infty)$.
For brevity, we denote these subregions by $\mathrm{R}_{0},\mathrm{R}_{1},...,\mathrm{R}_{3^{2n}-1}$,
where $\mathrm{R}_{0}:=\mathcal{L}_{\epsilon,n}^{2n}$ and $\mathrm{R}_{1},\mathrm{R}_{2},...,\mathrm{R}_{3^{2n}-1}$
denote the remaining subregions.  For $\mathrm{R}_{k},0\le k\le3^{2n}-1$,
we use $I_{i}^{(k)}\in\left\{ \mathcal{L}_{\epsilon,n}^{-},\mathcal{L}_{\epsilon,n},\mathcal{L}_{\epsilon,n}^{+}\right\} ,1\le i\le2n$
to denote the $i$th component of $\mathrm{R}_{k}$. That is, $\mathrm{R}_{k}=I_{1}^{(k)}\times I_{2}^{(k)}\times...\times I_{2n}^{(k)}$.
Furthermore, observe that $\widetilde{P}_{\left[X\right]^{n}+U^{n},\left[Y\right]^{n}+V^{n}}$
is supported on $\mathrm{R}_{0}$. Hence for $1\le k\le3^{2n}-1$,
$\widehat{P}(\cdot|\mathrm{R}_{k})=\pi_{XY}^{n}(\cdot|\mathrm{R}_{k})$.
This implies that $\widehat{P}(x^{n},y^{n}|\mathrm{R}_{k})=\prod_{i=1}^{n}\pi_{XY}(x_{i},y_{i}|I_{i}^{(k)}\times I_{n+i}^{(k)})$,
i.e., $\left(X_{i},Y_{i}\right),1\le i\le n$ are i.i.d. under the
distribution $\widehat{P}(\cdot|\mathrm{R}_{k})$ for $1\le k\le3^{2n}-1$.

Next we derive upper bounds on $T_{\mathrm{Exact}}(\widehat{P}_{X^{n}Y^{n}|\left(X^{n},Y^{n}\right)\in\mathrm{R}_{k}})$
for $0\le k\le3^{2n}-1$, by using dyadic decomposition schemes proposed
in \cite{li2017distributed}.

\subsection{Dyadic Decomposition Schemes for $\widehat{P}(\cdot|\mathrm{R}_{k})$,
$0\le k\le3^{2n}-1$}

We first consider $k=0$. Denote $I_{\Delta_{n}}^{2n}$ as a $2n$-cube
in $\mathrm{R}_{0}$
\begin{equation}
I_{\Delta_{n}}^{2n}:=\left(\left[x\right]^{n},\left[y\right]^{n}\right)+[0,\Delta_{n}]^{2n}\subseteq\mathrm{R}_{0}\label{eq:-123}
\end{equation}
for $\frac{\left[x\right]_{i}}{\Delta_{n}},\frac{\left[y\right]_{i}}{\Delta_{n}}\in\left[-\frac{\sqrt{n\left(1+\epsilon\right)}}{\Delta_{n}}:\frac{\sqrt{n\left(1+\epsilon\right)}}{\Delta_{n}}-1\right],1\le i\le n$.
By derivations similar to \eqref{eq:-47}-\eqref{eq:-90}, we obtain
that for a $2n$-cube $I_{\Delta_{n}}^{2n}\subseteq\mathrm{R}_{0}$
and for the distribution $\widehat{P}_{X^{n}Y^{n}}$, 
\begin{align}
 & I_{\widehat{P}}\left(X^{i};X_{i+1}^{n}Y^{n}|\left(X^{n},Y^{n}\right)\in I_{\Delta_{n}}^{2n}\right)\nonumber \\
 & \leq\frac{4\left(n\Delta_{n}L_{\epsilon,n}-\log P_{V}(1)+\epsilon_{n}\right)}{\delta_{n}}+o\left(1\right).
\end{align}
for every $1\le i\le n$, where $o\left(1\right)$ denotes a term
tending to zero as $\epsilon_{n}',\epsilon_{n},n\Delta_{n}L_{\epsilon,n},\delta_{n}\to0$.

By Lemma \ref{lem:logconcave}, $\widehat{P}_{X^{n}Y^{n}}\left(\cdot|I_{\Delta_{n}}^{2n}\right)$
is log-concave. On the other hand, for a log-concave distribution
$\pi_{Z^{m}}$, the dyadic decomposition scheme in \cite{li2017distributed}
realizes exactly generating $Z^{m}$ in a distributed way (with $Z_{i}$
realized at the $i$th terminal, $1\le i\le m$) as long as the rate
of common randomness $R\ge I^{(D)}\left(Z^{m}\right)+m^{2}+9\left(\log2\right)m\log m$
bits/symbol, where the dual total correlation 
\begin{align}
I^{(D)}\left(Z^{m}\right) & :=h(Z^{m})-\sum_{i=1}^{m}h\left(Z_{i}|Z^{i-1}Z_{i+1}^{m}\right)\\
 & =\sum_{i=1}^{m}h\left(Z_{i}|Z^{i-1}\right)-\sum_{i=1}^{n}h\left(Z_{i}|Z^{i-1}Z_{i+1}^{m}\right)\\
 & =\sum_{i=1}^{m}I\left(Z_{i};Z_{i+1}^{m}|Z^{i-1}\right)\\
 & \leq\sum_{i=1}^{m}I\left(Z^{i};Z_{i+1}^{m}\right).
\end{align}
That is, the exact common information $T_{\mathrm{Exact}}(\pi_{Z^{m}})\leq I^{(D)}\left(Z^{m}\right)+m^{2}+9\left(\log2\right)m\log m$.

Substituting $\widehat{P}_{X^{n}Y^{n}}\left(\cdot|I_{\Delta_{n}}^{2n}\right)$
into the dual total correlation, we have 
\begin{align}
 & I_{\widehat{P}}^{(D)}\left(X^{n}Y^{n}|\left(X^{n},Y^{n}\right)\in I_{\Delta_{n}}^{2n}\right)\nonumber \\
 & \leq\sum_{i=1}^{n}I_{\widehat{P}}\left(X^{i};X_{i+1}^{n}Y^{n}|\left(X^{n},Y^{n}\right)\in I_{\Delta_{n}}^{2n}\right)\nonumber \\
 & \qquad+\sum_{i=1}^{n}I_{\widehat{P}}\left(Y^{i};Y_{i+1}^{n}X^{n}|\left(X^{n},Y^{n}\right)\in I_{\Delta_{n}}^{2n}\right)\\
 & \leq2n\left(\frac{4\left(n\Delta_{n}L_{\epsilon,n}-\log P_{V}(1)+\epsilon_{n}\right)}{\delta_{n}}+o\left(1\right)\right).\label{eq:-146}
\end{align}

Now we consider the subregions $\mathrm{R}_{k},1\le k\le3^{2n}-1$.
Since $\left(X_{i},Y_{i}\right),1\le i\le n$ are i.i.d. under the
distribution $\widehat{P}(\cdot|\mathrm{R}_{k})$, we have 
\begin{align}
 & I_{\widehat{P}}\left(X_{i};Y_{i}|\left(X^{n},Y^{n}\right)\in\mathrm{R}_{k}\right)\nonumber \\
 & =I_{\pi}\left(X_{i};Y_{i}|\left(X_{i},Y_{i}\right)\in I_{i}^{(k)}\times I_{n+i}^{(k)}\right).
\end{align}
By Lemma \ref{lem:conditionalMI}, we further have 
\begin{equation}
I_{\widehat{P}}\left(X_{i};Y_{i}|\left(X^{n},Y^{n}\right)\in\mathrm{R}_{k}\right)\leq\Upsilon_{\pi}\left(I_{i}^{(k)},I_{n+i}^{(k)}\right).
\end{equation}

For $m=2$, the dyadic decomposition scheme in \cite{li2017distributed}
realizes exactly generating $Z^{2}$ in a distributed way as long
as the rate $R\ge I\left(Z_{1};Z_{2}\right)+24\log2$ nats/symbol.
Applying this to the distribution $\widehat{P}_{X_{i}Y_{i}|\left(X^{n},Y^{n}\right)\in\mathrm{R}_{k}}$,
we have that the exact common information 
\begin{align}
 & T_{\mathrm{Exact}}(\widehat{P}_{X_{i}Y_{i}|\left(X^{n},Y^{n}\right)\in\mathrm{R}_{k}})\nonumber \\
 & \leq I_{\widehat{P}}\left(X_{i};Y_{i}|\left(X^{n},Y^{n}\right)\in\mathrm{R}_{k}\right)+24\log2\\
 & \leq\Upsilon_{\pi}\left(I_{i}^{(k)},I_{n+i}^{(k)}\right)+24\log2.
\end{align}
Since $\widehat{P}_{X^{n}Y^{n}|\left(X^{n},Y^{n}\right)\in\mathrm{R}_{k}}$
is a product distribution, we have for $1\le k\le3^{2n}-1$,
\begin{align}
 & T_{\mathrm{Exact}}(\widehat{P}_{X^{n}Y^{n}|\left(X^{n},Y^{n}\right)\in\mathrm{R}_{k}})\nonumber \\
 & \leq\frac{1}{n}\sum_{i=1}^{n}I_{\widehat{P}}\left(X_{i};Y_{i}|\left(X^{n},Y^{n}\right)\in\mathrm{R}_{k}\right)+24\log2\\
 & \leq\frac{1}{n}\sum_{i=1}^{n}\Upsilon_{\pi}\left(I_{i}^{(k)},I_{n+i}^{(k)}\right)+24\log2.\label{eq:-147}
\end{align}

\subsection{\label{subsec:Exact-Synthesis-Scheme}Exact Synthesis Scheme for
$\pi_{XY}$}

Now we construct an exact synthesis scheme for the distribution $\pi_{XY}$,
which is similar as that in Appendix \ref{sec:equivalence}. Our scheme
is a mixture of the dyadic decomposition schemes above and the modified
fixed-length $\infty$-R\'enyi code $\left(\widetilde{P}_{M},\widetilde{P}_{X^{n}|M},\widetilde{P}_{Y^{n}|M}\right)$
constructed in Subsection \ref{subsec:A-Modified-Version}. The encoder
first generates a Bernoulli random variable $U$ with $P_{U}(1)=e^{-\epsilon_{n}'}$,
compresses it with $1$ bit, and transmits it to the two generators.
If $U=1$, then the encoder and two generators use the modified $\infty$-R\'enyi
code $\left(\widetilde{P}_{M},\widetilde{P}_{X^{n}|M},\widetilde{P}_{Y^{n}|M}\right)$
with rate $R$ to generate $\widetilde{P}_{X^{n}Y^{n}}$. Then by
quantizing $\left(X^{n},Y^{n}\right)$ and adding uniform random variables
to them, the generators obtain $\widetilde{P}_{\left[X\right]^{n}+U^{n},\left[Y\right]^{n}+V^{n}}$.
If $U=0$, then the encoder generates $\left(X^{n},Y^{n}\right)\sim\widehat{P}_{X^{n}Y^{n}}$,
uses $\frac{1}{n}\log\left(3^{2n}\right)+\frac{1}{n}\log\left(\frac{\sqrt{n\left(1+\epsilon\right)}}{\Delta_{n}}\right)^{n}$
rate to encode the index of the subregion $\mathrm{R}_{k}$ and the
$2n$-cube (if $\left(X^{n},Y^{n}\right)\in\mathrm{R}_{0}$) that
$\left(X^{n},Y^{n}\right)$ belongs to, and uses the dyadic decomposition
scheme in \cite{li2017distributed} to generate $\widehat{P}_{X^{n}Y^{n}|\left(X^{n},Y^{n}\right)\in I_{\Delta_{n}}^{2n}}$
with rate $I_{\widehat{P}}^{(D)}\left(X^{n}Y^{n}|\left(X^{n},Y^{n}\right)\in I_{\Delta_{n}}^{2n}\right)+4n^{2}+18\left(\log2\right)n\log\left(2n\right)$
if $\left(X^{n},Y^{n}\right)$ belongs to some $2n$-cube $I_{\Delta_{n}}^{2n}\subseteq\mathrm{R}_{0}$;
to generate $\widehat{P}_{X^{n}Y^{n}|\left(X^{n},Y^{n}\right)\in\mathrm{R}_{k}}$
with rate $T_{\mathrm{Exact}}(\widehat{P}_{X^{n}Y^{n}|\left(X^{n},Y^{n}\right)\in\mathrm{R}_{k}})$
if $\left(X^{n},Y^{n}\right)$ belongs to some subregion $\mathrm{R}_{k}$
for $1\le k\le3^{2n}-1$. The distribution generated by such a mixed
code is $e^{-\epsilon_{n}'}\widetilde{P}_{\left[X\right]^{n}+U^{n},\left[Y\right]^{n}+V^{n}}\left(x^{n},y^{n}\right)+\left(1-e^{-\epsilon_{n}'}\right)\widehat{P}_{X^{n}Y^{n}}\left(x^{n},y^{n}\right)$,
i.e., $\pi_{XY}^{n}\left(x^{n},y^{n}\right)$. The total code rate
is no larger than \eqref{eq:-77}-\eqref{eq:-34} (given on page \pageref{eq:-77}),
where the sum $\sum_{I_{\Delta_{n}}^{2n}\subseteq\mathrm{R}_{0}}$
is taken over all $2n$-cubes $I_{\Delta_{n}}^{2n}\subseteq\mathrm{R}_{0}$
(see \eqref{eq:-123}); \eqref{eq:-77} follows from \eqref{eq:-146},
\eqref{eq:-147}, and the fact that $\pi_{XY}^{n}\left(\mathrm{R}_{k}\right)=\left(1-e^{-\epsilon_{n}'}\right)\widehat{P}_{X^{n}Y^{n}}\left(\mathrm{R}_{k}\right)$
(since $\pi_{XY}^{n}\left(x^{n},y^{n}\right)=\left(1-e^{-\epsilon_{n}'}\right)\widehat{P}_{X^{n}Y^{n}}\left(x^{n},y^{n}\right)$
for $\left(x^{n},y^{n}\right)\notin\mathcal{L}_{\epsilon,n}^{2n}$;
see \eqref{eq:Phat}); and \eqref{eq:-34} follows since on one hand,
\begin{align}
 & \sum_{k=1}^{3^{2n}-1}\pi_{XY}^{n}\left(\mathrm{R}_{k}\right)\left\{ \frac{1}{n}\sum_{i=1}^{n}\Upsilon_{\pi}\left(I_{i}^{(k)},I_{n+i}^{(k)}\right)+24\log2\right\} \nonumber \\
 & =\sum_{k=1}^{3^{2n}-1}\pi_{XY}^{n}\left(\mathrm{R}_{k}\right)\left\{ \Upsilon_{\pi}\left(I_{1}^{(k)},I_{n+1}^{(k)}\right)+24\log2\right\} \label{eq:-148}\\
 & =\sum_{k=1}^{3^{2n}-1}\pi_{XY}^{n}\left(\mathrm{R}_{k}\right)\nonumber \\
 & \qquad\times\biggl(\Upsilon_{\pi}\left(I_{1}^{(k)},I_{n+1}^{(k)}\right)1\left\{ \left(I_{1}^{(k)},I_{n+1}^{(k)}\right)\neq\left(\mathcal{L}_{\epsilon,n},\mathcal{L}_{\epsilon,n}\right)\right\} \nonumber \\
 & \qquad+\Upsilon_{\pi}\left(I_{1}^{(k)},I_{n+1}^{(k)}\right)1\left\{ \left(I_{1}^{(k)},I_{n+1}^{(k)}\right)=\left(\mathcal{L}_{\epsilon,n},\mathcal{L}_{\epsilon,n}\right)\right\} \nonumber \\
 & \qquad+24\log2\biggr)\\
 & \leq\sum_{k=1}^{3^{2n}-1}\pi_{XY}^{n}\left(\mathrm{R}_{k}\right)\nonumber \\
 & \qquad\times\biggl(\Upsilon_{\pi}\left(I_{1}^{(k)},I_{n+1}^{(k)}\right)1\left\{ \left(I_{1}^{(k)},I_{n+1}^{(k)}\right)\neq\left(\mathcal{L}_{\epsilon,n},\mathcal{L}_{\epsilon,n}\right)\right\} \nonumber \\
 & \qquad+\frac{I_{\pi}\left(X;Y\right)+o(1)}{\pi_{XY}(\mathcal{L}_{\epsilon,n}\times\mathcal{L}_{\epsilon,n})}+24\log2\biggr)\\
 & =\sum_{k=1}^{3^{2n}-1}\pi_{XY}^{n}\left(\mathrm{R}_{k}\right)\Upsilon_{\pi}\left(I_{1}^{(k)},I_{n+1}^{(k)}\right)\nonumber \\
 & \qquad\times1\left\{ \left(I_{1}^{(k)},I_{n+1}^{(k)}\right)\neq\left(\mathcal{L}_{\epsilon,n},\mathcal{L}_{\epsilon,n}\right)\right\} +o(1)\label{eq:-50}
\end{align}
and on the other hand, 
\begin{align}
 & \sum_{k=1}^{3^{2n}-1}\pi_{XY}^{n}\left(\mathrm{R}_{k}\right)\Upsilon_{\pi}\left(I_{1}^{(k)},I_{n+1}^{(k)}\right)\nonumber \\
 & \qquad\times1\left\{ \left(I_{1}^{(k)},I_{n+1}^{(k)}\right)\neq\left(\mathcal{L}_{\epsilon,n},\mathcal{L}_{\epsilon,n}\right)\right\} \nonumber \\
 & =\sum_{k=0}^{3^{2n}-1}\pi_{XY}^{n}\left(\mathrm{R}_{k}\right)\Upsilon_{\pi}\left(I_{1}^{(k)},I_{n+1}^{(k)}\right)\nonumber \\
 & \qquad\times1\left\{ \left(I_{1}^{(k)},I_{n+1}^{(k)}\right)\neq\left(\mathcal{L}_{\epsilon,n},\mathcal{L}_{\epsilon,n}\right)\right\} \label{eq:-149}\\
 & =\sum_{x^{n},y^{n}}\pi_{XY}^{n}\left(x^{n},y^{n}\right)\sum_{I^{2n}\in\left\{ \mathcal{L}_{\epsilon,n}^{-},\mathcal{L}_{\epsilon,n},\mathcal{L}_{\epsilon,n}^{+}\right\} ^{2n}}\nonumber \\
 & \qquad1\left\{ \left(x^{n},y^{n}\right)\in I^{2n}\right\} \nonumber \\
 & \qquad\times\Upsilon_{\pi}\left(I_{1},I_{n+1}\right)1\left\{ \left(I_{1},I_{n+1}\right)\neq\left(\mathcal{L}_{\epsilon,n},\mathcal{L}_{\epsilon,n}\right)\right\} \label{eq:-41}\\
 & =\sum_{x_{1},y_{1}}\pi_{XY}\left(x_{1},y_{1}\right)\sum_{\substack{I_{1},I_{n+1}\in\left\{ \mathcal{L}_{\epsilon,n}^{-},\mathcal{L}_{\epsilon,n},\mathcal{L}_{\epsilon,n}^{+}\right\} ,\\
\left(I_{1},I_{n+1}\right)\neq\left(\mathcal{L}_{\epsilon,n},\mathcal{L}_{\epsilon,n}\right)
}
}\nonumber \\
 & \qquad1\left\{ \left(x_{1},y_{1}\right)\in I_{1}\times I_{n+1}\right\} \Upsilon_{\pi}\left(I_{1},I_{n+1}\right)\\
 & =\sum_{\substack{I_{1},I_{n+1}\in\left\{ \mathcal{L}_{\epsilon,n}^{-},\mathcal{L}_{\epsilon,n},\mathcal{L}_{\epsilon,n}^{+}\right\} ,\\
\left(I_{1},I_{n+1}\right)\neq\left(\mathcal{L}_{\epsilon,n},\mathcal{L}_{\epsilon,n}\right)
}
}\pi_{XY}\left(I_{1}\times I_{n+1}\right)\Upsilon_{\pi}\left(I_{1},I_{n+1}\right)\\
 & =o(1).
\end{align}
Here \eqref{eq:-148} follows by symmetry: $\Upsilon_{\pi}\left(I_{1}^{(k)},I_{n+1}^{(k)}\right)=\Upsilon_{\pi}\left(I_{i}^{(k)},I_{n+i}^{(k)}\right)$
for all $i$; \eqref{eq:-50} follows since $\sum_{k=1}^{3^{2n}-1}\pi_{XY}^{n}\left(\mathrm{R}_{k}\right)=1-\pi_{XY}^{n}\left(\mathrm{R}_{0}\right)=o(1)$;
and \eqref{eq:-149} follows since for $\mathrm{R}_{0}$, $\left(I_{1}^{(k)},I_{n+1}^{(k)}\right)=\left(\mathcal{L}_{\epsilon,n},\mathcal{L}_{\epsilon,n}\right)$.
\begin{figure*}
\begin{align}
 & \frac{1}{n}+e^{-\epsilon_{n}'}R+(1-e^{-\epsilon_{n}'})\Biggl\{\frac{1}{n}\log\left(3^{2n}\right)+\frac{1}{n}\log\left(\frac{\sqrt{n\left(1+\epsilon\right)}}{\Delta_{n}}\right)^{n}\nonumber \\
 & \quad+\sum_{I_{\Delta_{n}}^{2n}\subseteq\mathrm{R}_{0}}\widehat{P}_{X^{n}Y^{n}}\left(I_{\Delta_{n}}^{2n}\right)\left(I_{\widehat{P}}^{(D)}\left(X^{n}Y^{n}|\left(X^{n},Y^{n}\right)\in I_{\Delta_{n}}^{2n}\right)+4n+18\left(\log2\right)\log\left(2n\right)\right)\nonumber \\
 & \quad+\sum_{k=1}^{3^{2n}-1}\widehat{P}_{X^{n}Y^{n}}\left(\mathrm{R}_{k}\right)T_{\mathrm{Exact}}(\widehat{P}_{X^{n}Y^{n}|\left(X^{n},Y^{n}\right)\in\mathrm{R}_{k}})\Biggr\}\nonumber \\
 & \leq\frac{1}{n}+e^{-\epsilon_{n}'}R+(1-e^{-\epsilon_{n}'})\left\{ \log\left(\frac{9\sqrt{n\left(1+\epsilon\right)}}{\Delta_{n}}\right)+\frac{8\left(n\Delta_{n}L_{\epsilon,n}-\log P_{V}(1)+\epsilon_{n}\right)}{\delta_{n}}+o\left(1\right)+4n+18\left(\log2\right)\log\left(2n\right)\right\} \nonumber \\
 & \quad+\sum_{k=1}^{3^{2n}-1}\pi_{XY}^{n}\left(\mathrm{R}_{k}\right)\left\{ \frac{1}{n}\sum_{i=1}^{n}\Upsilon_{\pi}\left(I_{i}^{(k)},I_{n+i}^{(k)}\right)+24\log2\right\} \label{eq:-77}\\
 & \sim\frac{1}{n}+e^{-\epsilon_{n}'}R+(1-e^{-\epsilon_{n}'})\left\{ \log\left(\frac{\sqrt{n\left(1+\epsilon\right)}}{\Delta_{n}}\right)+\frac{8\left(n\Delta_{n}L_{\epsilon,n}-\log P_{V}(1)+\epsilon_{n}\right)}{\delta_{n}}+4n\right\} +o\left(1\right).\label{eq:-34}
\end{align}

\hrulefill{}
\end{figure*}

Observe that $\pi_{V}(0)\to0$ exponentially fast, and by the data
processing inequality, $P_{V}(0)\leq\pi_{V}(0)e^{\epsilon_{n}}\to0$
exponentially fast. Hence if $n\Delta_{n}L_{\epsilon,n},\delta_{n}\to0$,
then $\epsilon_{n}'$ (defined in \eqref{eq:-52}) satisfies $\epsilon_{n}'\to0$
as $n\to\infty$. On the other hand, 
\begin{align}
H_{\widehat{P}}(V) & \leq\log2.
\end{align}
Hence to ensure \eqref{eq:-34} converges to $R$, we only require
\begin{equation}
n\Delta_{n}L_{\epsilon,n},\delta_{n}\to0
\end{equation}
and
\begin{align}
 & \left(1-e^{-\epsilon_{n}'}\right)\Biggl(\log\left(\frac{\sqrt{n\left(1+\epsilon\right)}}{\Delta_{n}}\right)\nonumber \\
 & \qquad+\frac{8\left(n\Delta_{n}L_{\epsilon,n}-\log P_{V}(1)+\epsilon_{n}\right)}{\delta_{n}}+4n\Biggr)\to0.\label{eq:-58}
\end{align}
Note that \eqref{eq:-58} is equivalent to \eqref{eq:-177} (given
on page \pageref{eq:-177}).
\begin{figure*}
\begin{align}
 & \left(1-e^{-\left(n\Delta_{n}L_{\epsilon,n}-\log P_{V}(1)+\epsilon_{n}+\delta_{n}\right)}\right)\left(\log\left(\frac{\sqrt{n\left(1+\epsilon\right)}}{\Delta_{n}}\right)+\frac{8\left(n\Delta_{n}L_{\epsilon,n}-\log P_{V}(1)+\epsilon_{n}\right)}{\delta_{n}}+4n\right)\nonumber \\
 & \sim\left(n\Delta_{n}L_{\epsilon,n}-\log P_{V}(1)+\epsilon_{n}+\delta_{n}\right)\left(4n-\log\Delta_{n}+\frac{8\left(n\Delta_{n}L_{\epsilon,n}-\log P_{V}(1)+\epsilon_{n}\right)}{\delta_{n}}\right)\to0.\label{eq:-177}
\end{align}

\hrulefill{}
\end{figure*}

Choose $\delta_{n}=n\Delta_{n}L_{\epsilon,n}-\log P_{V}(1)+\epsilon_{n}$,
then we only require 
\begin{equation}
\left(n\Delta_{n}L_{\epsilon,n}-\log P_{V}(1)+\epsilon_{n}\right)\left(4n-\log\Delta_{n}\right)\to0.
\end{equation}

Observe that $\pi_{V}(0)\to0$ exponentially fast, and by the data
processing inequality, $-\log P_{V}(1)=-\log\left(1-P_{V}(0)\right)\sim P_{V}(0)\to0$
exponentially fast. Choose $\Delta_{n}=\frac{1}{\left(nL_{\epsilon,n}\right)^{3}}$,
then $n\Delta_{n}L_{\epsilon,n}\to0$ and 
\begin{align}
 & \left(\frac{1}{\left(nL_{\epsilon,n}\right)^{2}}-\log P_{V}(1)+\epsilon_{n}\right)\left(4n+3\log n+3\log L_{\epsilon,n}\right)\nonumber \\
 & \sim\left(\frac{1}{\left(nL_{\epsilon,n}\right)^{2}}+P_{V}(0)+\epsilon_{n}\right)\left(4n+\log L_{\epsilon,n}\right)\\
 & =\left(P_{V}(0)+\epsilon_{n}\right)\left(4n+\log L_{\epsilon,n}\right)+o(1)\\
 & =P_{V}(0)\log L_{\epsilon,n}+\epsilon_{n}\left(4n+\log L_{\epsilon,n}\right)+o(1).
\end{align}
Hence we only require 
\begin{align}
 & P_{V}(0)\log L_{\epsilon,n}\to0\\
 & \epsilon_{n}\left(n+\log L_{\epsilon,n}\right)\to0.
\end{align}
That is, $\epsilon_{n}=o\left(\frac{1}{n+\log L_{\epsilon,n}}\right)$
and $\log L_{\epsilon,n}$ is sub-exponentially growing in $n$. These
are the assumptions given in the lemma. Hence the proof is complete.

\section{\label{sec:Gaussian}Proof of Theorem \ref{thm:Gaussian}}

In this section, we extend the proof in Appendix \ref{subsec:achievability}
to the Gaussian case by combining it with discretization techniques.

Define $Q_{W}=\mathcal{N}(0,\rho),Q_{X|W}(\cdot|w)=\mathcal{N}(w,1-\rho),Q_{Y|W}(\cdot|w)=\mathcal{N}(w,1-\rho)$.
Then $Q_{XY}=\pi_{XY}$. For $\epsilon>0$, we define the distributions
\begin{align}
P_{W^{n}}\left(w^{n}\right) & \propto Q_{W}^{n}\left(w^{n}\right)1\left\{ w^{n}\in\mathcal{A}_{\frac{\epsilon}{2}}^{\left(n\right)}\left(Q_{W}\right)\right\} ,\\
P_{X^{n}|W^{n}}\left(x^{n}|w^{n}\right) & \propto Q_{X|W}^{n}\left(x^{n}|w^{n}\right)\nonumber \\
 & \qquad\times1\left\{ x^{n}\in\mathcal{A}_{\epsilon}^{\left(n\right)}\left(Q_{WX}|w^{n}\right)\right\} ,\\
P_{Y^{n}|W^{n}}\left(y^{n}|w^{n}\right) & \propto Q_{Y|W}^{n}\left(y^{n}|w^{n}\right)\nonumber \\
 & \qquad\times1\left\{ y^{n}\in\mathcal{A}_{\epsilon}^{\left(n\right)}\left(Q_{WY}|w^{n}\right)\right\} .
\end{align}
According to the definition of weakly typical sets, 
\begin{equation}
\mathcal{A}_{\frac{\epsilon}{2}}^{\left(n\right)}\left(Q_{W}\right)=\left\{ w^{n}\in\mathbb{R}^{n}:\left|\frac{\left\Vert w^{n}\right\Vert ^{2}}{n\rho}-1\right|\leq\epsilon\right\} 
\end{equation}
and 
\begin{align}
 & \mathcal{A}_{\epsilon}^{\left(n\right)}\left(Q_{WX}\right)=\mathcal{A}_{\epsilon}^{\left(n\right)}\left(Q_{WY}\right)\\
 & =\left\{ \left(w^{n},x^{n}\right)\in\mathbb{R}^{2n}:\begin{array}{c}
\left|\frac{\left\Vert w^{n}\right\Vert ^{2}}{n\rho}-1\right|\leq2\epsilon\\
\left|\frac{\left\Vert x^{n}\right\Vert ^{2}}{n}-1\right|\leq2\epsilon\\
\left|\frac{\left\Vert w^{n}\right\Vert ^{2}}{n\rho}+\frac{\left\Vert x^{n}-w^{n}\right\Vert ^{2}}{n\left(1-\rho\right)}-2\right|\leq2\epsilon
\end{array}\right\} .
\end{align}
Hence for $\left(w^{n},x^{n}\right)\in\mathcal{A}_{\epsilon}^{\left(n\right)}\left(Q_{WX}\right)$,
\begin{equation}
\left|\frac{\left\Vert x^{n}-w^{n}\right\Vert ^{2}}{n\left(1-\rho\right)}-1\right|\leq4\epsilon
\end{equation}
and 
\begin{align}
 & \left|\frac{1}{n}(x^{n}-w^{n})^{\top}w^{n}\right|\nonumber \\
 & =\left|\frac{1}{2n}\left(\left\Vert w^{n}\right\Vert ^{2}+\left\Vert x^{n}-w^{n}\right\Vert ^{2}-\left\Vert x^{n}\right\Vert ^{2}\right)\right|\\
 & \leq\frac{1}{2}\left|\rho\left(1+2\epsilon\right)+\left(1-\rho\right)\left(1+4\epsilon\right)-\left(1-2\epsilon\right)\right|\\
 & =\left(3-\rho\right)\epsilon.\label{eq:-91}
\end{align}

Define 
\begin{align}
 & \delta_{0,n}:=1-Q_{W}^{n}\left(\mathcal{A}_{\frac{\epsilon}{2}}^{\left(n\right)}\left(Q_{W}\right)\right)\\
 & \delta_{1,n}:=1-\inf_{w^{n}\in\mathcal{A}_{\frac{\epsilon}{2}}^{\left(n\right)}\left(Q_{W}\right)}Q_{X|W}^{n}\left(\mathcal{A}_{\epsilon}^{\left(n\right)}\left(Q_{WX}|w^{n}\right)|w^{n}\right)\\
 & \delta_{2,n}:=1-\inf_{w^{n}\in\mathcal{A}_{\frac{\epsilon}{2}}^{\left(n\right)}\left(Q_{W}\right)}Q_{Y|W}^{n}\left(\mathcal{A}_{\epsilon}^{\left(n\right)}\left(Q_{WY}|w^{n}\right)|w^{n}\right).
\end{align}
Then $\delta_{0,n},\delta_{1,n},\delta_{2,n}\to0$ exponentially
fast, as shown in the following lemma.
\begin{lem}[Gaussian Typicality Lemma]
\label{lem:-exponentially-fast.}  $\delta_{0,n},\delta_{1,n},\delta_{2,n}\to0$
exponentially fast.
\end{lem}
\begin{IEEEproof}[Proof of Lemma \ref{lem:-exponentially-fast.}]
 By large deviation theory, we know that $\delta_{0,n}\to0$ exponentially
fast. Next we prove $\delta_{1,n}\to0$ exponentially fast. (That
$\delta_{2,n}\to0$ exponentially fast follows by symmetry.)

Under the condition $w^{n}\in\mathcal{A}_{\frac{\epsilon}{2}}^{\left(n\right)}\left(Q_{W}\right)$,
\begin{equation}
\left|\frac{\left\Vert w^{n}\right\Vert ^{2}}{n\rho}-1\right|\leq\epsilon
\end{equation}
is satisfied automatically. Denote $Z^{n}=X^{n}-w^{n}$. Then $Z_{i}$'s
are i.i.d., and $Z_{i}\sim Q_{Z}=\mathcal{N}(0,1-\rho)$. By large
deviation theory, 
\begin{equation}
\mathbb{P}_{Z^{n}\sim Q_{Z}^{n}}\left(\left|\frac{\left\Vert w^{n}\right\Vert ^{2}}{n\rho}+\frac{\left\Vert Z^{n}\right\Vert ^{2}}{n\left(1-\rho\right)}-2\right|\leq2\epsilon\right)\to1
\end{equation}
exponentially fast.

Now we consider the condition $\left|\frac{\left\Vert X^{n}\right\Vert ^{2}}{n}-1\right|\leq2\epsilon$,
which is equivalent to $\left|\frac{\left\Vert w^{n}+Z^{n}\right\Vert ^{2}}{n}-1\right|\leq2\epsilon$.
Observe that 
\begin{equation}
\left\Vert w^{n}+Z^{n}\right\Vert ^{2}=\left\Vert w^{n}\right\Vert ^{2}+\left\Vert Z^{n}\right\Vert ^{2}+2\sum_{i=1}^{n}w_{i}Z_{i}.
\end{equation}
By the large deviation theory, for $\epsilon'>0$, 
\begin{equation}
\mathbb{P}_{Z^{n}\sim Q_{Z}^{n}}\left(\left|\frac{\left\Vert Z^{n}\right\Vert ^{2}}{n\left(1-\rho\right)}-1\right|\leq\epsilon'\right)\to1
\end{equation}
exponentially fast. On the other hand, observe that $\frac{1}{n}\sum_{i=1}^{n}w_{i}\text{\ensuremath{Z_{i}}}\sim\mathcal{N}(0,\frac{1}{n^{2}}\left\Vert w^{n}\right\Vert ^{2}\left(1-\rho\right))$.
Hence 
\begin{equation}
\mathbb{P}_{Z^{n}\sim Q_{Z}^{n}}\left(\left|\frac{1}{n}\sum_{i=1}^{n}w_{i}\text{\ensuremath{Z_{i}}}\right|\leq\epsilon'\right)=1-2\mathsf{Q}\left(\frac{n\epsilon'}{\left\Vert w^{n}\right\Vert \sqrt{1-\rho}}\right),
\end{equation}
where $\mathsf{Q}$ is the Q-function for the standard normal distribution.
Since $\mathsf{Q}(x)\leq e^{-\frac{x^{2}}{2}},\,x>0$, we have 
\begin{align}
 & \mathbb{P}_{Z^{n}\sim Q_{Z}^{n}}\left(\left|\frac{1}{n}\sum_{i=1}^{n}w_{i}\text{\ensuremath{Z_{i}}}\right|\leq\epsilon'\right)\nonumber \\
 & \geq1-2\exp\left(-\frac{1}{2}\left(\frac{n\epsilon'}{\left\Vert w^{n}\right\Vert \sqrt{1-\rho}}\right)^{2}\right)\\
 & \geq1-2\exp\left(-\frac{1}{2}\frac{n\epsilon'^{2}}{\rho\left(1-\epsilon\right)\left(1-\rho\right)}\right)\\
 & \to1
\end{align}
exponentially fast. Hence $\inf_{w^{n}\in\mathcal{A}_{\frac{\epsilon}{2}}^{\left(n\right)}\left(Q_{W}\right)}Q_{Z}^{n}\left(\mathcal{B}_{\epsilon'}\left(w^{n}\right)\right)\to1$
exponentially fast, where
\begin{equation}
\mathcal{B}_{\epsilon'}\left(w^{n}\right):=\left\{ z^{n}\in\mathbb{R}^{n}:\begin{array}{c}
\left|\frac{\left\Vert z^{n}\right\Vert ^{2}}{n\left(1-\rho\right)}-1\right|\leq\epsilon'\\
\left|\frac{1}{n}\sum_{i=1}^{n}w_{i}\text{\ensuremath{z_{i}}}\right|\leq\epsilon'
\end{array}\right\} .
\end{equation}

Now we claim that for sufficiently small $\epsilon'$, if $w^{n}\in\mathcal{A}_{\frac{\epsilon}{2}}^{\left(n\right)}\left(Q_{W}\right)$
and $z^{n}=x^{n}-w^{n}\in\mathcal{B}_{\epsilon'}\left(w^{n}\right)$,
then $\left(w^{n},x^{n}\right)\in\mathcal{A}_{\epsilon}^{\left(n\right)}\left(Q_{WX}\right)$.
Since $\inf_{w^{n}\in\mathcal{A}_{\frac{\epsilon}{2}}^{\left(n\right)}\left(Q_{W}\right)}Q_{Z}^{n}\left(\mathcal{B}_{\epsilon'}\left(w^{n}\right)\right)\to1$
exponentially fast, this claim implies that $\delta_{1,n}\to0$ exponentially
fast as well. Hence the rest is to prove this claim.

Observe that for $w^{n}\in\mathcal{A}_{\frac{\epsilon}{2}}^{\left(n\right)}\left(Q_{W}\right)$
and $z^{n}=x^{n}-w^{n}\in\mathcal{B}_{\epsilon'}\left(w^{n}\right)$,
we have 
\begin{align}
 & \left|\frac{\left\Vert x^{n}\right\Vert ^{2}}{n}-1\right|\nonumber \\
 & =\left|\frac{\left\Vert w^{n}+z^{n}\right\Vert ^{2}}{n}-1\right|\\
 & =\left|\frac{\left\Vert w^{n}\right\Vert ^{2}+\left\Vert z^{n}\right\Vert ^{2}+2\sum_{i=1}^{n}w_{i}z_{i}}{n}-1\right|\\
 & \leq\max\Bigl\{\left|\rho\left(1+\epsilon\right)+\left(1-\rho\right)\left(1+\epsilon'\right)+2\epsilon'-1\right|,\nonumber \\
 & \qquad\left|\rho\left(1-\epsilon\right)+\left(1-\rho\right)\left(1-\epsilon'\right)-2\epsilon'-1\right|\Bigr\}\\
 & =\rho\epsilon+\left(1-\rho\right)\epsilon'+2\epsilon',
\end{align}
and 
\begin{align}
 & \left|\frac{\left\Vert w^{n}\right\Vert ^{2}}{n\rho}+\frac{\left\Vert x^{n}-w^{n}\right\Vert ^{2}}{n\left(1-\rho\right)}-2\right|\nonumber \\
 & =\left|\frac{\left\Vert w^{n}\right\Vert ^{2}}{n\rho}+\frac{\left\Vert z^{n}\right\Vert ^{2}}{n\left(1-\rho\right)}-2\right|\\
 & \leq\epsilon+\epsilon'.
\end{align}
Now we choose 
\[
\epsilon'\leq\min\left\{ \epsilon,\frac{\left(2-\rho\right)\epsilon}{3-\rho}\right\} =\frac{\left(2-\rho\right)\epsilon}{3-\rho},
\]
then $\left|\frac{\left\Vert x^{n}\right\Vert ^{2}}{n}-1\right|\leq2\epsilon$
and $\left|\frac{\left\Vert w^{n}\right\Vert ^{2}}{n\rho}+\frac{\left\Vert x^{n}-w^{n}\right\Vert ^{2}}{n\left(1-\rho\right)}-2\right|\leq2\epsilon$.
Hence we complete the proof of the claim above.
\end{IEEEproof}
We set $\mathcal{C}_{n}=\left\{ W^{n}\left(m\right)\right\} _{m\in\calM_{n}}$
with $W^{n}\left(m\right),m\in\calM_{n}$ drawn independently for
different $m$'s and according to the same distribution $P_{W^{n}}$
such that $P_{W^{n}}$. Upon receiving $W^{n}\left(M\right)$, the
two generators respectively use random mappings $P_{X^{n}|W^{n}}$
and $P_{Y^{n}|W^{n}}$ to generate $X^{n}$ and $Y^{n}$. For a sequence
of positive numbers $\left\{ \Delta_{n}\right\} $, we quantize $X^{n}$
and $Y^{n}$ as $\left[X\right]^{n}=\Delta_{n}\left\lfloor \frac{X^{n}}{\Delta_{n}}\right\rfloor $
and $\left[Y\right]^{n}=\Delta_{n}\left\lfloor \frac{Y^{n}}{\Delta_{n}}\right\rfloor $.
Define $\left[\mathcal{A}_{\epsilon}^{(n)}\right]\times\left[\mathcal{A}_{\epsilon}^{(n)}\right]:=\left(\Delta\mathbb{Z}^{n}\cap\mathcal{A}_{\epsilon}^{(n)}\left(\pi_{X}\right)\right)\times\left(\Delta\mathbb{Z}^{n}\cap\mathcal{A}_{\epsilon}^{(n)}\left(\pi_{Y}\right)\right)$.
Define $U^{n},V^{n}\sim\mathrm{Unif}\left(I_{\Delta_{n}}^{n}\right)$
with $I_{\Delta_{n}}^{n}=[0,\Delta_{n}]^{n}$ are mutually independent,
and also independent of $\left[X\right]^{n},\left[Y\right]^{n}$.
For such a code, we have the following Gaussian version of distributed
R\'enyi-covering lemma. 
\begin{lem}[Distributed Gaussian R\'enyi-Covering]
\label{lem:Renyicovering-1} For the random code described above,
if 
\begin{equation}
R>\frac{1}{2}\log\left[\frac{1+\rho}{1-\rho}\right]+\frac{\rho}{1+\rho},\label{eq:-74-1}
\end{equation}
then there exists some $\alpha,\epsilon>0$ and some positive sequence
$\left\{ \Delta_{n}\right\} $ such that 
\begin{align}
 & \mathbb{P}_{\mathcal{C}_{n}}\left(D_{\infty}(P_{\left[X\right]^{n}+U^{n},\left[Y\right]^{n}+V^{n}|\mathcal{C}_{n}}\|\pi_{XY}^{n})\leq e^{-n\alpha}\right)\to1
\end{align}
doubly exponentially fast.
\end{lem}
This lemma implies that there exists a sequence of codebooks $\left\{ c_{n}\right\} $
with rate $R$ such that $D_{\infty}(P_{\left[X\right]^{n}+U^{n},\left[Y\right]^{n}+V^{n}|\mathcal{C}_{n}=c_{n}}\|\pi_{XY}^{n})\leq e^{-n\alpha}$
as long as $R>\frac{1}{2}\log\left[\frac{1+\rho}{1-\rho}\right]+\frac{\rho}{1+\rho}$.
 This completes the proof of $T_{\infty}(\pi_{XY})\le\frac{1}{2}\log\left[\frac{1+\rho}{1-\rho}\right]+\frac{\rho}{1+\rho}$.
Hence what we need to do is to prove Lemma \ref{lem:Renyicovering-1}.
The proof is provided in the following.
\begin{IEEEproof}[Proof of Lemma \ref{lem:Renyicovering-1}]
 Assume $\epsilon>0$ is a number such that 
\begin{equation}
R>\left(1+\epsilon\right)\left(\frac{1}{2}\log\left[\frac{1+\rho}{1-\rho}\right]+\frac{\rho}{1+\rho}\right)+3\epsilon.
\end{equation}
For brevity, in the following we denote $\mathsf{M}=e^{nR}$. According
to the definition of the R\'enyi divergence, we have 
\begin{align}
 & e^{D_{\infty}(P_{\left[X\right]^{n}\left[Y\right]^{n}}\|\pi_{\left[X\right]\left[Y\right]}^{n})}\nonumber \\
 & =\sup_{\left(x^{n},y^{n}\right)\in\left[\mathcal{A}_{\epsilon}^{(n)}\right]\times\left[\mathcal{A}_{\epsilon}^{(n)}\right]}\frac{P_{\left[X\right]^{n}\left[Y\right]^{n}}\left(x^{n},y^{n}\right)}{\pi_{\left[X\right]\left[Y\right]}^{n}\left(x^{n},y^{n}\right)}\\
 & =\sup_{\left(x^{n},y^{n}\right)\in\left[\mathcal{A}_{\epsilon}^{(n)}\right]\times\left[\mathcal{A}_{\epsilon}^{(n)}\right]}\widetilde{g}_{\left[X\right]^{n}\left[Y\right]^{n}|\mathcal{C}_{n}}(x^{n},y^{n}|\mathcal{C}_{n}),\label{eq:-151-1}
\end{align}
where 
\begin{align}
 & \widetilde{g}_{\left[X\right]^{n}\left[Y\right]^{n}|\mathcal{C}_{n}}(x^{n},y^{n}|\mathcal{C}_{n})\nonumber \\
 & :=\sum_{m\in\calM_{n}}\frac{g_{\left[X\right]^{n}\left[Y\right]^{n}|W^{n}}(x^{n},y^{n}|W^{n}(m))}{\mathsf{M}}
\end{align}
with 
\begin{align}
 & g_{\left[X\right]^{n}\left[Y\right]^{n}|W^{n}}(x^{n},y^{n}|w^{n})\nonumber \\
 & :=\frac{P_{\left[X\right]^{n}|W^{n}}\left(x^{n}|w^{n}\right)P_{\left[Y\right]^{n}|W^{n}}\left(y^{n}|w^{n}\right)}{\pi_{\left[X\right]\left[Y\right]}^{n}\left(x^{n},y^{n}\right)}.\label{eq:-59}
\end{align}

By the data processing inequality, 
\begin{align}
 & \sup_{\left(x^{n},y^{n}\right)\in\left[\mathcal{A}_{\epsilon}^{(n)}\right]\times\left[\mathcal{A}_{\epsilon}^{(n)}\right]}g_{\left[X\right]^{n}\left[Y\right]^{n}|W^{n}}(x^{n},y^{n}|w^{n})\nonumber \\
 & \leq\sup_{\left(x^{n},y^{n}\right)\in\mathcal{A}_{\epsilon}^{(n)}\times\mathcal{A}_{\epsilon}^{(n)}}g_{X^{n}Y^{n}|W^{n}}(x^{n},y^{n}|w^{n}).\label{eq:-70}
\end{align}

On the other hand, define 
\begin{align}
\mathcal{A} & :=\Bigl\{\left(x^{n},y^{n}\right):\exists w^{n}\textrm{ s.t. }\left(w^{n},x^{n}\right)\in\mathcal{A}_{\epsilon}^{\left(n\right)}\left(Q_{WX}\right),\nonumber \\
 & \qquad\left(w^{n},y^{n}\right)\in\mathcal{A}_{\epsilon}^{\left(n\right)}\left(Q_{WY}\right)\Bigr\}
\end{align}
and 
\begin{equation}
\delta_{12,n}:=\frac{1}{\left(1-\delta_{1,n}\right)\left(1-\delta_{2,n}\right)}.
\end{equation}
Since Lemma \ref{lem:-exponentially-fast.} shows that $\delta_{1,n},\delta_{2,n}\to0$
exponentially fast, we know that $\delta_{12,n}\to1$ exponentially
fast. Then similar to \eqref{eq:-93}, we can show that for $w^{n}\in\mathcal{A}_{\frac{\epsilon}{2}}^{\left(n\right)}\left(Q_{W}\right)$,
\eqref{eq:-178}-\eqref{eq:-71} (given on page \pageref{eq:-178})
hold, where \eqref{eq:-92} follows from \eqref{eq:-91}. Combining
\eqref{eq:-70} and \eqref{eq:-71}, we obtain 
\begin{align}
\sup_{\left(x^{n},y^{n}\right)\in\left[\mathcal{A}_{\epsilon}^{(n)}\right]\times\left[\mathcal{A}_{\epsilon}^{(n)}\right]}g_{\left[X\right]^{n}\left[Y\right]^{n}|W^{n}}(x^{n},y^{n}|w^{n}) & \leq\beta_{n}.
\end{align}
\begin{figure*}
\begin{align}
 & g_{X^{n}Y^{n}|W^{n}}(x^{n},y^{n}|w^{n})\nonumber \\
 & :=\frac{P_{X^{n}|W^{n}}\left(x^{n}|w^{n}\right)P_{Y^{n}|W^{n}}\left(y^{n}|w^{n}\right)}{\pi_{XY}^{n}\left(x^{n},y^{n}\right)}\label{eq:-178}\\
 & \leq\delta_{12,n}\sup_{\left(x^{n},y^{n}\right)\in\mathcal{A}}e^{4n\epsilon-nh\left(X|W\right)-nh\left(Y|W\right)+n\log2\pi\sqrt{1-\rho^{2}}+\frac{n\left(1+2\epsilon\right)-\rho\left(x^{n}{}^{\top}y^{n}\right)}{1-\rho^{2}}}\\
 & \leq\delta_{12,n}\sup_{\left(x^{n},y^{n}\right)\in\mathcal{A}}e^{4n\epsilon-nh\left(X|W\right)-nh\left(Y|W\right)+n\log2\pi\sqrt{1-\rho^{2}}+\frac{1}{1-\rho^{2}}\left[n\left(1+2\epsilon\right)-\rho\left(\left\Vert w^{n}\right\Vert ^{2}+(x^{n}-w^{n})^{\top}(y^{n}-w^{n})-2n\left(3-\rho\right)\epsilon\right)\right]}\label{eq:-92}\\
 & \leq\delta_{12,n}e^{4n\epsilon-nh\left(X|W\right)-nh\left(Y|W\right)+n\log2\pi\sqrt{1-\rho^{2}}+\frac{1}{1-\rho^{2}}\left[n\left(1+2\epsilon\right)-\rho\left(n\rho\left(1-2\epsilon\right)-n(1-\rho)\left(1+4\epsilon\right)-2n\left(3-\rho\right)\epsilon\right)\right]}\\
 & =\delta_{12,n}e^{\frac{6-10\rho}{1-\rho^{2}}n\epsilon-nh\left(X|W\right)-nh\left(Y|W\right)+n\left(\log2\pi\sqrt{1-\rho^{2}}+\frac{\left(1-\rho\right)\left(1+2\rho\right)}{1-\rho^{2}}\right)}\\
 & =\delta_{12,n}e^{n\left(\frac{1}{2}\log\frac{1+\rho}{1-\rho}+\frac{\rho}{1+\rho}+\frac{6-10\rho}{1-\rho^{2}}\epsilon\right)}\\
 & =:\beta_{n},\label{eq:-71}
\end{align}

\hrulefill{}
\end{figure*}

Continuing \eqref{eq:-151-1}, we get for any $\epsilon'>0$, 
\begin{align}
 & \mathbb{P}_{\mathcal{C}_{n}}\left(e^{D_{\infty}(P_{\left[X\right]^{n}\left[Y\right]^{n}}\|\pi_{\left[X\right]\left[Y\right]}^{n})}\geq1+\epsilon'\right)\nonumber \\
 & =\mathbb{P}_{\mathcal{C}_{n}}\Biggl(\sup_{\left(x^{n},y^{n}\right)\in\left[\mathcal{A}_{\epsilon}^{(n)}\right]\times\left[\mathcal{A}_{\epsilon}^{(n)}\right]}\widetilde{g}_{\left[X\right]^{n}\left[Y\right]^{n}|\mathcal{C}_{n}}(x^{n},y^{n}|\mathcal{C}_{n})\nonumber \\
 & \qquad\geq1+\epsilon'\Biggr)\\
 & \leq\left|\left[\mathcal{A}_{\epsilon}^{(n)}\right]\times\left[\mathcal{A}_{\epsilon}^{(n)}\right]\right|\sup_{\left(x^{n},y^{n}\right)\in\left[\mathcal{A}_{\epsilon}^{(n)}\right]\times\left[\mathcal{A}_{\epsilon}^{(n)}\right]}\nonumber \\
 & \qquad\mathbb{P}_{\mathcal{C}_{n}}\left(\widetilde{g}_{\left[X\right]^{n}\left[Y\right]^{n}|\mathcal{C}_{n}}(x^{n},y^{n}|\mathcal{C}_{n})\geq1+\epsilon'\right),\label{eq:-152-1}
\end{align}
where \eqref{eq:-152-1} follows from the union bound. If the probability
in \eqref{eq:-152-1} vanishes doubly exponentially fast and $\left|\left[\mathcal{A}_{\epsilon}^{(n)}\right]\times\left[\mathcal{A}_{\epsilon}^{(n)}\right]\right|$
is growing much slower, then $\max_{x^{n},y^{n}}\widetilde{g}(x^{n},y^{n}|\mathcal{C}_{n})<1+\epsilon'$
with high probability as $n\to\infty$. To this end, we use similar
techniques used in Appendix \ref{subsec:achievability} to bound the
probability. Define $I_{\Delta}^{n}:=[0,\Delta]^{n}$. Observe that
$g(x^{n},y^{n}|W^{n}(m)),m\in\calM_{n}$ are i.i.d. random variables
with mean $\mu_{\epsilon,n}$ given in \eqref{eq:-179}-\eqref{eq:-180}
(given on page \pageref{eq:-179}) and variance 
\begin{align}
 & \mathrm{Var}_{W^{n}}\left[g_{\left[X\right]^{n}\left[Y\right]^{n}|W^{n}}(x^{n},y^{n}|W^{n})\right]\nonumber \\
 & \leq\mathbb{E}_{W^{n}}\left[g_{\left[X\right]^{n}\left[Y\right]^{n}|W^{n}}(x^{n},y^{n}|W^{n})^{2}\right]\\
 & \leq\beta_{n}\mu_{\epsilon,n}.
\end{align}
\begin{figure*}
\begin{align}
\mu_{\epsilon,n} & :=\mathbb{E}_{W^{n}}\left[g_{\left[X\right]^{n}\left[Y\right]^{n}|W^{n}}(x^{n},y^{n}|w^{n})\right]\label{eq:-179}\\
 & =\int\frac{Q_{W}^{n}\left(w^{n}\right)1\left\{ w^{n}\in\mathcal{A}_{\frac{\epsilon}{2}}^{\left(n\right)}\left(Q_{W}\right)\right\} }{Q_{W}^{n}\left(\mathcal{A}_{\frac{\epsilon}{2}}^{\left(n\right)}\left(Q_{W}\right)\right)}\nonumber \\
 & \qquad\times\frac{\int_{\left[x\right]^{n}+I_{\Delta}^{n}}\frac{Q_{X|W}^{n}\left(x'^{n}|w^{n}\right)1\left\{ x'^{n}\in\mathcal{A}_{\epsilon}^{\left(n\right)}\left(Q_{WX}|w^{n}\right)\right\} }{Q_{X|W}^{n}\left(\mathcal{A}_{\epsilon}^{\left(n\right)}\left(Q_{WX}|w^{n}\right)|w^{n}\right)}\mathrm{d}x'^{n}\int_{\left[y\right]^{n}+I_{\Delta}^{n}}\frac{Q_{Y|W}^{n}\left(y'^{n}|w^{n}\right)1\left\{ y'^{n}\in\mathcal{A}_{\epsilon}^{\left(n\right)}\left(Q_{WY}|w^{n}\right)\right\} }{Q_{Y|W}^{n}\left(\mathcal{A}_{\epsilon}^{\left(n\right)}\left(Q_{WY}|w^{n}\right)|w^{n}\right)}\mathrm{d}y'^{n}}{\int_{\left(\left[x\right]^{n}+I_{\Delta}^{n}\right)\times\left(\left[y\right]^{n}+I_{\Delta}^{n}\right)}\pi_{XY}^{n}\left(x'^{n},y'^{n}\right)\mathrm{d}x'^{n}\mathrm{d}y'^{n}}\mathrm{d}w^{n}\label{eq:-72}\\
 & \leq\sup_{\left(x^{n},y^{n}\right)\in\left(\left[x\right]^{n}+I_{\Delta}^{n}\right)\times\left(\left[y\right]^{n}+I_{\Delta}^{n}\right)}\int\frac{Q_{W}^{n}\left(w^{n}\right)1\left\{ w^{n}\in\mathcal{A}_{\frac{\epsilon}{2}}^{\left(n\right)}\left(Q_{W}\right)\right\} }{Q_{W}^{n}\left(\mathcal{A}_{\frac{\epsilon}{2}}^{\left(n\right)}\left(Q_{W}\right)\right)}\nonumber \\
 & \qquad\times\frac{\frac{Q_{X|W}^{n}\left(x^{n}|w^{n}\right)1\left\{ x^{n}\in\mathcal{A}_{\epsilon}^{\left(n\right)}\left(Q_{WX}|w^{n}\right)\right\} }{Q_{X|W}^{n}\left(\mathcal{A}_{\epsilon}^{\left(n\right)}\left(Q_{WX}|w^{n}\right)|w^{n}\right)}\frac{Q_{Y|W}^{n}\left(y^{n}|w^{n}\right)1\left\{ y^{n}\in\mathcal{A}_{\epsilon}^{\left(n\right)}\left(Q_{WY}|w^{n}\right)\right\} }{Q_{Y|W}^{n}\left(\mathcal{A}_{\epsilon}^{\left(n\right)}\left(Q_{WY}|w^{n}\right)|w^{n}\right)}}{\pi_{XY}^{n}\left(x^{n},y^{n}\right)}\mathrm{d}w^{n}\label{eq:-69}\\
 & \leq\frac{1}{\left(1-\delta_{0,n}\right)\left(1-\delta_{1,n}\right)\left(1-\delta_{2,n}\right)}\label{eq:-68}\\
 & \to1\textrm{ exponentially fast}.\label{eq:-180}
\end{align}

\hrulefill{}
\end{figure*}
Here \eqref{eq:-69} follows by the following inequality. For two
functions $f\left(x\right)\geq0,g\left(x\right)>0$, 
\begin{equation}
\frac{\int f\left(x\right)\mathrm{d}x}{\int g\left(x\right)\mathrm{d}x}\leq\sup_{x}\frac{f\left(x\right)}{g\left(x\right)}.
\end{equation}

Following steps similar to \eqref{eq:-94}-\eqref{eq:-96} (but with
a lower bound on the exponent of $\frac{1}{\left(1-\delta_{0,n}\right)\left(1-\delta_{1,n}\right)\left(1-\delta_{2,n}\right)}-1$
can be obtained in the proof of Lemma \ref{lem:-exponentially-fast.},
which was derived by the large deviation theory, instead of the method
of types), we get that there exists $\epsilon'_{n}\to0$ exponentially
fast such that \eqref{eq:-152-1} with $\epsilon'$ replaced by $\epsilon'_{n}$
converges to zero doubly exponentially fast, as long as $\left|\left[\mathcal{A}_{\epsilon}^{(n)}\right]\times\left[\mathcal{A}_{\epsilon}^{(n)}\right]\right|$
is growing slower than doubly exponentially fast. Hence 
\begin{equation}
\mathbb{P}_{\mathcal{C}_{n}}\left(e^{D_{\infty}(P_{\left[X\right]^{n}\left[Y\right]^{n}}\|\pi_{\left[X\right]\left[Y\right]}^{n})}\geq1+\epsilon'_{n}\right)\to0\label{eq:-98}
\end{equation}
doubly exponentially fast, as long as $\left|\left[\mathcal{A}_{\epsilon}^{(n)}\right]\times\left[\mathcal{A}_{\epsilon}^{(n)}\right]\right|$
is growing slower than doubly exponentially fast. Obviously, \eqref{eq:-98}
implies there exists a codebook such that $D_{\infty}(P_{\left[X\right]^{n}\left[Y\right]^{n}}\|\pi_{\left[X\right]\left[Y\right]}^{n})\to0$
exponentially fast.

On the other hand, as shown in Remark \ref{rem:Gaussian}, for the
Gaussian source, 
\begin{align}
L_{\epsilon,n} & =\frac{\sqrt{n\left(1+\epsilon\right)}}{1-\rho}.
\end{align}
Similarly to \eqref{eq:-55}, for $\left(x^{n},y^{n}\right)\in\left[\mathcal{A}_{\epsilon}^{(n)}\right]\times\left[\mathcal{A}_{\epsilon}^{(n)}\right]\subseteq\mathcal{L}_{\epsilon,n}^{n}\times\mathcal{\mathcal{L}}_{\epsilon,n}^{n}$,
and $\left|x_{i}-\hat{x}_{i}\right|,\left|y_{i}-\hat{y}_{i}\right|\leq\Delta_{n},\forall i$,
we have 
\begin{align}
\frac{\pi_{XY}^{n}\left(x^{n},y^{n}\right)}{\pi_{XY}^{n}\left(\hat{x}^{n},\hat{y}^{n}\right)} & \leq\exp\left(n\Delta_{n}L_{\epsilon,n}\right).
\end{align}
Choose $\Delta_{n}=\frac{e^{-n\delta}}{n\frac{\sqrt{n\left(1+\epsilon\right)}}{1-\rho}}$
for some $\delta>0$, then 
\begin{align}
\left|\left[\mathcal{A}_{\epsilon}^{(n)}\right]\times\left[\mathcal{A}_{\epsilon}^{(n)}\right]\right| & =\left(\frac{\sqrt{n\left(1+\epsilon\right)}}{\Delta_{n}}\right)^{2n}\\
 & =\left(\frac{n^{2}\left(1+\epsilon\right)}{\left(1-\rho\right)e^{-n\delta}}\right)^{2n}\\
 & =e^{2n^{2}\delta+2n\log\frac{n^{2}\left(1+\epsilon\right)}{1-\rho}},
\end{align}
which grows much slower than doubly exponentially fast. Hence the
doubly exponential convergence of \eqref{eq:-98} is guaranteed.

Define $U^{n},V^{n}\sim\mathrm{Unif}\left(I_{\Delta_{n}}^{n}\right)$
with $I_{\Delta_{n}}^{n}=[0,\Delta_{n}]^{n}$ are mutually independent,
and also independent of $\left[X\right]^{n},\left[Y\right]^{n}$.
Then 
\begin{align}
 & e^{D_{\infty}(P_{\left[X\right]^{n}+U^{n},\left[Y\right]^{n}+V^{n}}\|\pi_{XY}^{n})}\nonumber \\
 & =\sup_{x^{n},y^{n}}\frac{P_{\left[X\right]^{n}+U^{n},\left[Y\right]^{n}+V^{n}}\left(x^{n},y^{n}\right)}{\pi_{\left[X\right]^{n}+U^{n},\left[Y\right]^{n}+V^{n}}^{n}\left(x^{n},y^{n}\right)}\nonumber \\
 & \qquad\times\frac{\pi_{\left[X\right]^{n}+U^{n},\left[Y\right]^{n}+V^{n}}^{n}\left(x^{n},y^{n}\right)}{\pi_{XY}^{n}\left(x^{n},y^{n}\right)}\\
 & \leq\sup_{x^{n},y^{n}}\frac{P_{\left[X\right]^{n}+U^{n},\left[Y\right]^{n}+V^{n}}\left(x^{n},y^{n}\right)}{\pi_{\left[X\right]^{n}+U^{n},\left[Y\right]^{n}+V^{n}}^{n}\left(x^{n},y^{n}\right)}\nonumber \\
 & \qquad\times\sup_{\left(x'^{n},y'^{n}\right)\in\left(\left[x\right]^{n}+I_{\Delta_{n}}^{n}\right)\times\left(\left[y\right]^{n}+I_{\Delta_{n}}^{n}\right)}\frac{\pi_{XY}^{n}\left(x'^{n},y'^{n}\right)}{\pi_{XY}^{n}\left(x'^{n},y'^{n}\right)}\\
 & \leq\exp\left(n\Delta_{n}L_{\epsilon,n}\right)e^{D_{\infty}(P_{\left[X\right]^{n}\left[Y\right]^{n}}\|\pi_{\left[X\right]\left[Y\right]}^{n})}
\end{align}
and hence 
\begin{align}
 & D_{\infty}(P_{\left[X\right]^{n}+U^{n},\left[Y\right]^{n}+V^{n}}\|\pi_{XY}^{n})\nonumber \\
 & \leq n\Delta_{n}L_{\epsilon,n}+D_{\infty}(P_{\left[X\right]^{n}\left[Y\right]^{n}}\|\pi_{\left[X\right]\left[Y\right]}^{n})\\
 & =e^{-n\delta}+D_{\infty}(P_{\left[X\right]^{n}\left[Y\right]^{n}}\|\pi_{\left[X\right]\left[Y\right]}^{n})\to0
\end{align}
exponentially fast.
\end{IEEEproof}

\section{\label{sec:exactrenyi}Proof of Proposition \ref{prop:exactRenyi}}

For $\alpha=0$ and $1$, by definition, one can easily obtain that
$G_{0}(\pi_{XY})=\log\mathrm{rank}^{+}(\pi_{XY})$ and $G_{1}(\pi_{XY})=G(\pi_{XY})$.
 Next we consider the case of $\alpha=\infty$. For this case, 
\begin{align}
H_{\infty}(W) & =-\log\max_{w}P_{W}(w).\label{eq:-40-1-1}
\end{align}
Hence 
\begin{align}
 & G_{\infty}(\pi_{XY})\\
 & =\min_{P_{W}P_{X|W}P_{Y|W}:P_{XY}=\pi_{XY}}-\log\max_{w}P_{W}(w)\\
 & =-\log\max_{P_{W}P_{X|W}P_{Y|W}:P_{XY}=\pi_{XY}}\max_{w}P_{W}(w)\\
 & =-\log\max_{w}\max_{P_{W}P_{X|W}P_{Y|W}:P_{XY}=\pi_{XY}}P_{W}(w)\\
 & \geq-\log\max_{w}\max_{\substack{P_{X|W}P_{Y|W}:\\
P_{W}(w)P_{X|W}(x|w)P_{Y|W}(y|w)\\
\le\pi_{XY}(x,y),\forall(x,y)
}
}P_{W}(w)\\
 & \geq\min_{w}\min_{P_{X|W=w},P_{Y|W=w}}D_{\infty}(P_{X|W=w}P_{Y|W=w}\|\pi_{XY})\\
 & \geq\min_{Q_{X},Q_{Y}}D_{\infty}(Q_{X}Q_{Y}\|\pi_{XY}).\label{eq:-111}
\end{align}
On the other hand, denote $\left(Q_{X}^{*},Q_{Y}^{*}\right)$ as an
optimal pair of distributions attaining the minimum in the optimization
problem $\min_{Q_{X},Q_{Y}}D_{\infty}(Q_{X}Q_{Y}\|\pi_{XY})$. Let
$\epsilon:=D_{\infty}(Q_{X}^{*}Q_{Y}^{*}\|\pi_{XY})$.  By Lemma
\ref{lem:mixturedecomposition}, we can decompose $\pi_{XY}$ as 
\begin{equation}
\pi_{XY}=e^{-\epsilon}Q_{X}^{*}Q_{Y}^{*}+\left(1-e^{-\epsilon}\right)\widehat{P},\label{eq:-100-1-1}
\end{equation}
where 
\begin{equation}
\widehat{P}:=\begin{cases}
\textrm{any distribution} & \epsilon=0\\
\frac{e^{\epsilon}Q_{X}^{*}Q_{Y}^{*}-\pi_{XY}}{e^{\epsilon}-1} & \epsilon\in(0,\infty)\\
Q_{X}^{*}Q_{Y}^{*} & \epsilon=\infty
\end{cases}.
\end{equation}
Then set $\mathcal{W}:=\left(\mathcal{X}\times\mathcal{Y}\right)\cup\left\{ w_{0}\right\} $
with some $w_{0}\notin\mathcal{X}\times\mathcal{Y}$, and choose
\begin{equation}
P_{W}(w):=\begin{cases}
e^{-\epsilon} & w=w_{0}\\
\left(1-e^{-\epsilon}\right)\widehat{P}(x',y') & w=\left(x',y'\right)\in\mathcal{X}\times\mathcal{Y}
\end{cases},
\end{equation}
and 
\begin{align}
P_{X|W}(x|w) & :=\begin{cases}
Q_{X}^{*} & w=w_{0}\\
1\left\{ x=x'\right\}  & w=\left(x',y'\right)\in\mathcal{X}\times\mathcal{Y}
\end{cases},\\
P_{Y|W}(y|w) & :=\begin{cases}
Q_{Y}^{*} & w=w_{0}\\
1\left\{ y=y'\right\}  & w=\left(x',y'\right)\in\mathcal{X}\times\mathcal{Y}
\end{cases}.
\end{align}
It is easy to verify that such a distribution $P_{W}P_{X|W}P_{Y|W}$
satisfies
\begin{align}
P_{XY} & =\pi_{XY},\\
H_{\infty}(W) & \leq\epsilon.
\end{align}
Hence 
\begin{equation}
G_{\infty}(\pi_{XY})\leq\min_{Q_{X},Q_{Y}}D_{\infty}(Q_{X}Q_{Y}\|\pi_{XY}).\label{eq:-112}
\end{equation}
Combining \eqref{eq:-111} and \eqref{eq:-112} we conclude that 
\begin{equation}
G_{\infty}(\pi_{XY})=\min_{Q_{X},Q_{Y}}D_{\infty}(Q_{X}Q_{Y}\|\pi_{XY}).\label{eq:-112-1}
\end{equation}

Now we claim 
\begin{equation}
G_{\infty}(\pi_{XY}^{n})=nG_{\infty}(\pi_{XY}).
\end{equation}
This follows since on one hand, by choosing $Q_{X^{n}},Q_{Y^{n}}$
as product distributions, we have 
\begin{equation}
G_{\infty}(\pi_{XY}^{n})\leq nG_{\infty}(\pi_{XY}).
\end{equation}
On the other hand, 
\begin{align}
 & D_{\infty}(Q_{X^{n}}Q_{Y^{n}}\|\pi_{XY}^{n})\nonumber \\
 & =\max_{x^{n},y^{n}}\log\frac{Q_{X^{n}}\left(x^{n}\right)Q_{Y^{n}}\left(y^{n}\right)}{\pi_{XY}^{n}\left(x^{n},y^{n}\right)}\\
 & =\max_{x^{n},y^{n}}\sum_{i=1}^{n}\log\frac{Q_{X_{i}|X^{i-1}}\left(x_{i}|x^{i-1}\right)Q_{Y_{i}|Y^{i-1}}\left(y_{i}|y^{i-1}\right)}{\pi_{XY}\left(x_{i},y_{i}\right)}\\
 & =\max_{x^{n-1},y^{n-1}}\Biggl(\sum_{i=1}^{n-1}\log\frac{Q_{X_{i}|X^{i-1}}\left(x_{i}|x^{i-1}\right)Q_{Y_{i}|Y^{i-1}}\left(y_{i}|y^{i-1}\right)}{\pi_{XY}\left(x_{i},y_{i}\right)}\nonumber \\
 & \qquad+\max_{x_{n},y_{n}}\log\frac{Q_{X_{n}|X^{n-1}}\left(x_{i}|x^{i-1}\right)Q_{Y_{i}|Y^{i-1}}\left(y_{i}|y^{i-1}\right)}{\pi_{XY}\left(x_{i},y_{i}\right)}\Biggr)\\
 & \geq\max_{x^{n-1},y^{n-1}}\Biggl(\sum_{i=1}^{n-1}\log\frac{Q_{X_{i}|X^{i-1}}\left(x_{i}|x^{i-1}\right)Q_{Y_{i}|Y^{i-1}}\left(y_{i}|y^{i-1}\right)}{\pi_{XY}\left(x_{i},y_{i}\right)}\nonumber \\
 & \qquad+G_{\infty}(\pi_{XY})\Biggr)\\
 & \geq\max_{x^{n-2},y^{n-2}}\Biggl(\sum_{i=1}^{n-2}\log\frac{Q_{X_{i}|X^{i-1}}\left(x_{i}|x^{i-1}\right)Q_{Y_{i}|Y^{i-1}}\left(y_{i}|y^{i-1}\right)}{\pi_{XY}\left(x_{i},y_{i}\right)}\nonumber \\
 & \qquad+2G_{\infty}(\pi_{XY})\Biggr)\\
 & \vdots\nonumber \\
 & \geq nG_{\infty}(\pi_{XY}).
\end{align}

Therefore,
\begin{align}
T_{\mathrm{Exact}}^{(\infty)}(\pi_{XY}) & =\lim_{n\to\infty}\frac{1}{n}\min_{Q_{X^{n}},Q_{Y^{n}}}G_{\infty}(\pi_{XY}^{n})\\
 & =G_{\infty}(\pi_{XY}).
\end{align}

\subsection*{Acknowledgements}
The authors would like to thank the Associate Editor Prof.\  Maxim Raginsky and reviewers for their extensive, constructive and helpful feedback to improve the manuscript.

\bibliographystyle{unsrt}
\bibliography{ref}

\begin{IEEEbiographynophoto}{Lei Yu} received the B.E. and Ph.D. degrees, both in electronic engineering, from University of Science and Technology of China (USTC) in 2010 and 2015, respectively. From 2015 to 2017, he was a postdoctoral researcher at the Department of Electronic Engineering and Information Science (EEIS), USTC. From 2017 to 2019, he was a research fellow at the Department of Electrical and Computer Engineering, National University of Singapore. Currently, he is a postdoc at the Department of Electrical Engineering and Computer Sciences, University of California, Berkeley. His research interests lie in the intersection of information theory, probability theory, and combinatorics.  \end{IEEEbiographynophoto}

\begin{IEEEbiographynophoto}{Vincent Y.\ F.\ Tan} (S'07-M'11-SM'15) was born in Singapore in 1981. He is currently a Dean's Chair Associate Professor in the Department of Electrical and Computer Engineering and the Department of Mathematics at the National University of Singapore (NUS). He received the B.A.\ and M.Eng.\ degrees in Electrical and Information Sciences from Cambridge University in 2005 and the Ph.D.\ degree in Electrical Engineering and Computer Science (EECS) from the Massachusetts Institute of Technology (MIT) in 2011. His research interests include information theory, machine learning, and statistical signal processing.
Dr.\ Tan received the MIT EECS Jin-Au Kong outstanding doctoral thesis prize in 2011, the NUS Young Investigator Award in 2014, the NUS Engineering Young Researcher Award in 2018, and the Singapore National Research Foundation (NRF) Fellowship (Class of 2018). He is also an IEEE Information Theory Society Distinguished Lecturer for 2018/9. He has authored a research monograph on {\em ``Asymptotic Estimates in Information Theory with Non-Vanishing Error Probabilities''} in the Foundations and Trends in Communications and Information Theory Series (NOW Publishers). He is currently serving as an Associate Editor of the IEEE Transactions on Signal Processing.  \end{IEEEbiographynophoto} 
\end{document}